\documentclass[trackchanges, twocolumn]{aastex701}
\usepackage{amsmath}

\newcommand\latex{La\TeX}

\begin{document}

\title{Discovery of 2~km ultra-fast rotating asteroid in Rubin DP2 catalog via varying Fourier order method.}

\author[orcid=0009-0007-1972-5975]{Dmitrii E. Vavilov}
\affiliation{DiRAC Institute and the Department of Astronomy, University of Washington, 3910 15th Ave NE, Seattle, WA, 98195, U.S.A.}
\affiliation{Institute of Applied Astronomy of Russian Academy of Sciences, Kutuzova emb. 10, St. Petersburg, 191187, Russia}
\email[show]{vavilov@uw.edu; vavilov@iaaas.ru}

\author[orcid=0000-0002-1398-6302]{Siegfried Eggl}
\affiliation{The Grainger College of Engineering, Department of Aerospace Engineering, University of Illinois Urbana-Champaign, Urbana, U.S.A.}
\affiliation{Department of Astronomy, University of Illinois Urbana-Champaign, Urbana, IL 61801, U.S.A.}
\email[show]{eggl@illinois.edu}

\author[orcid=0000-0002-4439-1539]{Sarah Greenstreet}
\affiliation{Rubin Observatory/NSF NOIRLab, 950 N. Cherry Ave, Tucson, AZ, 85719, U.S.A.}
\affiliation{DiRAC Institute and the Department of Astronomy, University of Washington, 3910 15th Ave NE, Seattle, WA, 98195, U.S.A.}
\email[show]{sarah.greenstreet@noirlab.edu}

\begin{abstract}
Time-resolved photometry from large surveys can yield rotation periods for
huge numbers of asteroids, but only if the reliability of each period is
assessed automatically, since sparse and irregular sampling produces many
false periodogram minima. We present an improved method that determines asteroid
rotation periods with a Fourier series whose order is chosen for each object
individually. This has substantial advantages over lower order methods, especially when unique solutions for the spin period cannot be reliably determined. In such cases, the method reports the set of alternative
periods together with the corresponding probability estimates.

We apply the improved spin period determination method to the NSF-DOE Vera C. Rubin Observatory Data Preview 2 (DP2) catalog of solar system objects and
determine reliable periods for a large sample of asteroids. Among them we find
66 super-fast rotators, which spin faster than the classical $2.2$~h spin
barrier. Two of
them are remarkable: asteroids 491384 and 566130 are both larger than 2~km, yet they
complete a full rotation in only about five minutes. To survive such a fast spin, they must have
significant internal cohesion. At a diameter of about $2.5$~km, asteroid 491384 is the largest ultra-fast rotator known so far.

\end{abstract}




\section{Introduction}
\label{sec:intro}

Time-resolved photometry is one of the simplest and yet most powerful tools 
available for the physical study of asteroids. As a non-spherical body rotates, 
the sunlight scattered from its surface produces a periodic variation of 
brightness -- a lightcurve -- from which the rotation period and, through the 
amplitude, a first estimate of the elongation of the body can be recovered. 
The overwhelming majority of the roughly thirty thousand rotation periods 
compiled in the Asteroid Lightcurve Database \citep[LCDB;][]{Warner2009Icarus} 
have been obtained in this way. The distribution of these periods carries a 
sharp imprint of the internal structure of asteroids: bodies larger than 
$\approx0.15$~km almost never rotate faster than $\approx2.2$~hours 
\citep{harris1996, pravecharris2000, 2007Icar..190..250P, Carbognani_2017}. 
This ``spin barrier'' is naturally explained if such asteroids are 
gravitationally bound ``rubble piles'' with negligible tensile strength, which 
would be disrupted by centrifugal forces at faster spins. Objects that lie 
beyond the barrier therefore require an additional explanation -- a monolithic 
structure, or, more commonly invoked for larger bodies, a non-zero cohesive 
strength provided by van der Waals forces between regolith grains 
\citep{holsapple2007, sanchez2014strength}. The spin state is not static: the 
Yarkovsky--O'Keefe--Radzievskii--Paddack (YORP) effect \citep{1969JGR....74.4379P} 
can gradually spin small asteroids up or down, so that the position of an object 
relative to the spin barrier, taken together with its lightcurve amplitude, 
becomes a diagnostic of its density, cohesion, and collisional history.


For a long time such studies were restricted to relatively bright, individually 
targeted objects. This situation has changed with the arrival of wide-field 
time-domain surveys, which allow to derive rotation periods for large, 
statistically meaningful samples. The Sloan Digital Sky Survey 
\citep{2000AJ....120.1579Y}, the Palomar Transient Factory \citep[PTF;][]{waszczak2015}, the Zwicky Transient Facility, the 
Asteroid Terrestrial-impact Last Alert System \citep[ATLAS;][]{erasmus2020}, 
the {\it WISE}/NEOWISE mission \citep{2011ApJ...741...68M, 2014ApJ...792...30M}, 
the {\it TESS} mission \citep{pal2020}, and the {\it Gaia} mission 
\citep{2023A_26A...674A..12T} have together provided photometry, colors, and 
rotational information for many thousands of asteroids, and have begun to 
populate both extremes of the period distribution -- from super-slow rotators 
with periods of hundreds of hours to the fast-rotating tail.
It is precisely 
these surveys that revealed the first large super-fast rotators (SFRs): 
sub-kilometer to kilometer-sized bodies spinning faster than the barrier, such 
as the $\approx0.7$~km near-Earth asteroid 2001~OE$_{84}$ 
\citep{pravec2002} and a small number of main-belt objects subsequently 
confirmed with dedicated campaigns \citep{chang2014, chang2016}. Their existence 
is commonly interpreted as evidence for internal cohesion 
\citep{holsapple2007, sanchez2014strength}. Nevertheless, the number of such 
objects known remains small, and, importantly, the very rapid rotators found so 
far are almost exclusively near-Earth objects; the faint, sub-km main-belt 
regime, where dense sampling and sufficient depth are simultaneously required, 
has stayed only sparsely explored \citep{2020MNRAS.495.3990M, 2021MNRAS.502.5277H, 
2024AJ....168..184S}.

The NSF--DOE Vera C. Rubin Observatory \citep{ivezic2019} is expected to 
transform this picture. Over the next decade its Legacy Survey of Space and Time 
(LSST), which began in mid-2026, will deliver deep, multi-band 
photometry for up to five million small bodies of the Solar System, the vast 
majority of them main-belt asteroids \citep{kurlander2025tno}. A first indication 
of how powerful these data will be for rotational studies has already come from 
the commissioning phase. Using the $\sim2{,}100$ asteroids, $\sim1{,}500$ of which were new discoveries, that Rubin detected
in the ``Rubin First Look'' (RFL) commissioning field near M49 -- imaged over nine 
nights in April-May 2025 with an irregular, commissioning-driven cadence that nonetheless 
provided dense, multi-band sampling -- \citet{greenstreet2026} determined 
reliable rotation periods and colors for 76 objects, with periods spanning 
$0.031$--$21.3$~hours. Remarkably, 19 of these objects were found to rotate 
faster than the $2.2$~hr spin barrier, and three of them turned out to be 
``ultra-fast rotators'' with periods shorter than five minutes, the fastest 
completing one rotation in only $\simeq1.9$~minutes. With diameters $d>0.5$~km 
for most of them, these main-belt bodies would require cohesive strengths 
comparable to those of clay or rock to avoid disruption, and asteroid 2025~MN45 is now 
the fastest-spinning asteroid with $H<20$ ever reported. 

The main challenge of such surveys is, however, not the fitting technique itself but the assessment of the reliability of the obtained solutions. The sparse and irregular sampling produces many false periodogram minima: diurnal aliases, harmonics of the true period, and deep spurious solutions of high-order models in the poorly covered parts of the light curve. In \citet{greenstreet2026} the final decision about the reliability of every solution was evaluated by a visual inspection of the light curves and periodograms produced by two independent methods, and only 76 objects out of the more than 2000 asteroids with a sufficient number of observations were classified as reliable. Such a manual approach is clearly not scalable: LSST is expected to provide light curves for millions of asteroids, and even the ongoing data releases will contain hundreds of thousands of objects. An automated procedure is needed that not only finds the best-fitting period, but also decides how trustworthy this period is and reports the possible alternative solutions.

In this work we present such a procedure. We substantially develop the
model selection of the pipeline of \citet{2025A&A...693A..66V}: the number
of Fourier harmonics and the choice between the periodogram minima are now
determined by Fisher and Vuong statistical tests adapted to the specific
properties of the light-curve fitting, the alternative periods are reported
together with the probabilities with which they can compete with the
adopted solution, and the reliability of the result is quantified by a set
of automatically computed criteria.

We apply this method to the Rubin Observatory Data Preview 2 (DP2) catalog \citep{10.71929/rubin/3382528, RTN-115},
selecting all asteroids with more than 80 photometric measurements, which
yields a sample of more than 9,000 objects. DP2 intersects with the Rubin First
Look (RFL) field near M49, for which we 
already reported reliable rotation periods and colors for 76 objects among
$\sim$~2,100 candidates \citep{greenstreet2026}. We reprocess the same asteroids as part of our larger
DP2 sample for two reasons. First, DP2 now provides a number of additional
observations for objects in this field beyond what was available to
\citet{greenstreet2026}. Second, since the RFL sample already contains a
set of periods vetted by an independent analysis, reprocessing it with our
substantially revised method lets us check that the pipeline still
recovers the objects already established as reliable, in addition to
extending the analysis to a much larger and more homogeneous data set.

The paper is organized as follows. Section~\ref{sec:methods} fully describes the method of the period determination by Fourier method with varying order, the way to choose the best model and credibility of the result; 
Section~\ref{sec:results} presents and discusses the results, where in Section~\ref{sec:fastrot} we present the results for 66 found super-fast rotators, and a discovery of a larger than 2~km in size ultra-fast rotator, in Section~\ref{sec:order_need} we show the importance of the proposed technique with a variable Fourier order and in Section~\ref{sec:baddata} we discuss flagging the problematic visits of Rubin observatory.  
Section~\ref{sec:conclusions} summarizes our conclusions.

\section{Method}
\label{sec:methods}

We present an improved rotation period determination method based on \citet{2025A&A...693A..66V} and utilized in \citet{greenstreet2026}, which now can automatically decide about the credibility of the result. Below we describe the method, paying the main attention to the model selection procedure, which was substantially developed for the present work: the number of Fourier harmonics and the possible alternative periods are now determined by the combination of Fisher and Vuong statistical tests.

\subsection{Distance and light-speed corrections}

The observed magnitude of an asteroid depends not only on its rotation, but also on the positions of the asteroid, the Sun and the observer. Therefore, before the period search, the observations are reduced to the standard geometry. The reduced magnitude is
\begin{equation}
m_{\rm red}=m_{\rm obs}-5\log_{10}\big(r\,\Delta\big), \label{eq:reduction}
\end{equation}
where $m_{\rm obs}$ is the observed magnitude and $r$ and $\Delta$ are the heliocentric and topocentric distances of the asteroid at the observation epoch, computed with the JPL Horizons API\footnote{https://ssd.jpl.nasa.gov/horizons}. The observation epochs are corrected for the light-travel time, $t=t_{\rm obs}-\Delta/c$, i.e. each observation is referred to the moment when the light left the asteroid; without this correction the rotation phases of the observations taken at different distances would be systematically shifted.

The reduced magnitudes are then modeled as
\begin{equation}
\begin{split}
m_{\rm red}(t)={}&\sum_{k=1}^{n}\big[a_k\cos(2\pi k f t)+b_k\sin(2\pi k f t)\big] +\\
&+ M_b+\Phi(\alpha(t)),
\end{split}
\label{eq:model}
\end{equation}
where $M_b$ is the mean absolute magnitude of the asteroid in the photometric band $b$, $\Phi(\alpha)$ is the phase function describing the dependence of the brightness on the solar phase angle $\alpha$, $f$ is the rotation frequency, $n$ is the order of the Fourier series, and $a_k$, $b_k$ are the Fourier amplitudes. The mean magnitudes $M_b$ are fitted separately for each photometric band, which absorbs the color of the object, so the observations of all bands are used together in one fit.

\subsection{Least-squares fit}

The model (\ref{eq:model}) is linear in all $\{a_i, b_i\, M_j\}$ parameters. For a fixed $f$ the parameters are found by weighted least squares with the weights $1/\sigma_i^2$, where $\sigma_i$ are the reported photometric uncertainties. The frequency is scanned over a uniform grid starting from $2/T$, where $T$ is the total time span of the observations; the number of grid points is set to $50\,T(f_{\max}-f_{\min})$, i.e. fifty frequencies per peak width of the periodogram. The upper limit of the grid is constructed following \citet{2025A&A...693A..66V}. First, the maximal frequency $f_{Ny}$ that can be probed with the given observation times is computed as an analogue of the Nyquist limit adapted to irregular sampling \citep{1999A&AS..135....1E}: if $p$ is the largest value such that every observation epoch can be written as $t_i=t_0+n_i p$ with integer $n_i$ (we allow a 2\% tolerance for this equality), then $f_{Ny}=1/(2p)$. For our data $f_{Ny}$ is typically close to 2000 cycles per day. Second, the limit is reduced for the higher Fourier orders: the order-$n$ series contains the term $\cos(2\pi f n t)$, whose own frequency is $nf$, so the highest harmonic must itself stay below the Nyquist limit, and the fundamental frequency of the order-$n$ model is scanned up to $\min(f_{Ny}/n,\ f_{\rm lim})$, where $f_{\rm lim}=1000$ cycles per day is an additional cut we impose: it corresponds to the rotation period of 86 seconds. This is done for every Fourier order $n$ from 1 to $n_{\max}=8$.

For each trial fit the pipeline rejects outliers iteratively. First, the standard deviation of the residuals is estimated from the fit itself. Then all observations whose residuals exceed $\theta=3$ such standard deviations are rejected, and the model is fitted again; this is repeated until no new points are rejected. 

We also estimate the number of local maxima of the fit, and if this number is
one we double the period. In practice we compute the local
maxima of the derivative of the fit rather than of the fit itself, since a light
curve can have a single maximum while its derivative has two. The reason for this
choice is explained briefly in Sect.~\ref{sec:results} and in more detail in a
forthcoming paper.

\subsection{Phase function}
\label{sec:phase}

The parameters of the phase function $\Phi(\alpha(t))$, which describes the change of the brightness with the solar phase angle $\alpha$, are fitted together with the Fourier amplitudes $\{a_k, b_k\}$ and the mean magnitudes $M_b$ for every trial frequency. The pipeline offers a choice of the phase-function model: the classical $H,G$ magnitude system \citep{1989aste.conf..524B}, the three-parameter $H,G_1,G_2$ system and its two-parameter version $H,G_{12}$ \citep{2010Icar..209..542M}, or a polynomial of the phase angle of an arbitrary degree (quadratic by default). For any of these models, a priori constraints can be imposed on the parameters: a constraint is a pair of the expected value and its standard deviation, which enters the least-squares system as one more pseudo-observation.

The models differ in speed and in the physical meaning of the result. The polynomial is the fastest option: the model stays fully linear in all parameters, so one least-squares solution per trial frequency is enough. The $H,G$, $H,G_1,G_2$ and $H,G_{12}$ functions are not linear in their slope parameters, and the fit requires several iterations to converge, which multiplies the cost of the frequency scan accordingly. On the other hand, the polynomial coefficients have no direct physical interpretation: this option does not provide the absolute magnitudes, and it cannot be extrapolated safely when the phase angles of the observations change significantly. The choice is therefore a trade-off between the speed and the physical content of the phase-function parameters.

In the present work we use the $H,G$ system. The reason for the change is that the phase angle of an object varies only by a few degrees within our observation windows. On such a short arc the slope parameter $G$ is constrained by the data very weakly and can degenerate with the mean magnitudes $M_b$. To keep the solution stable, we impose the a priori constraint described above: the expected value $G=0.15$, which is the typical slope of asteroids, with the standard deviation $0.05$. If the data really constrain the slope, the fitted $G$ moves away from 0.15; if they do not, the solution simply stays near this typical value instead of becoming degenerate, and the absolute magnitudes remain well defined.

\subsection{Model selection for the same order}

As a measure of the fit quality we use the weighted standard deviation of the residuals,
\begin{equation}
\sigma=\sqrt{\frac{\sum_i r_i^2/\sigma_i^2}{\sum_i 1/\sigma_i^2}\cdot\frac{N_u}{N_u-p}}, \label{eq:sigma}
\end{equation}
where $r_i$ are the residuals of the $N_u$ observations used in the fit and $p$ is the total number of free parameters of the model. This quantity has the units of magnitude and directly reflects the mean quality of the fit and of the observations. The periodograms shown in this work are the plots of $\sigma$ as a function of the trial frequency.

As opposed to \citet{2025A&A...693A..66V}, in this method we do not use $\sigma$ to choose the best model within one order. Instead, the best trial frequency is selected by the Bayesian information criterion, which additionally penalizes the models that reject many observations. We compute it as
\begin{equation}
\mathrm{BIC}=N_u\ln\!\big(\chi^2/N_u\big)+R+p\ln N, \label{eq:bic}
\end{equation}
where $\chi^2=\sum_i (r_i/\sigma_i)^2$ is the weighted sum of squared residuals over the $N_u$ used observations, $N=N_u+N_r$ is the total number of observations with $N_r$ of them rejected, $p$ is the number of free parameters, and $R$ is the rejection penalty defined below.

The first term is the Gaussian log-likelihood written for the case when the common scale of the uncertainties is unknown and is estimated from the fit; in this form the criterion does not change when all $\sigma_i$ are multiplied by a constant. The third term is the standard complexity penalty \citep{2011ApJS..197...31S}.

The rejection penalty $R$ is the price a model pays for discarding observations, and it is the main modification with respect to a plain information criterion. Its purpose is twofold. First, without any such term a model could lower its BIC simply by rejecting the observations that do not fit it; a penalty is needed so that the rejection of a normal point never pays off, while the rejection of a genuine outlier still does. A single rejected point deviates from the model by more than $\theta$ standard deviations, so, if it were kept, it would add at least $\theta^2$ to $\chi^2$; the natural price for removing it is therefore $\theta^2$, and for $N_r$ independent rejections the penalty would be $\theta^2 N_r$.

Second, the rejected observations of asteroid light curves are, however, often not independent. A single cause -- a background star entering the photometric aperture, or a short calibration problem in one exposure series -- typically spoils several consecutive measurements in the same band at once. Charging $\theta^2$ for each of them would overpenalize a solution that correctly identifies such a contaminated stretch as a single event. We therefore group the rejected points into clusters: within a given photometric band, all rejected observations that fall inside a time window of $\Delta t_{\rm cl}$ (we use 1~hour) of the first point of the cluster are treated as one event. The penalty of a cluster of $k$ rejected points grows linearly for small clusters and only logarithmically for large ones,
\begin{equation}
R = \theta^2 \sum_{\rm clusters}
\begin{cases}
k, & k \le k_0,\\[2pt]
k_0 + \alpha\,\ln(k/k_0), & k > k_0,
\end{cases}
\label{eq:rejpenalty}
\end{equation}
where the sum runs over all clusters and $k$ is the number of rejected points in each. For $k\le k_0$ (we use $k_0=4$) the penalty is the full $\theta^2$ per point, so that removing a few scattered points is charged in the same way as before; for larger clusters the marginal cost of each additional rejected point decreases, reflecting the fact that a long contaminated stretch carries the information of essentially one event rather than of $k$ independent ones. The parameter $\alpha$ (we use $\alpha=k_0$) sets the slope of this logarithmic regime.

This penalty is also the reason why $\sigma$ itself is not used for the selection: two models that rejected different numbers of observations have their $\sigma$ computed on different data sets, and the model that rejected more points would obtain a smaller $\sigma$ almost for free. The term $R$ restores the comparability between models with different rejections.

We additionally reject the solutions whose fitted amplitude exceeds twice the difference between the maximal and minimal observed magnitude, since such a large amplitude in a poorly sampled phase range is a sign of a spurious high-order feature rather than a real signal.

\subsection{Model selection among different orders}
\label{subsec:diff_order_selection}

We do not use the BIC to compare models with different numbers of Fourier harmonics; however, we still provide its values for reference, as an additional diagnostic of the solution.
The best models of different orders are  compared with each other by statistical tests in pairs, and here two situations must be distinguished. If the best-fitting frequencies of the two orders coincide, the lower-order model is a special case of the higher-order one: it is enough to set the extra amplitudes to zero. Such models are called nested, and for them we apply the F-test of \citet{1989Icar...77..171H},

\begin{equation}
F=\frac{(\chi^2_{l}-\chi^2_{h})/(p_h-p_l)}{\chi^2_{h}/(N_u-p_h)}, \label{eq:ftest}
\end{equation}
where the indices $l$ and $h$ refer to the lower- and the higher-order
model. Rather than restricting the comparison to the observations
accepted by both models, we compute $\chi^2_l$ and $\chi^2_h$ over the
union of the two models' accepted sets, of size $N_u$. For a point
accepted by a given model, its actual normalized residual contributes to
that model's $\chi^2$ as usual; for a point that this same model rejects,
but the other model keeps, we substitute the same minimal rejection cost
already used in Eq.~(\ref{eq:bic}), $\theta^2$, but here rescaled by the
model's own reduced $\chi^2$, $\hat s_l^2$ or $\hat s_h^2$, so that the
substituted cost reflects the true noise level of that particular fit
rather than the nominal, reported uncertainties. This construction charges
each model fairly for the points it discards, instead of simply dropping
them from the comparison, consistently with how rejected points are
already priced in the single-order selection.

If the best-fitting frequencies are different, the harmonics of one frequency cannot be expressed through the harmonics of the other, so neither model is a special case of the other one. Such models are not nested, and the F-test is not applicable to them; for example, the $\chi^2$ of the model with more parameters can then be larger than the $\chi^2$ of the simpler model, which is impossible for nested models. For such pairs we use the test of \citet{Vuong1989}, which is designed exactly for non-nested models. Two frequencies are considered coinciding when the difference between the two model curves, accumulated over the whole time span because of the frequency difference, stays below the noise level; this gives the tolerance $|f_1-f_2|<\sigma/(2\pi A T)$, where $A$ is the light-curve semi-amplitude estimated from the second order fit. The tolerance is never smaller than 3 grid steps and never larger than $0.2/T$.

The idea of the Vuong test is to compare the two models point by point instead of comparing their total $\chi^2$. For every observation $i$ the difference $d_i=\big[r_i^2(B)-r_i^2(A)\big]/2$ shows how much better model $A$ describes this particular point than model $B$. The test statistic is
\begin{equation}
Z=\frac{\sum_i d_i - C/2}{\sqrt{m}\,s_d},\qquad P(A\ \mathrm{over}\ B)=\Phi(Z), \label{eq:vuong}
\end{equation}
where $m$ is the number of observations accepted by both models, $s_d$ is the standard deviation of the individual $d_i$, $\Phi$ is the normal cumulative distribution function, and $C$ is the correction term described below. In words: model $A$ is preferred only if its total advantage is larger than the natural scatter of the point-wise advantages, so the advantage must be consistent over the data and not produced by a few points. No assumption about nesting is needed here.

The correction term $C$ combines two separate effects, each expressed in
the units of $d_i$, that is, in units of half a $\chi^2$ contribution.
The first, $C_{\rm BIC}$, is standard: a model with more free parameters
improves its $\chi^2$ even on pure noise, by about one unit per extra
parameter, and this false advantage is removed by the term
$\Delta p\ln m$ proposed already by \citet{Vuong1989}. The
second, $C_{\rm rej}$, is specific to our pipeline: the two models may
reject different observations, and, as in the single-order selection,
this must be paid for.

To construct $C_{\rm rej}$ we reuse the same rejection penalty $R$ defined
in Eq.~(\ref{eq:rejpenalty}), but a point rejected by \emph{both} models is
a consensus outlier: it carries no information about which of the two
models is preferable, since neither model is penalized relative to the
other for discarding it. We therefore first remove from the comparison
every point rejected by both models, leaving for each model only the
points it rejects exclusively; $R_A$ and $R_B$ are the clustered rejection
penalty of Eq.~(\ref{eq:rejpenalty}) evaluated on these exclusive rejects
for models $A$ and $B$, respectively.

Both $C_{\rm BIC}$ and $R_A$, $R_B$ are defined in units of the
fit-estimated standard deviations, through the threshold $\theta$; these
units differ from the units of $d_i$ by the true error scale, which we do
not know a priori. We rescale $C_{\rm BIC}$ by $\hat s^2=\min(\hat
s_A^2,\hat s_B^2)$, the smaller (better) of the two models' own reduced
$\chi^2$, since this term does not distinguish between the models. The
rejection penalties, however, are rescaled individually: $R_A$ by model
$A$'s own reduced $\chi^2$, $\hat s_A^2$, and $R_B$ by model $B$'s own,
$\hat s_B^2$, since the cost of a model's rejected points should be set by
that model's own noise level rather than by a value borrowed from its
competitor. The reason rescaling is needed at all is the following: the
rejection threshold is set in the standard deviations estimated from the
fit, so if, for example, the reported uncertainties are two times smaller
than the real ones, the real cost of a rejected point in the units of
$d_i$ is four times larger. Combining the two rescaled terms,
\begin{equation}
C=\Delta p\,\ln m\,\hat s^{2}
\;+\; \Big(\hat s_A^{2}R_A-\hat s_B^{2}R_B\Big). \label{eq:corr}
\end{equation}

The test above assumes that the two models share no frequency content,
which holds whenever the harmonics of one model do not coincide, even
partially, with those of the other. This is not always the case: two
models can share some, but not all, of their frequency
content -- for instance when the higher harmonics of one solution happen
to match those of the other while the lower ones do not -- without either
model being a special case of the other. \citet{Vuong1989} calls such a
pair \emph{overlapping} models, as distinct from the nested case (handled
by the F-test) and the strictly non-nested case with no shared frequency
content (handled by the point-wise test above); for overlapping models,
Vuong shows that the log-likelihood-ratio statistic is, under the null
hypothesis, no longer normally distributed: its asymptotic distribution is
instead a weighted difference of two independent $\chi^2$ variables, with
degrees of freedom equal to the number of frequency components unique to
each model, i.e. not shared between the two.

We determine the number of unique components on each side, $p$ and $q$,
by comparing the two fitted frequency sets within a small tolerance, and
form the bias-corrected $\chi^2$ difference
\begin{equation}
\mathrm{LR}=\frac{(\chi^2_A-\chi^2_B)+C_{rej}}{\hat s^2},
\label{eq:lr_overlap}
\end{equation}
using the same correction $C_{rej}$ as in Eq.~(\ref{eq:corr}) -- the
parameter-count term and the clustered rejection penalty. Under the null hypothesis $\mathrm{LR}$ is distributed
as $\chi^2_q-\chi^2_p$, and we obtain $P(A\ \mathrm{over}\ B)$ from the
corresponding tail probability of this distribution. If the two models
share their entire frequency content ($p=q=0$), or if too few points are
accepted by both models relative to their number of free parameters to
constrain the comparison, we do not attempt to discriminate between them
and set $P=0.5$.

From these pairwise probabilities we form the selection matrix: the pairs with coinciding frequencies are judged by the F-test, all other pairs by the Vuong tests. The adopted model is the model of the lowest order $n\ge2$ that no higher-order model beats with the probability above the threshold. The threshold depends on the difference $g$ between the orders of the two models: $p_{\rm thr}=\min\!\big[0.99,\,0.95+0.01(g-1)\big]$. For the neighbouring orders it equals the conventional 95\% level of the sequential F-test \citep{1989Icar...77..171H}; for the distant pairs it grows to 99\%, because a model with many extra parameters can find a deep minimum somewhere on the frequency grid just by chance, and a stronger evidence is required from it.

\subsection{Uncertainty of the result}
\label{sec:uncertainty}

The adopted solution is not always unique, and the pipeline reports the alternative periods together with a quantitative measure of their plausibility. This is done in two steps: first the alternatives of the same Fourier order as the adopted solution are identified and used to define the uncertainty interval of the period, then the alternatives coming from the other orders are examined.

Within the adopted order, a minimum of the periodogram is taken as a candidate if its BIC exceeds the global minimum of this order by less than $\Delta\mathrm{BIC_c}=20$, which is a more conservative than the standard boundary of the ``very strong evidence'' in favour of the best model \citep{KassRaftery1995}. If the number of such candidates exceeds seven, we consider the fit itself unreliable: with so many comparable minima the periodogram does not single out one solution, and we do not attempt to identify individual alternatives. In this case the reported period is only the interval between the smallest and the largest period among the candidates, and no other information is given.

If there are seven or fewer candidates, the uncertainty interval of the period is defined as the range between the smallest and the largest period among them for which the BIC does not exceed the global minimum by more than $\Delta\mathrm{BIC_c}$. The remaining candidates, that is, those lying outside this interval, are treated as separate alternative solutions and are compared with the adopted solution one by one using the same Vuong test (Sect.~\ref{subsec:diff_order_selection}); the F-test cannot be used here, because the two periods correspond to different frequencies by construction. If the adopted solution beats a candidate with a probability above 0.95, the candidate is discarded; otherwise it is kept and reported together with its probability, so that the reliability of every alternative can be judged directly.


When an alternative comes from a Fourier order different from that of the adopted solution, comparing it directly with the adopted model would be unfair: the two models are then allowed a different number of harmonics, and the
alternative may win simply because its order is more flexible, not because its frequency is better. To remove this asymmetry, we refit the adopted frequency at every Fourier order up to that of the alternative (only the orders whose
highest harmonic still fits below the Nyquist limit are used), duel the alternative against each of these refits by the Vuong test, and keep the outcome most favourable to the adopted frequency. In this way the frequencies, rather than the orders, are compared.

In addition to the tests described above, we also compute the sigma-based significance level of \citet{2025A&A...693A..66V}, converted to the same BIC-based scale,
\begin{equation}
\sigma_{\rm thr}=\sigma_{\min}\exp\!\left(\frac{\Delta\mathrm{BIC_c}}{2\,N_u}\right),
\label{eq:sigthr}
\end{equation}
where $\sigma_{\min}$ is the standard deviation of the adopted solution (see Equation~(\ref{eq:sigma})) and $N_u$ is the number of observations used in its fit. This threshold is not used to identify the alternative periods, since it does not properly account for the different number of rejected observations at different frequencies; nevertheless, it remains a reasonably accurate estimate and is convenient for plotting on the periodograms shown throughout this paper.

\subsection{Phase coverage}
\label{sec:coverage}

An additional way in which a light-curve fit can be misleading is through gaps in the phase coverage. High-order models can produce large-amplitude, unphysical features in the parts of the phase curve where no observations constrain the fit; conversely, an apparently good fit can rely on a single passage through the light curve, with no independent confirmation from a different rotation cycle. To detect both situations, we compute the phase coverage of every adopted solution using the observations retained by the fit.

We slide a window of width $w=0.07$ in rotation phase across the light curve in steps of $1/100$ of the period, so that 100 overlapping windows cover the full cycle. For every window we count the number of observations it contains, $N_w$, and the number of independent rotation cycles they belong to, $N_c$; two observations are considered to belong to different cycles if their absolute times differ by more than half the period. A window is called \emph{covered} if $N_w\ge3$ and \emph{confirmed} if $N_c\ge2$.

From these window counts we define three coverage quantities. The \emph{light-curve coverage} is the fraction of the rotation phase that lies in at least one covered window, computed as one minus the total length of the union of all uncovered windows (overlapping uncovered windows are merged before their lengths are summed, so that no part of the phase is counted twice). The \emph{group coverage} is simply the fraction of the 100 sliding windows that are confirmed. Finally, we report the length of the longest contiguous arc of the phase that is not covered by any window, which  (unlike the light-curve coverage alone) distinguishes a light curve with several small gaps from one with a single large gap, the latter being far more prone to spurious high-order fitting features.

Since the observations are not evenly distributed in phase, even a perfectly uniform sampling cannot always reach full coverage, especially when the number of observations is small. For a random and uniform distribution of $N$ observations in phase, the number of observations falling in a window of width $w$ follows a Poisson distribution with mean $\lambda=wN$, and the probability that a given window contains fewer than three observations is $e^{-\lambda}(1+\lambda+\lambda^2/2)$. We use this expression to define the coverage \emph{ceiling},
\begin{equation}
C_{\max}(N)=1-e^{-\lambda}\left(1+\lambda+\frac{\lambda^2}{2}\right),\qquad \lambda=wN,
\label{eq:ceiling}
\end{equation}
which represents the coverage expected on average for a uniform sample of $N$ observations. We report this quantity together with the light-curve coverage for every object, and, as described in Sect.~\ref{sec:credibility}, we compare the observed coverage with $\min(0.8,\,C_{\max})$ rather than with a fixed threshold, so that light curves with few observations are not penalized for a limit imposed by their sample size rather than by the quality of the sampling.

\subsection{Reliability assessment}
\label{sec:credibility}

For the objects with a determined period, we perform a set of automatic checks
to assess how much the result can be trusted. None of these checks alone proves
that the period is correct; together, however, they identify the classes of
unreliable solutions discussed in the previous sections. We use them first to
decide whether the fit is usable at all, and then to sort the usable objects
into classes along two independent axes: the reliability of the period and the
quality of the phase coverage.

\textit{Base filter.} The fit is considered usable if it passes two
conditions. The first is the width of the period uncertainty interval defined in
Sect.~\ref{sec:uncertainty}: we require it to stay within a factor of two on
each side of the adopted period. A wider interval means that the periodogram
does not localize the frequency even to within a possible factor-of-two
(half/double) ambiguity, and the period cannot be used. The second is the phase
coverage of Sect.~\ref{sec:coverage}: we require the light-curve coverage to
exceed $\min(0.8,\,C_{\max})$, comparing with the coverage ceiling rather than
with a fixed threshold so that light curves with few observations are not
penalized for a limit imposed by their sample size. At the window width we
adopt, this second condition is also a strong statement about the period
itself: if the period were wrong, the observations would not fold into a compact
light curve and the coverage would break down, so a good coverage acts as a
first confirmation of the period.

An object that fails either condition is not classified further and is not used
in the analysis at all: neither its period nor the rest of its light-curve model
is trusted. The possibility of still using a fit for the color and elongation
analysis, discussed at the end of this section, applies only to objects that
pass the base filter but whose period is left ambiguous by a surviving
alternative -- not to objects that fail the filter itself.

\textit{Coverage axis.} For the objects that pass the base filter we grade the
phase coverage into four levels, using the light-curve coverage $C_{\rm lc}$ and
the group coverage $C_{\rm grp}$ defined in Sect.~\ref{sec:coverage}. The
highest level, \emph{full+groups}, requires a complete light-curve coverage,
$C_{\rm lc}\ge0.99$, together with a group coverage $C_{\rm grp}\ge0.9$: the
phase is covered everywhere and, in addition, confirmed by more than one
rotation cycle over almost the whole curve. The second level, \emph{full},
requires the same complete light-curve coverage, $C_{\rm lc}\ge0.99$, but
without the condition on the group coverage. The third level, \emph{above base},
gathers the objects with $C_{\rm lc}\ge0.9$, and the fourth level, \emph{base},
those with $0.8\le C_{\rm lc}<0.9$, that is, a coverage that only just clears the
base filter. 


\textit{Period axis.} The period is graded from the alternative solutions
reported in Sect.~\ref{sec:uncertainty}. Among the surviving alternatives we
first separate those that are genuine, i.e. that correspond to a physically
different period, from those that are merely harmonics (a half, double, or
other low-integer ratio up to 1/4) of the adopted period. A harmonic alternative does
not represent a different rotation: it only reflects an ambiguity in the number
of maxima of the light curve, while its shape, and therefore the elongation and
the colors, remain unaffected. The near-duplicate alternatives produced by the
fine frequency grid are merged before this step. 

We then assign five period classes in order of decreasing reliability:
(i)~a single period, with a narrow uncertainty interval (2\% of the period) and no surviving
alternative of any kind; (ii)~a narrow interval with harmonic alternatives only,
so that the shape of the light curve is fixed and only the number of maxima is
uncertain; (iii)~a narrow interval with genuine alternatives, but such that the
adopted solution is preferred over all of them with a combined probability above
a chosen threshold ($>$80\% probability); (iv)~a narrow interval with genuine alternatives that the
adopted solution does not beat confidently ($<$80\% probability); and (v)~a period whose uncertainty
interval is itself too wide (but still within the factor of two of the base
filter), in which case the individual alternatives are not examined. For the
combined probability in classes (iii) and (iv) we multiply the per-alternative
Vuong probabilities, which gives the probability that the adopted period is
preferred over the whole set of distinct alternatives at once; the harmonic
alternatives are excluded from this product for the reason given above.

\textit{Overfitting check.} Independently of the two axes, we also flag the
solutions whose reduced $\chi^2$ is too low, since such a fit most likely
follows the photometric noise rather than a real signal. Under the null
hypothesis that the model is correct and the reported uncertainties are
accurate, $\chi^2/\nu$ has an expectation of unity and a standard deviation of
$\sqrt{2/\nu}$, where $\nu=N_u-p$ is the number of degrees of freedom. We define
the floor value as the $3\sigma$ lower bound of this distribution,
\begin{equation}
\left(\chi^2/\nu\right)_{\rm floor}=\max\!\left(0,\ 1-3\sqrt{2/\nu}\right),
\label{eq:chi2floor}
\end{equation}
and flag the solutions for which the reduced $\chi^2$ falls below this value as
a sign of overfitting.

The two axes address different questions and should not be confused. A light
curve can have an excellent fit (a reduced $\chi^2$ close to unity, a good
phase coverage, and a narrow uncertainty interval) and still admit a viable
alternative period, for instance when two periods are related by a one-day alias
that the data cannot resolve. Such a solution is strong on the coverage axis and
weak on the period axis at the same time, which is precisely why the two axes are
reported independently: a sample can then be selected by combining them according
to the needs of a given study, requiring the strongest period classes where the
rotation itself is the subject, or relaxing the period class where it is not.

\subsection{Taxonomic classification}
\label{sec:taxonomy}

The absolute magnitudes $M_b$ obtained from the light-curve fit in each photometric band provide the colors of an asteroid, which we use to estimate its taxonomic class with the \texttt{classy} package \citep{2022A&A...665A..26M}. Before the colors are computed, the solar contribution must be removed, so that the resulting quantity is a reflectance rather than a color with respect to the Sun: for two bands $b$ and $b'$ with solar magnitudes $M_{b,\odot}$ and $M_{b',\odot}$, the reflectance relative to the band $b'$ is
\begin{equation}
r_b = 10^{-0.4\left[(M_b-M_{b'})-(M_{b,\odot}-M_{b',\odot})\right]},
\label{eq:reflectance}
\end{equation}
and \texttt{classy} classifies the resulting spectrum $r_b(\lambda_b)$, where $\lambda_b$ is the effective wavelength of the band $b$.

The effective wavelengths $\lambda_b$ and the solar absolute magnitudes
$M_{b,\odot}$ we adopt for the six LSST bands are collected in
Table~\ref{tab:lsst_bands}. The effective wavelengths are taken from the
SVO Filter Profile Service\footnote{\url{https://svo2.cab.inta-csic.es/theory/fps/index.php?mode=browse&gname=LSST}}
for the full LSST system (filter, optics, detector, and atmosphere), and the
solar magnitudes in the AB system are taken from \citet{2018ApJS..236...47W}.

\begin{table}
\centering
\caption{Effective wavelengths and solar absolute magnitudes (AB system)
adopted for the six LSST bands.}
\label{tab:lsst_bands}
\begin{tabular}{lcc}
\hline
Band & $\lambda_b$ ($\mu$m) & $M_{b,\odot}$ (AB) \\
\hline
$u$ & 0.3751 & 6.27 \\
$g$ & 0.4741 & 5.06 \\
$r$ & 0.6172 & 4.64 \\
$i$ & 0.7501 & 4.52 \\
$z$ & 0.8679 & 4.51 \\
$y$ & 0.9712 & 4.50 \\
\hline
\end{tabular}
\end{table}

A direct application of Eq.~(\ref{eq:reflectance}), however, faces a problem specific to our fitting procedure. As discussed in Sect.~\ref{sec:phase}, the phase angle of every object in our sample changes only by a few degrees over the observation window, so the slope of the phase function is poorly constrained and is stabilized with a soft a priori constraint. As a consequence, the mean magnitude level of the light curve (and hence the absolute magnitude $M_b$ of every band individually) carries a substantial uncertainty coming from the phase-function degeneracy. This uncertainty is, however, nearly identical for all bands, because it originates from the same poorly constrained phase-angle dependence common to the whole light curve; in the covariance matrix of the fitted magnitudes, $\mathrm{Cov}(M_b,M_{b'})$, it therefore appears as a large, strongly correlated component shared by all bands. The colors $M_b-M_{b'}$ that enter Eq.~(\ref{eq:reflectance}) are far less affected, since this common component cancels in the difference; using the individual uncertainties $\sigma(M_b)$ as if they were independent would overestimate the true uncertainty of the colors, sometimes by a large factor.

To account for this, we propagate the full covariance matrix of the fitted magnitudes into the color uncertainties, rather than treating the bands as independent. For a color $c_b=M_b-M_{\rm ref}$ defined with respect to a chosen reference band, the variance is
\begin{equation}
\begin{split}
\sigma^2(c_b)=\mathrm{Cov}(M_b,M_b)+\mathrm{Cov}(M_{\rm ref},M_{\rm ref})- \\ - 2\,\mathrm{Cov}(M_b,M_{\rm ref}),
\end{split}
\label{eq:colorvar}
\end{equation}
which reduces to the naive sum of variances only if $M_b$ and $M_{\rm ref}$ are uncorrelated; in our fits the covariance term in Eq.~(\ref{eq:colorvar}) removes most of the common phase-function uncertainty and typically shrinks $\sigma(c_b)$ by a large factor compared to $\sqrt{\sigma^2(M_b)+\sigma^2(M_{\rm ref})}$.

Since \texttt{classy} takes as an input a set of reflectance points with individual uncertainties rather than a full covariance matrix, we cast the problem into a form in which the color uncertainties are effectively independent. We designate one band as the reference and assign it a formal color uncertainty of zero, which is equivalent to expressing the whole spectrum relative to this band rather than treating all bands symmetrically; the propagated uncertainty of every other band is then given entirely by Eq.~(\ref{eq:colorvar}). The reference band is chosen, for every object individually, as the one that minimizes the sum of the resulting color variances over all other bands,
\begin{equation}
\begin{split}
 b_{\rm ref}=\operatorname*{arg\,min}_{b_0}\sum_{b\neq b_0}\Big[\mathrm{Cov}(M_b,M_b)+\mathrm{Cov}(M_{b_0},M_{b_0})- \\ -2\,\mathrm{Cov}(M_b,M_{b_0})\Big],   
\end{split}
\label{eq:refband}
\end{equation}
rather than simply the band with the smallest individual uncertainty $\sigma(M_b)$, since a band with a small variance but a small covariance with the others does not necessarily minimize the propagated color uncertainties. The reflectance uncertainty used by \texttt{classy} for the band $b$ is then obtained from Eq.~(\ref{eq:colorvar}) by standard error propagation through Eq.~(\ref{eq:reflectance}),
\begin{equation}
\sigma(r_b) = r_b\,\frac{\ln 10}{2.5}\,\sigma(c_b),
\label{eq:reflerr}
\end{equation}
with $\sigma(r_{b_{\rm ref}})=0$ by construction.

This procedure yields, for every object with a reliable light-curve fit as defined in Sect.~\ref{sec:credibility}, a taxonomic class and its formal probability from \texttt{classy}, together with the probabilities of other classes.

\subsection{Elongation, absolute magnitude, and size}
\label{sec:size}

For every object with a determined light curve we also estimate its elongation,
its absolute magnitude in the $V$ band, and, using the taxonomic class, its
diameter.

The elongation of the body is obtained from the light-curve amplitude as in 
\citet{greenstreet2026} following the approach of \citet{1990A&A...231..548Z}. The peak-to-peak amplitude $A$ measured at a solar phase
angle $\alpha$ is first reduced to zero phase angle,
\begin{equation}
A_0=\frac{A}{1+m\,\alpha},
\label{eq:amp0}
\end{equation}
which removes the linear brightening of the amplitude with phase angle; we adopt the slope $m=0.02\ \mathrm{deg}^{-1}$. Assuming the
brightness variation is dominated by the projected cross-section of a triaxial
ellipsoid observed equatorially, the ratio of the two long axes is
\begin{equation}
a/b=10^{0.4\,A_0}.
\label{eq:elong}
\end{equation}
This is a lower limit on the true elongation, since a non-equatorial aspect angle
reduces the observed amplitude.

The absolute magnitude in the $V$ band is obtained from the mean magnitudes $M_b$
fitted in the individual LSST bands. Each $M_b$ is converted to $V$ by removing
the solar color, $H_{V,b}=M_b-(M_{b,\odot}-M_{V,\odot})$, and the values from all
available bands are combined into a single $H_V$ by a weighted average, using the
fitted magnitude uncertainties as weights. The solar magnitudes are those of
Table~\ref{tab:lsst_bands}, with $M_{V,\odot}$ taken from \citet{2018ApJS..236...47W}.

\begin{figure*}
\centering
\includegraphics[width=0.90\textwidth]{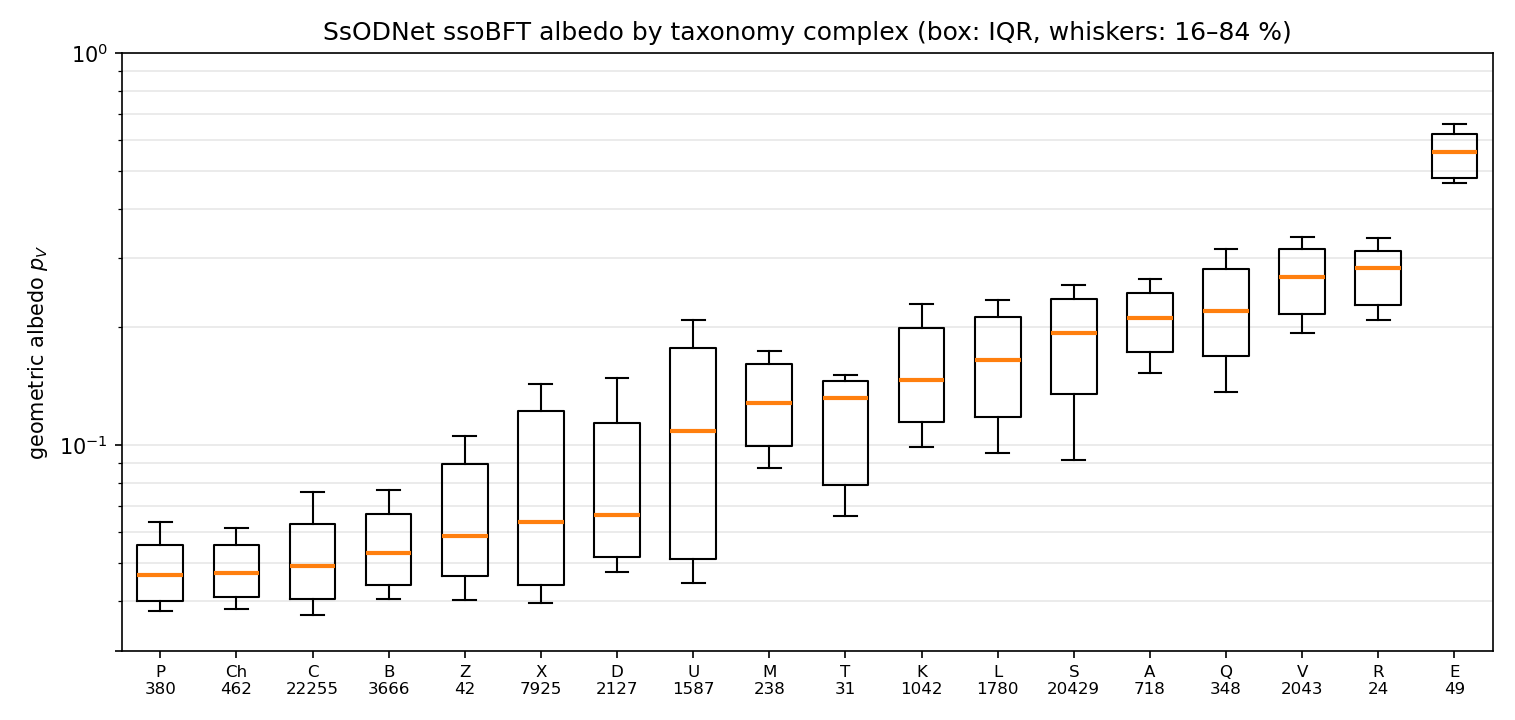}

\caption{Asteroid albedo statistics per complex as derived from SsODNet data \citep{2023A&A...671A.151B}. Complex categories including sample size are plotted against geometric albedo ($p_V$). The orange bars represent median values and the boxes show interquartile ranges (IQR). Whiskers denote 16$^{\text{th}}$ and 84$^{\text{th}}$ percentiles, respectively. }
\label{fig:asteroid_albedo}
\end{figure*}

Finally, the diameter is estimated from $H_V$ and an assumed geometric albedo
\citep{1989aste.conf..524B, 2007Icar..190..250P},
\begin{equation}
D=\frac{1329\ \mathrm{km}}{\sqrt{p_V}}\,10^{-H_V/5}.
\label{eq:diameter}
\end{equation}
As pointed out in \citet{2021PSJ.....2...32M}, the constant 1329 is band dependent and varies with albedo $p_V$. Fitting $H_V$ across all available bands mitigates the single band issue, but the albedo $p_V$ is not measured in our data. Instead, we infer albedo from the taxonomic class determined in Sect.~\ref{sec:taxonomy}. More precisely, we determine albedo statistics derived from SsODNet data for each individual class \citet{2023A&A...671A.151B}. We calculate the median albedo of every class, together with its lower and upper bounds, based on best values for every characterized asteroid as provided by SsODNet, see Figure \ref{fig:asteroid_albedo}. When a single class is
determined with high confidence ($\ge90\%$), we use its median albedo for the
nominal diameter and its lower and upper bounds to obtain the corresponding
upper and lower limits on $D$. When no single class reaches this confidence, we
instead combine the most probable classes until their probabilities sum to at
least $90\%$, and take the probability-weighted mean of their median albedos,
with the extreme bounds over this set of classes defining the diameter limits.
Since the diameter scales as $p_V^{-1/2}$, the albedo remains the dominant
source of uncertainty in $D$, at the level of a factor of about two between the
dark and bright albedo scales.

From the elongation and the diameter we further estimate the individual
axes $a$ and $b$ of the body, modeled as a prolate ellipsoid with
$a>b=c$. We fix the ratio $a/b$ from Eq.~(\ref{eq:elong}) and use the
diameter of Eq.~(\ref{eq:diameter}) to set the overall scale. The scale is
set by requiring that the mean of the brightest and faintest viewing
orientations reproduce the projected area of the equivalent sphere: the
broadside projection has area $\tfrac{\pi}{4}ab$ and the pole-on (edge-on)
projection $\tfrac{\pi}{4}b^2$, so that
\begin{equation}
\frac{1}{2}\left(\frac{\pi}{4}ab+\frac{\pi}{4}b^2\right)=\frac{\pi}{4}D^2,
\label{eq:axes}
\end{equation}
where $D$ is the diameter of Eq.~(\ref{eq:diameter}). Together with the
fixed $a/b$ ratio, this determines the axes $a$ and $b$ for every object.

We also estimate the cohesive strength
required to prevent rotational disruption, following
\citet{sanchez2014strength}. Disruption occurs when the rotation
frequency exceeds the critical value $\omega_{\rm crit}$ given by
\begin{equation}
\omega_{\rm crit}^2 \le \omega_\alpha^2
+\frac{2}{\rho\,\alpha^2}\left(\frac{3-\sin\phi}{3+\sin\phi}\right)\sigma_Y,
\label{eq:wcrit}
\end{equation}
where $\rho$ is the bulk density, $\alpha$ the largest semi-axis of the
best-fitting triaxial ellipsoid ($\alpha\ge\beta\ge\gamma$), $\phi$ the
internal friction angle of the material, $\sigma_Y$ its cohesive (yield)
strength, and $\omega_\alpha^2$ the gravitational contribution of the
ellipsoidal mass distribution,
\begin{equation}
\omega_\alpha^2=\frac{3}{2}\,GM
\int_0^\infty\frac{dx}
{(\alpha^2+x)\sqrt{(\alpha^2+x)(\beta^2+x)(\gamma^2+x)}} .
\label{eq:wa2}
\end{equation}
Unlike in the previous work \citep{greenstreet2026}, where we further simplified
Eq.~(\ref{eq:wcrit}) to the spherical case, we keep the prolate shape
derived above, with semi-axes $\alpha=a/2$ and $\beta=\gamma=b/2$, for
which the integral in Eq.~(\ref{eq:wa2}) has the closed form
\begin{equation}
\omega_\alpha^2=\frac{3}{2}\,GM\,\frac{1}{\alpha^3 e^2}
\left[\frac{1}{e}\ln\!\frac{1+e}{1-e}-2\right], 
\label{eq:wa2_prolate}
\end{equation}
with $M=\tfrac{4}{3}\pi\rho\,\alpha\beta^2$ and eccentricity $e=\sqrt{1-(\beta/\alpha)^2}$. Solving Eq.~(\ref{eq:wcrit})
for $\sigma_Y$ then gives the minimum cohesive strength consistent with
the observed spin. We adopt a bulk density $\rho=1700\ \mathrm{kg\,m^{-3}}$,
typical of rocky asteroids, and the conservative choice $\phi=\pi/2$; when
self-gravity alone is sufficient to hold the body together, the required
strength is zero.

\begin{table*}
\centering
\caption{Number of objects in each combination of the period class (rows) and
the coverage class (columns), ordered from the best to the worst along both
axes. The \emph{below base} column and the objects with a period interval wider
than the base filter are those that fail the base filter of
Sect.~\ref{sec:credibility}; they are listed for completeness but are not used
in the analysis. The last row and column give the marginal totals.}
\label{tab:classification}
\begin{tabular}{lcccccc}
\hline
 & \multicolumn{5}{c}{Coverage class} & \\
\cline{2-6}
Period class & full+groups & full & above base & base & below base & Total \\
\hline
1. unique                 & 260 &  61 &  82 & 136 & 628  & 1167 \\
2. harmonic               &  44 &   9 &  11 &  28 & 171  &  263 \\
3. rivals beaten          &  32 &  10 &  11 &  22 & 182  &  257 \\
4. rivals uncertain       & 159 &  60 &  63 & 139 & 1238 & 1659 \\
5. wide bracket           & 863 &  76 & 291 & 466 & 4375 & 6071 \\
\hline
Total                     & 1358 & 216 & 458 & 791 & 6594 & 9417 \\
\hline
\end{tabular}
\end{table*}

\section{Vera C. Rubin Observatory DP2}
\label{sec:results}

\begin{table*}[htbp]
\centering
\caption{Rotation periods, colors, and derived properties of the full sample. Columns are as in Table~\ref{tab:super_fast_rotators}, except that the alternative periods are omitted here and coverage is added. Table~\ref{tab:full} is published in its entirety, with all columns, in the machine-readable format. A portion, showing a subset of the columns for a few objects, is presented here for guidance regarding its form and content.}
\label{tab:full}
\begin{tabular}{lcccccccccccr}
\hline
Object & $P$ (h) & $P_{\min}$ (h) & $P_{\max}$ (h) & $n$ & cl & Class & $p_V$ & $D$ (km) & $a/b$ & Cov & $\chi^2_\nu$ & $N$ \\
\hline
100009 & 12.859 & 1.916 & 174.631 & 3 & 5 & S (1.00) & 0.22 & 1.19 & 1.24 & 0.48 & 3.55 & 123 (18) \\
100077 & 8.477 & 0.981 & 376.880 & 2 & 5 & L (0.88) & 0.19 & 2.90 & 1.04 & 0.65 & 1.28 & 93 (8) \\
100099 & 6.561 & 1.637 & 12.655 & 3 & 5 & Q (0.99) & 0.24 & 1.99 & 1.48 & 0.59 & 2.09 & 231 (49) \\
100231 & 108.242 & 108.152 & 108.557 & 8 & 1 & C (0.89) & 0.05 & 7.45 & 1.40 & 0.30 & 2.31 & 86 (1) \\
100272 & 68.452 & 63.430 & 69.465 & 2 & 5 & S (0.37) & 0.22 & 3.75 & 1.12 & 0.15 & 1.87 & 265 (17) \\
100361 & 25.601 & 25.595 & 25.607 & 5 & 4 & S (0.84) & 0.21 & 1.77 & 1.18 & 0.44 & 4.11 & 81 (8) \\
100389 & 6.324 & 6.309 & 6.351 & 7 & 4 & Q (0.50) & 0.21 & 1.38 & 1.07 & 0.75 & 1.57 & 348 (11) \\
100409 & 10.473 & 4.005 & 26.791 & 4 & 5 & X (1.00) & 0.08 & 2.74 & 1.19 & 0.66 & 3.19 & 94 (7) \\
100420 & 17.017 & 17.015 & 17.019 & 7 & 4 & C (0.59) & 0.05 & 2.88 & 3.49 & 0.37 & 3.56 & 85 (3) \\
100598 & 0.761 & 0.497 & 3.629 & 2 & 5 & K (0.43) & 0.12 & 2.42 & 1.03 & 1.00 & 2.37 & 200 (7) \\
\hline
\end{tabular}
\end{table*}

We apply the method to the asteroid photometry contained in the Rubin
Observatory Data Preview 2 (DP2) catalog \citep{10.71929/rubin/3382528, RTN-115}, which comprises
observations obtained with the LSST Camera between 2025 April and 2026
January. From the DP2 solar system object catalog we select all objects with at least 80
photometric measurements, so that the light curve is sampled densely
enough for the period search and the reliability checks to be meaningful.

DP2 does not fully overlap with the Rubin First Look (RFL) observations
analyzed by \citet{greenstreet2026}; at the same time, DP2 provides, for
many of the objects already processed in that earlier work, a number of
additional measurements beyond those available in the RFL data alone. It
is therefore natural to combine the two catalogs. We supplement the DP2
selection with the Rubin observations obtained between 2025 April 21 and
2026 January 6, downloaded from the Asteroid Institute (B612)
archive\footnote{\url{https://doi.org/10.5281/zenodo.17047589}}
\citep{b612_data} on 2026 August 16, which mirrors the astrometric and
photometric measurements submitted to the IAU Minor Planet Center.

We additionally exclude, from each combined light curve, any isolated
cluster of fewer than 10 observations separated from the rest of the data
by a gap much longer than the cluster itself spans. By the time such a
cluster is reached, the phase connection to the rest of the light curve
is, in practice, already lost; 
with fewer than 10 points, the cluster is too sparse to
meaningfully narrow the periodogram peak, and instead tends to split it
further into additional aliases rather than genuinely constrain the
period. Since the size of the frequency grid scales with the total time
span of the observations (Sect.~\ref{sec:methods}), keeping such a
distant, uninformative cluster in the fit substantially increases the
computation time without a corresponding benefit, and we therefore remove
it.


After these exclusions and the addition of the B612 observations, our
final sample consists of 9,417 asteroids with at least 80 photometric
measurements.
We determined the rotation periods of these objects with the method
described above and classified each solution along the two axes of
Sect.~\ref{sec:credibility}: the reliability of the period and the quality
of the phase coverage. The full results for all 9,417 objects -- adopted
periods, uncertainty intervals, phase coverage, colors, taxonomic classes,
derived sizes -- are given in Table~\ref{tab:full}, published in its
entirety in machine-readable form. Table~\ref{tab:classification} gives the
number of objects in every combination of the two classifications, with the
columns and rows ordered from the most to the least reliable. 1,221 objects
passed our base filter, with a light-curve coverage of at least 80\% and the
period uncertainty interval within a factor of two of the adopted period.
We report 273 asteroids with determined spectral classes at more than
90\% confidence: 215 of C class, 30 of S class, 7 of L class, 6 each of V
and Q class, 4 of A class, 3 of X class, and one each of B and R class.

\subsection{Super-fast rotators}
\label{sec:fastrot}

\begin{figure*}
\centering

\includegraphics[width=0.22\textwidth]{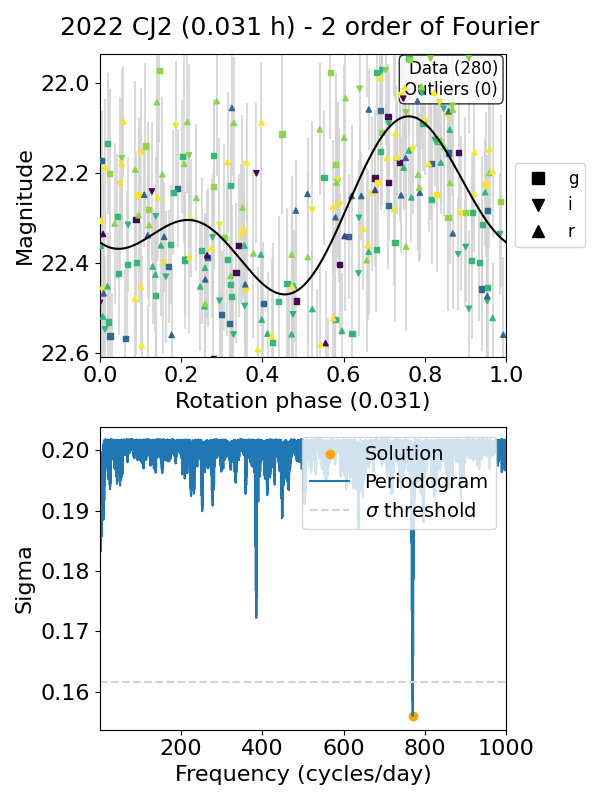}   \includegraphics[width=0.22\textwidth]{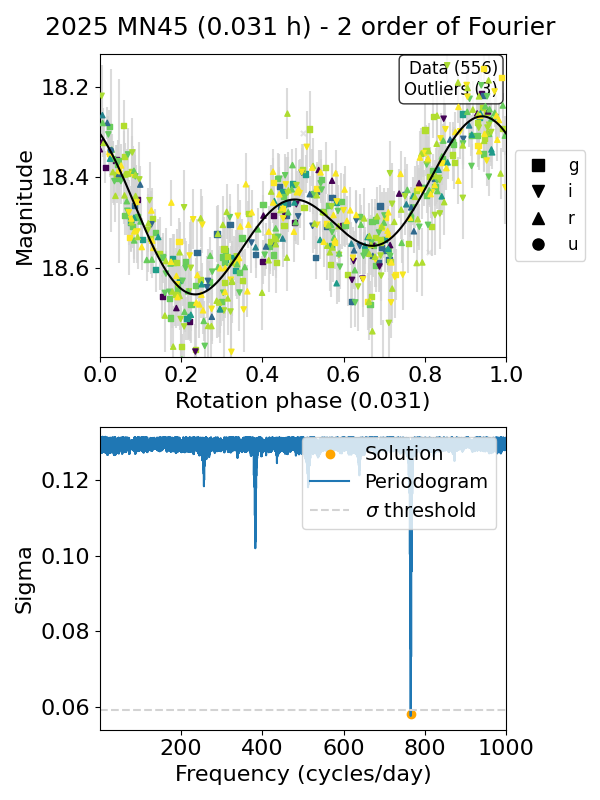} \includegraphics[width=0.22\textwidth]{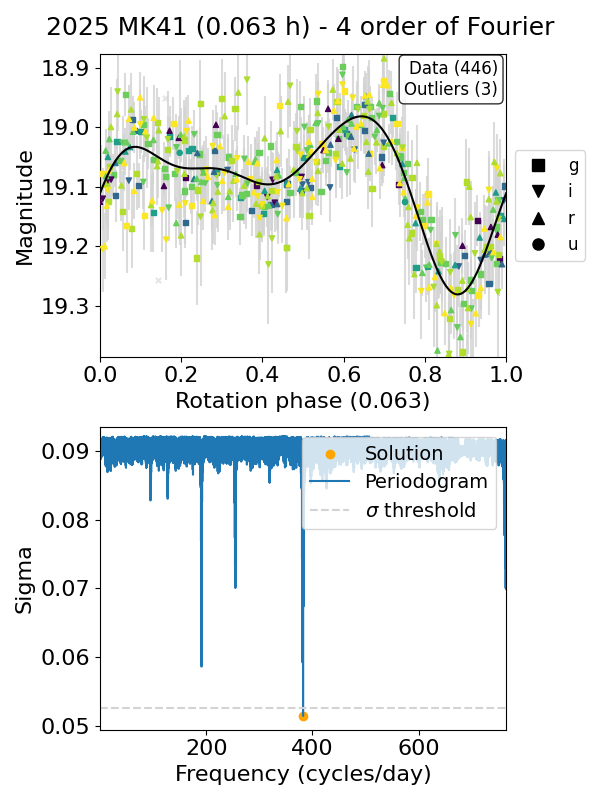} \\

\includegraphics[width=0.22\textwidth]{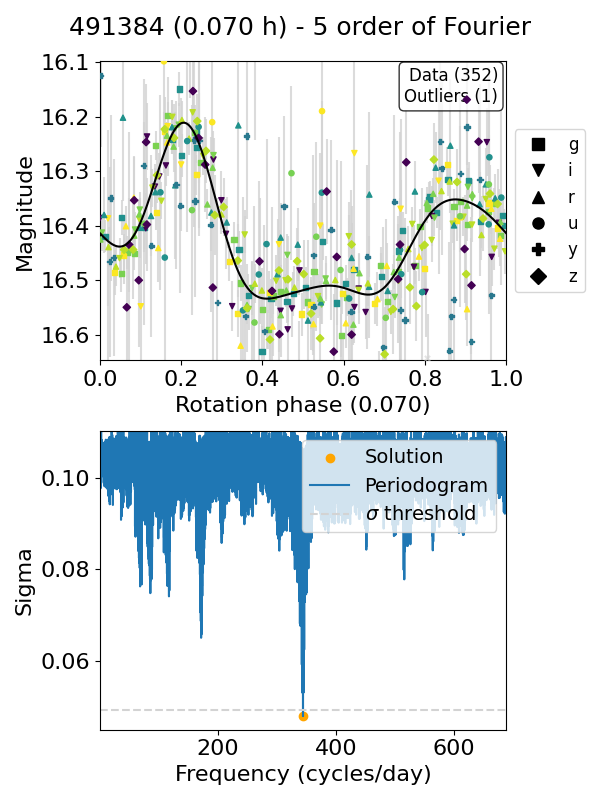}   \includegraphics[width=0.22\textwidth]{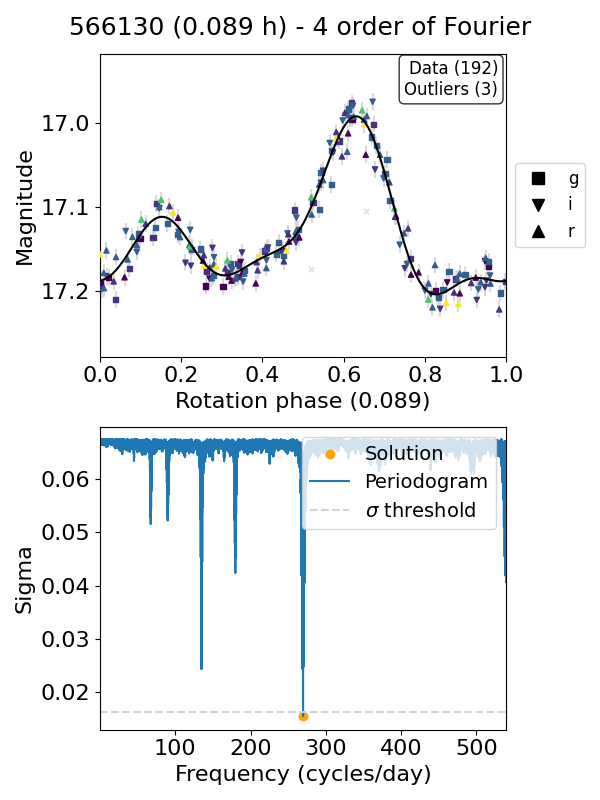} \includegraphics[width=0.22\textwidth]{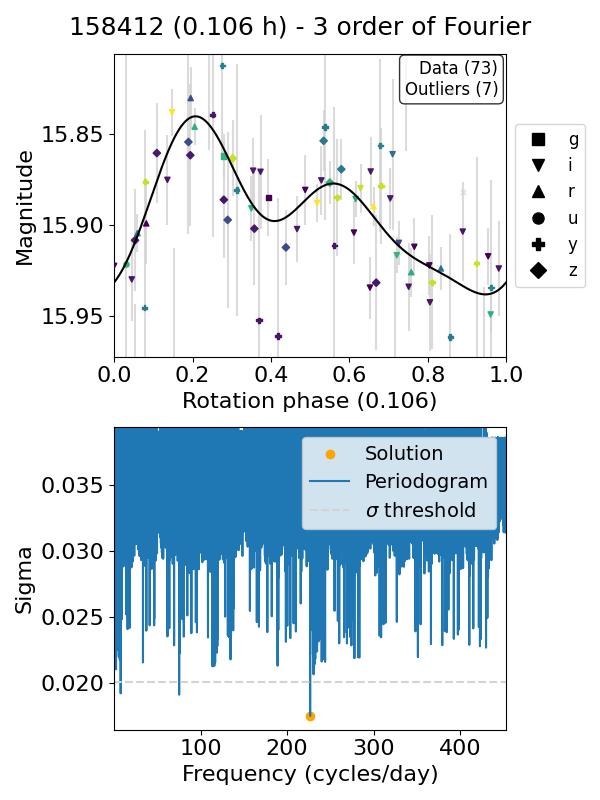} 
\includegraphics[width=0.22\textwidth]{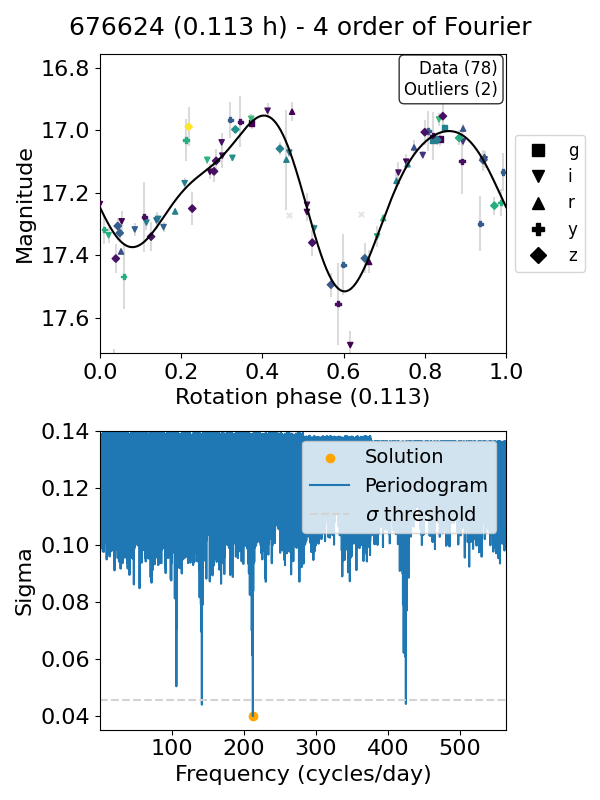}   \\ \includegraphics[width=0.22\textwidth]{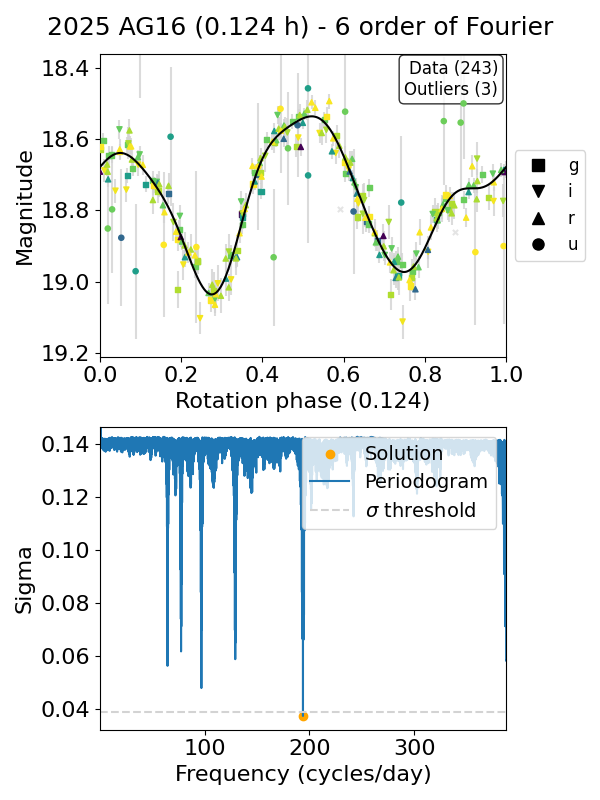} \includegraphics[width=0.22\textwidth]{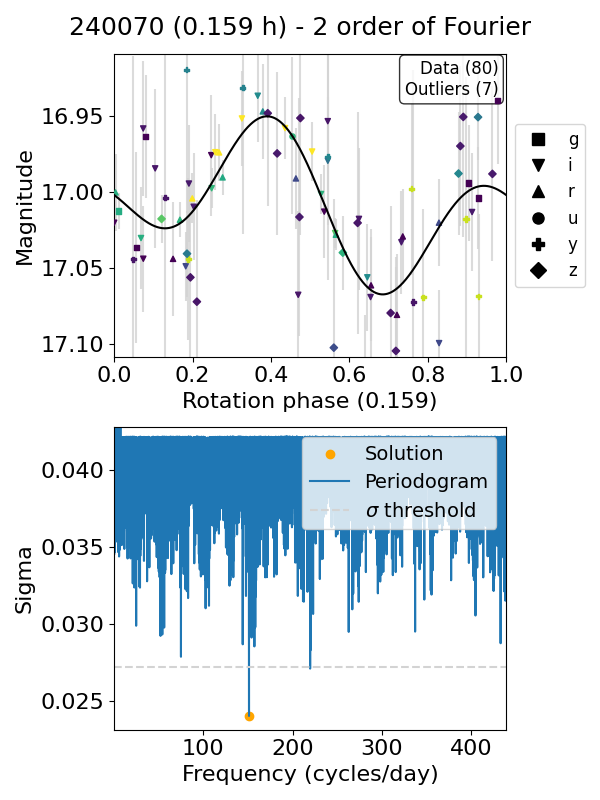}
\includegraphics[width=0.22\textwidth]{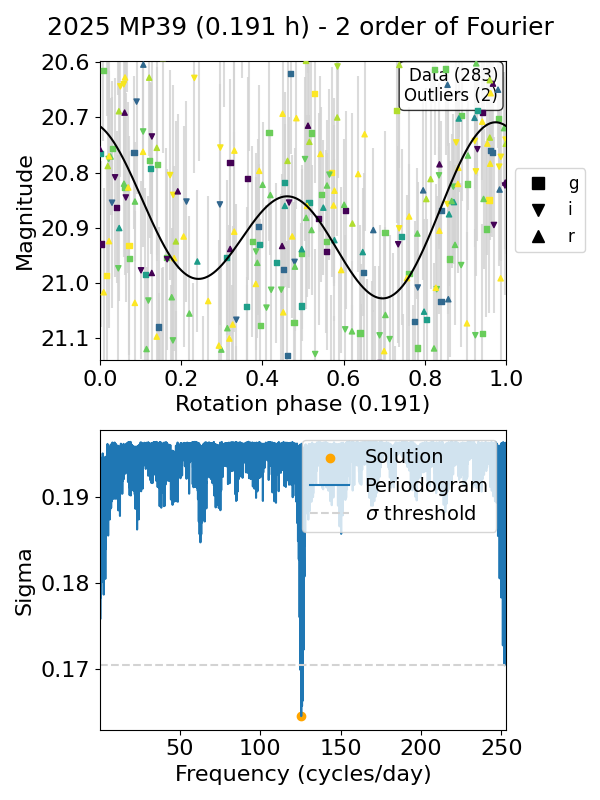}
\includegraphics[width=0.22\textwidth]{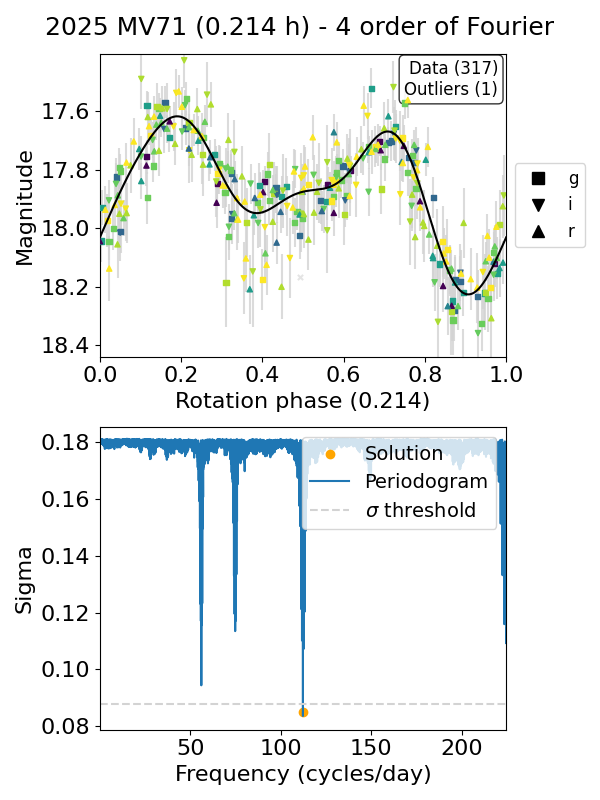} \\
\includegraphics[width=0.22\textwidth]{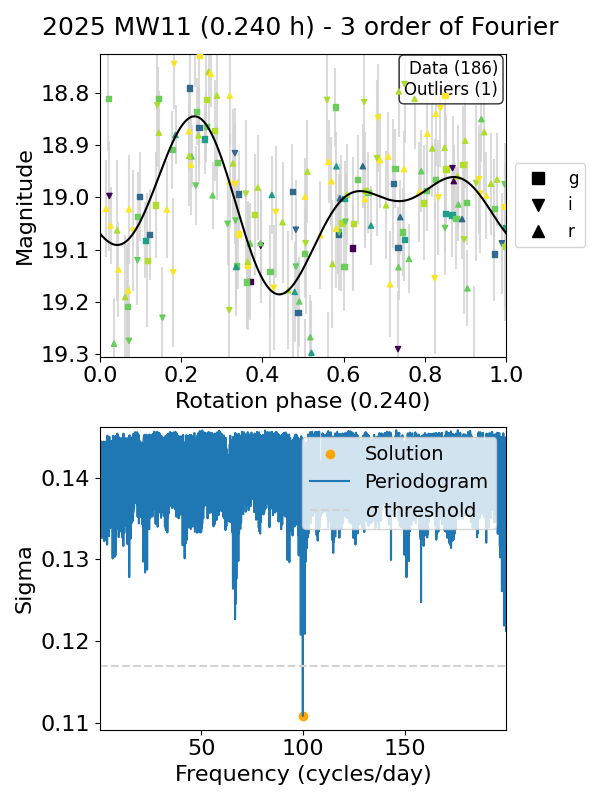}
\includegraphics[width=0.22\textwidth]{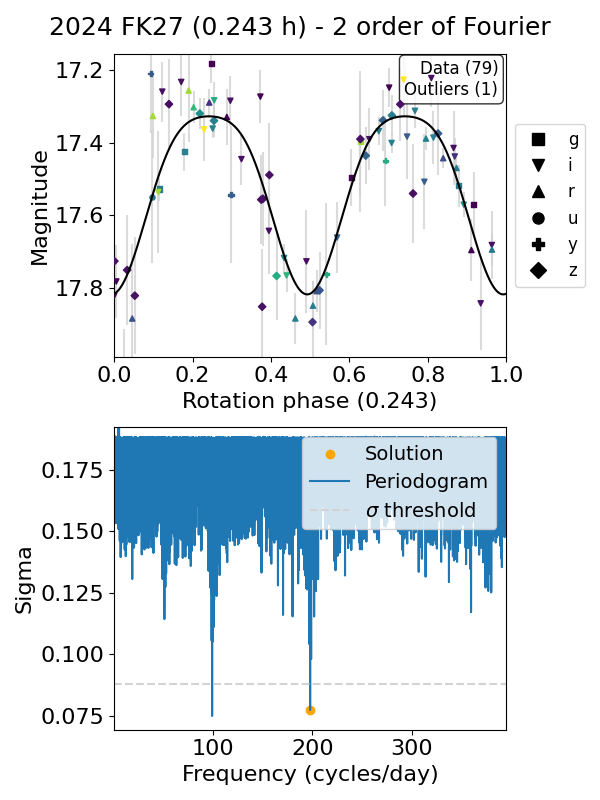}
\includegraphics[width=0.22\textwidth]{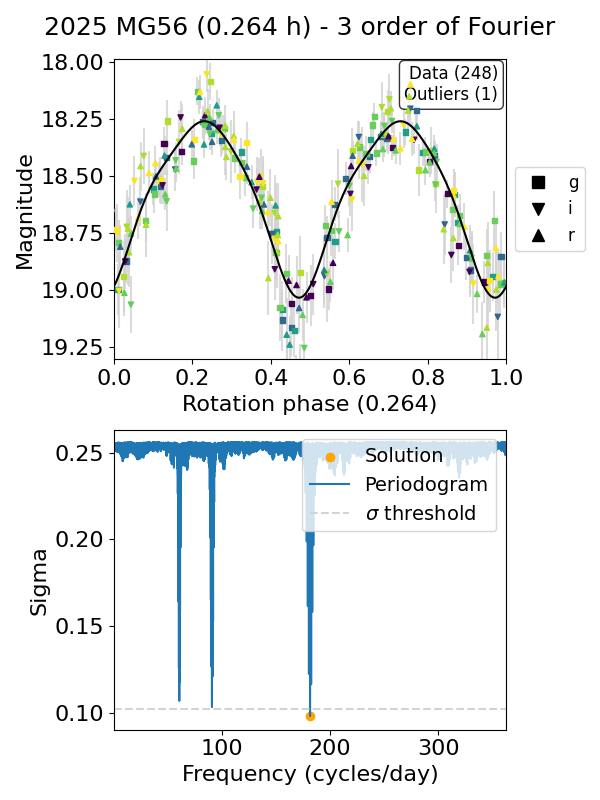}    
\caption{The 14 fastest rotators among the 66 super-fast rotator candidates
of Table~\ref{tab:super_fast_rotators}, ordered by increasing period; the slowest of
them completes a rotation in 16~minutes. Several of these objects were
already reported by \citet{greenstreet2026}, and our results agree with
theirs. Top panels show the light curve folded at the adopted period,
bottom panels the periodogram, with the adopted solution marked with an
orange dot and the $\sigma$ threshold of Eq.~(\ref{eq:sigthr}) shown
dashed. Symbols denote the LSST bands, point color encodes observation
time (darker points are earlier observations, yellow points are later
ones), and grey crosses are points rejected by the fit.}
\label{fig:fastests}
\end{figure*}

\begin{figure}
\centering
\includegraphics[width=0.49\textwidth]{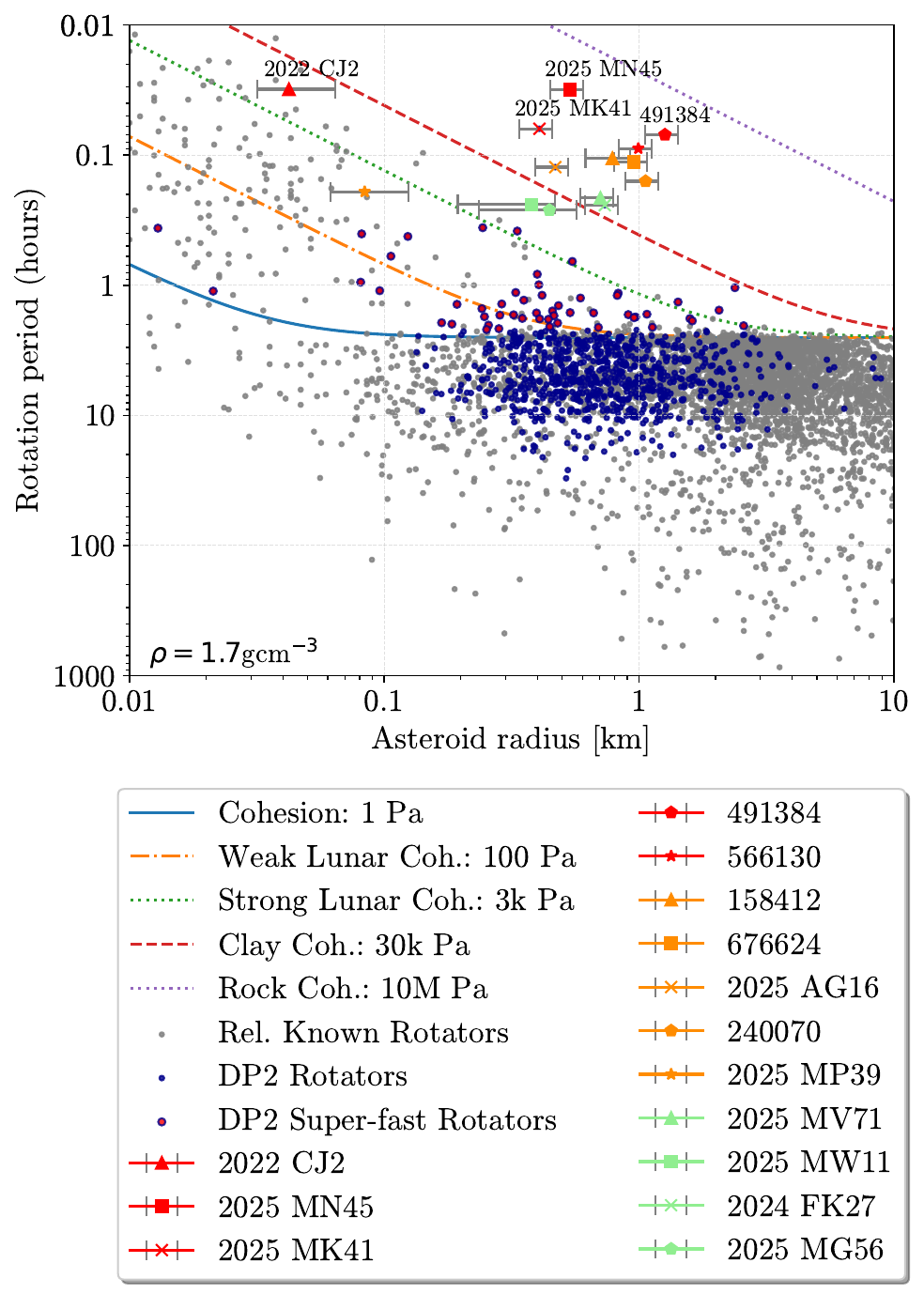}

\caption{Asteroid size versus rotation period for numbered asteroids with reliable periods (MPC LCDB quality codes $3+$, $3$ and $3-$, gray dots). The new Rubin DP2 fast (blue dots) and super-fast rotator (blue dots with red circles) candidates with well determined spin period uncertainties derived in this study are shown. The largest and fastest rotators are marked with individual symbols. Blue dots represent Rubin observed objects . The so-called “spin barrier” is indicated by the level curves for various values of cohesive strengths. Some rotators such as 2025 MN45, 2025 MK41 and 2022 CJ2 were already presented in \citet{greenstreet2026}, but have been better characterized in this study. Several new super-fast rotators such as 566130, 158412, 676624, 240070 and 2025 AG16 require cohesive strengths well beyond clay, while 491384 is the largerst super-fast rotator known to date.}
\label{fig:spinbarrier}
\end{figure}

Our pipeline identified 66 candidate super-fast rotators in the present
sample, selected among the objects whose light-curve coverage exceeds
$80\%$, whose period belongs to one of the three most reliable classes (a
single period, a period determined up to harmonics, or a period with the
probability of beating all rivals exceeding 50\%), and whose adopted
period is shorter than $2.2$~h. Almost all of them received coverage by
groups of more than 91\%; the two exceptions, with a group coverage of
only 84\%, are flagged as questionable and are discussed separately below.
Those are shown in Figures~\ref{fig:spinbarrier} and
\ref{fig:fast_rotator_orbits}.

For comparison, without any reliability filtering the sample contains
1487 asteroids with a nominal period shorter than $2.2$~h and 899 objects
with a period below $5$~min -- most of which are spurious minima produced
by aliases, harmonics, or noise. This large difference illustrates why
the reliability checks are essential.

Not all of these 66 objects are, however, equally convincing: for some of
them the fast period is only weakly preferred over a slower alternative,
and for a few there is a rival period that is in fact preferred over the
adopted one. The properties of all 66 objects are collected in
Table~\ref{tab:super_fast_rotators}. We discuss them below in three
groups: 14 objects that complete a rotation in less than 16~minutes
(Sect.~\ref{sec:veryfast}), a further 32 objects with a uniquely
determined period (Sect.~\ref{sec:uniqueperiod}), and the remaining 20
objects (Sect.~\ref{sec:remaining}), assessing for each group the
likelihood that its members are genuine super-fast rotators.

\subsubsection{The 14 fastest rotators}
\label{sec:veryfast}

\begin{figure}
\centering
\includegraphics[width=0.47\textwidth]{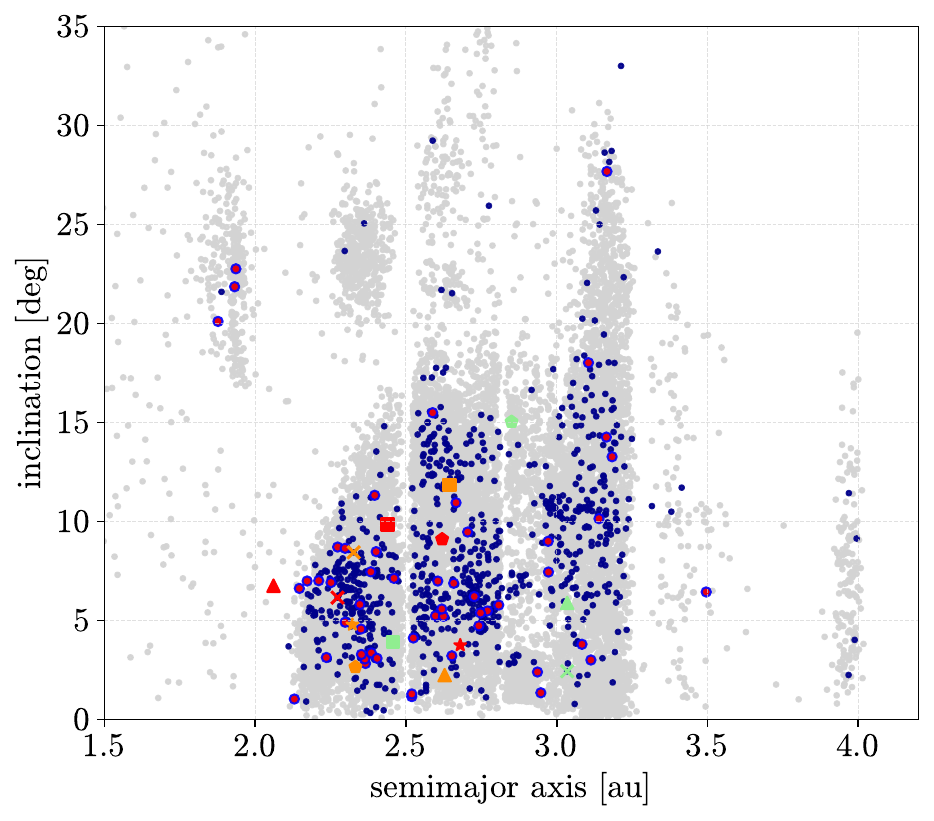}   
\includegraphics[width=0.47\textwidth]{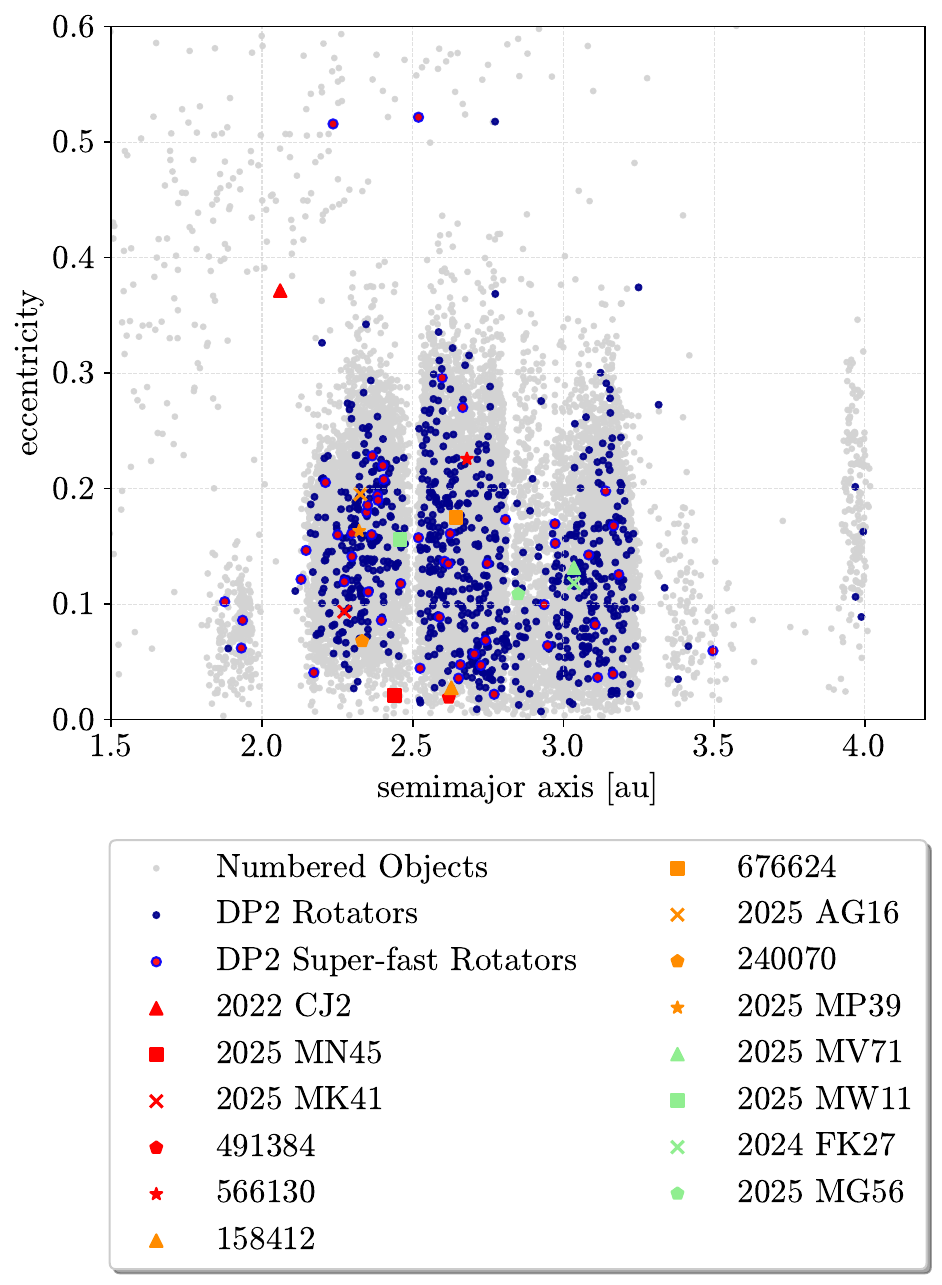}   

\caption{Semimajor axis, eccentricity, and inclination for the new, reliable DP2 rotators (blue) and super-fast rotator candidates (red dots). The gray dots correspond to numbered objects with reliable spin periods (MPC LCDB quality codes $3$ and $3-$). The fastest and largest rotators from this study are marked individually, see also Figure \ref{fig:spinbarrier}. TNOs are not shown.}
	\label{fig:fast_rotator_orbits}
\end{figure}

The 14 fastest objects of our sample complete a rotation in under
16~minutes; their light curves and periodograms are shown in
Fig.~\ref{fig:fastests}, and their orbital elements in
Fig.~\ref{fig:fast_rotator_orbits}.

Five of them were already reported by \citet{greenstreet2026}: 2022~CJ2
(previously designated 2025~MJ79), 2025~MN45, 2025~MK41, 2025~MV71, and
2025~MG56. For four of these (all but 2025~MV71) the present DP2-based
sample provides more photometry than was available in the
RFL discovery release (from 253 to 280 measurements for 2022~CJ2, 500 to
559 for 2025~MN45, 408 to 449 for 2025~MK41, and 213 to 249 for
2025~MG56); in every case the additional observations fold consistently
onto the previously reported period, confirming it. 
The RFL dataset used in \citet{greenstreet2026} lacked z-band observations, restricting their ability to provide taxonomic classifications to providing colors alone from the $g, r, i$ observations available. Our pipeline determines a taxonomic class for each object assuming albedo according to the taxonomic class: four of the five have a
confidently determined C-type classification (2025~MN45, 2025~MK41,
2025~MV71, and 2025~MG56, with probabilities of 0.96, 0.97, 0.93, and
0.79, respectively), while 2022~CJ2 is more likely S-type (probability
 0.69). 
In \citet{greenstreet2026} the diameter of 2025~MN45 was estimated at
$0.71$~km. Here, with a taxonomic class now available, we obtain a larger
value: $1.07$~km on average (ranging from $0.90$ to $1.20$~km across the
albedo bounds). Moreover, its elongation of $a/b\simeq1.32$ implies that,
even at the smallest end of this range, the long axis reaches
$a\simeq1.10$~km. This confirms 2025~MN45 -- already noted in
Sect.~\ref{sec:intro} as the fastest-spinning asteroid with $H<20$ ever
reported \citep{greenstreet2026} -- as a genuinely kilometer-sized
ultra-fast rotator.

Two of the remaining objects stand out as both very fast and
unusually large, and we discuss them individually.

\textit{{491384.}} This object has an adopted period of $0.0697$~h, i.e.
$4.18$~minutes, placing it among the ultra-fast rotators. The fit is
based on $353$ of its $354$ observations, using the fifth Fourier order,
and has a reduced $\chi^2$ of $1.09$, consistent with a well-behaved
solution. The taxonomic class is confidently C-type (probability $97\%$),
and from $H_V=16.91$ we obtain a nominal diameter of $2.53$~km, $a=2.94$~km,
$b=2.40$~km (diameter ranges from $2.11$~km to $2.85$~km) for the assumed albedo $p_V=0.0474$ -- the largest ultra-fast
rotator. The elongation is modest, $a/b\simeq1.23$. A
strengthless rubble pile of this size could not survive such a rotation;
following \citet{holsapple2007}, the cohesive strength required to
prevent centrifugal disruption is $\sim2.3$~MPa.

\textit{{566130.}} The adopted period of this object, $0.08899$~h or
$5.34$~minutes, narrowly misses our ultra-fast rotator threshold of five
minutes. The fit uses $195$ of its $198$ observations at the fourth
Fourier order and has a reduced $\chi^2$ of $2.40$, somewhat higher than
ideal. The object is confidently C-type (probability $96\%$), with a
diameter of $1.99$~km ($a=2.27$~km, $b=1.90$~km, $a/b\simeq1.20$) for the
assumed albedo. The required cohesive strength is correspondingly lower
than for 491384, $\sim0.9$~MPa.


\textit{{676624.}} This object rotates in $0.1129$~h, or $6.78$~minutes,
based on only $80$ observations, $2$ of which are rejected, fitted at the
fourth Fourier order; the reduced $\chi^2$ of $3.68$ is higher than
desirable but not alarming given the small sample. Two harmonic
alternatives survive, at $0.169$~h and $0.339$~h, both of which are also
fast but are preferred with lower probability ($0.89$ and $0.65$ probabilities that the selected period beats them,
respectively); the $0.169$~h alternative would require three maxima per
rotation and the longer one even more, so by Occam's razor we adopt the
shortest period of 0.1129 with 2 local maxima. The taxonomic class is confidently C-type (probability
$0.97$), and the amplitude of $0.56$~mag -- unusually large for this
sample -- implies an elongation of $a/b\simeq1.48$.

To resolve the residual harmonic ambiguity of 676624, we supplemented the
Rubin photometry with observations from the Pan-STARRS telescopes (MPC
observatory codes F51 and F52), taken in the $w$ filter between 2025 June 22
and 2025 September 18 and retrieved in ADES format from the Minor Planet
Center\footnote{\url{https://data.minorplanetcenter.net/api/get-obs}}.

We combined the two data sets following the approach of
\citet{2026Icar..45417074V}, in which the reduced $\chi^2$ of a fit measures
how well the assumed uncertainties match the actual scatter of the data, so
that rescaling each data set by the square root of its own reduced $\chi^2$
brings both to a common, empirically calibrated error scale. For the Rubin
data this factor is $\sqrt{3.68}$, taken from the Rubin-only fit, and we
inflated the reported Rubin uncertainties accordingly. To calibrate the
Pan-STARRS uncertainties, we temporarily down-weighted the Rubin data by a
factor of $10^8$, refitted at the same frequency, and found a reduced
$\chi^2$ of $0.91$ for the Pan-STARRS points; being close to unity, this
indicates that the MPC-reported uncertainties are already well scaled, and we
left them unchanged.

\begin{figure}
\centering
\includegraphics[width=0.3\textwidth]{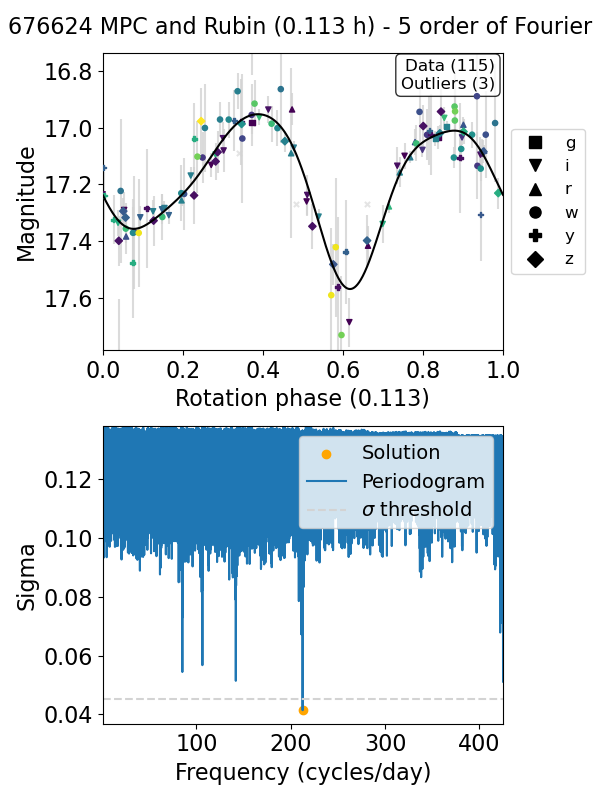}    

\caption{The folded lightcurve (top) and periodogram (bottom) for 676624 asteroid using data from Rubin observatory combined with Pan-STARRS 1 and 2 from June 22 until September 18 2025.}
	\label{fig:676624_MPC_and_Rubin}
\end{figure}

With both data sets included, the fit returns the same period as before,
$0.112924$~h, but now as a unique solution: the two harmonic alternatives
present in the Rubin-only fit no longer survive. The combined light curve and
periodogram are shown in Fig.~\ref{fig:676624_MPC_and_Rubin}. The addition of
the Pan-STARRS observations therefore removes the last remaining ambiguity,
and we can state with confidence that $1.9$~km asteroid 676624 rotates in $6.78$~min.

\clearpage
\startlongtable
\begin{deluxetable*}{lccclcccccr}
\tablecaption{The 66 super-fast rotator candidates ($P<2.2$~h) selected in this work, ordered by rotation period. Column $n$ is the number of Fourier harmonics of the adopted solution; ``cl'' gives the period class: 1 -- unique period, 2 -- harmonic alternatives only, 3 -- genuine alternatives that the adopted solution beats, 4 -- genuine alternatives that the adopted solution does not beat confidently. ``Alt.\ periods'' lists the surviving alternative periods in hours with, in parentheses, the Fourier order (when it differs from that of the adopted solution) and the Vuong probability with which the adopted solution is preferred. ``Class'' gives the two most probable taxonomic classes with their probabilities. $p_V$ is the geometric albedo assumed for the adopted class and $D$ the resulting diameter (Sect.~\ref{sec:size}); $a/b$ is the elongation from the light-curve amplitude; $\chi^2_\nu$ is the reduced $\chi^2$ of the fit; $N$ is the number of observations, with the number rejected by the fit in parentheses. \label{tab:super_fast_rotators}}
\tablehead{
\colhead{Object} & \colhead{$P$ (h)} & \colhead{$n$} & \colhead{cl} & \colhead{Alt.\ periods (h)} & \colhead{Class} & \colhead{$p_V$} & \colhead{$D$ (km)} & \colhead{$a/b$} & \colhead{$\chi^2_\nu$} & \colhead{$N$}
}
\startdata
2022 CJ2 & 0.031 & 2 & 1 & -- & S (0.69), M (0.19) & 0.20 & 0.08 & 1.27 & 1.19 & 280 (0) \\
2025 MN45 & 0.031 & 2 & 1 & -- & C (0.96), B (0.03) & 0.05 & 1.07 & 1.32 & 1.28 & 559 (3) \\
2025 MK41 & 0.063 & 4 & 1 & -- & C (0.97), B (0.02) & 0.05 & 0.81 & 1.23 & 1.11 & 449 (3) \\
491384 & 0.070 & 5 & 1 & -- & C (0.97), M (0.03) & 0.05 & 2.53 & 1.23 & 1.09 & 353 (1) \\
566130 & 0.089 & 4 & 1 & -- & C (0.96), B (0.02) & 0.05 & 1.99 & 1.20 & 2.40 & 195 (3) \\
158412 & 0.106 & 3 & 1 & -- & S (0.84), Q (0.11) & 0.22 & 1.57 & 1.07 & 0.87 & 80 (7) \\
676624 & 0.113 & 4 & 2 & 0.169\,(6, 0.89); 0.339\,(6, 0.65) & C (0.97), B (0.02) & 0.05 & 1.91 & 1.48 & 3.68 & 80 (2) \\
2025 AG16 & 0.124 & 6 & 1 & -- & C (0.95), P (0.03) & 0.05 & 0.94 & 1.40 & 1.64 & 246 (3) \\
240070 & 0.159 & 2 & 1 & -- & C (0.93), B (0.07) & 0.05 & 2.12 & 1.08 & 0.76 & 87 (7) \\
2025 MP39 & 0.191 & 2 & 1 & -- & Q (0.32), S (0.32) & 0.20 & 0.17 & 1.24 & 1.09 & 285 (2) \\
2025 MV71 & 0.214 & 4 & 1 & -- & C (0.93), K (0.03) & 0.05 & 1.41 & 1.56 & 1.28 & 318 (1) \\
2025 MW11 & 0.240 & 3 & 1 & -- & C (0.83), K (0.10) & 0.06 & 0.76 & 1.27 & 1.54 & 187 (1) \\
2024 FK27 & 0.243 & 2 & 1 & -- & C (1.00), S (0.00) & 0.05 & 1.47 & 1.43 & 1.56 & 80 (1) \\
2025 MG56 & 0.264 & 3 & 1 & -- & C (0.79), K (0.14) & 0.06 & 0.89 & 1.74 & 1.32 & 249 (1) \\
2025 MN25 & 0.361 & 2 & 1 & -- & C (0.90), P (0.06) & 0.05 & 0.49 & 1.28 & 1.74 & 343 (2) \\
2025 HZ6 & 0.364 & 4 & 3 & \shortstack[l]{0.545\,(0.92); 0.733\,(5, 0.84); \\ 0.909\,(5, 0.83); 1.096\,(6, 0.62)} & Q (0.82), S (0.08) & 0.24 & 0.03 & 2.80 & 309.62 & 113 (0) \\
673606 & 0.383 & 6 & 1 & -- & Q (0.75), S (0.25) & 0.23 & 0.67 & 1.32 & 0.92 & 131 (12) \\
2025 MU15 & 0.403 & 2 & 1 & -- & M (0.55), K (0.25) & 0.14 & 0.16 & 1.41 & 1.15 & 207 (0) \\
2025 KQ25 & 0.422 & 4 & 1 & -- & S (1.00), C (0.00) & 0.22 & 0.25 & 1.56 & 1.77 & 430 (2) \\
2025 KD30 & 0.598 & 2 & 3 & 0.594\,(0.90) & M (0.60), K (0.22) & 0.14 & 0.21 & 1.43 & 0.99 & 131 (0) \\
557484 & 0.656 & 3 & 1 & -- & S (0.67), X (0.17) & 0.19 & 1.10 & 1.15 & 1.60 & 249 (8) \\
2025 MU8 & 0.823 & 4 & 2 & 1.646\,(6, 0.12) & C (0.64), K (0.27) & 0.07 & 0.80 & 1.50 & 0.98 & 317 (0) \\
2025 ME68 & 0.947 & 3 & 2 & 1.894\,(4, 0.34) & S (0.36), L (0.32) & 0.22 & 0.16 & 1.51 & 0.96 & 185 (2) \\
526703 & 0.991 & 5 & 3 & 1.239\,(6, 0.80) & X (0.61), L (0.30) & 0.12 & 0.81 & 2.13 & 2.58 & 171 (1) \\
146127 & 1.044 & 8 & 1 & -- & C (0.53), B (0.45) & 0.06 & 4.76 & 2.86 & 81.70 & 87 (4) \\
2025 MM81 & 1.101 & 7 & 1 & -- & K (0.42), M (0.25) & 0.16 & 0.19 & 2.26 & 1.09 & 404 (2) \\
2025 JL & 1.109 & 7 & 1 & -- & Q (0.66), S (0.25) & 0.23 & 0.04 & 2.13 & 28.07 & 253 (6) \\
453009 & 1.137 & 4 & 1 & -- & Q (0.47), S (0.44) & 0.23 & 0.66 & 1.21 & 2.06 & 149 (2) \\
2016 UO174 & 1.151 & 4 & 1 & -- & C (0.94), B (0.04) & 0.05 & 1.66 & 1.39 & 1.13 & 434 (0) \\
2017 QW90 & 1.194 & 4 & 1 & -- & S (0.67), Q (0.22) & 0.22 & 0.83 & 1.62 & 1.40 & 414 (1) \\
2025 MZ78 & 1.198 & 3 & 1 & -- & C (0.79), P (0.10) & 0.05 & 1.64 & 1.54 & 0.87 & 253 (1) \\
461805 & 1.254 & 2 & 1 & -- & L (0.45), S (0.21) & 0.19 & 1.18 & 1.12 & 1.59 & 99 (3) \\
216708 & 1.302 & 3 & 3 & \shortstack[l]{2.604\,(4, 0.31); 4.675\,(5, 0.88); \\ 5.208\,(8, 0.01)} & S (0.71), Q (0.20) & 0.23 & 0.71 & 1.15 & 2.07 & 87 (6) \\
111058 & 1.345 & 2 & 3 & 1.274\,(0.89); 2.547\,(3, 0.92) & K (0.59), M (0.20) & 0.15 & 2.85 & 1.08 & 3.14 & 86 (2) \\
849371 & 1.393 & 2 & 1 & -- & Q (0.70), S (0.26) & 0.23 & 0.58 & 1.27 & 1.60 & 158 (1) \\
2014 HR185 & 1.401 & 4 & 1 & -- & S (0.65), Q (0.23) & 0.22 & 0.39 & 1.25 & 1.41 & 394 (4) \\
2025 MC8 & 1.433 & 3 & 4 & \shortstack[l]{1.351\,(0.77); 1.525\,(0.88); \\ 2.867\,(4, 0.14); 4.578\,(0.90)} & C (0.92), P (0.04) & 0.05 & 0.97 & 1.26 & 1.27 & 141 (0) \\
2025 MZ70 & 1.513 & 4 & 1 & -- & S (0.52), Q (0.39) & 0.23 & 0.48 & 1.37 & 0.92 & 186 (1) \\
30991 & 1.555 & 3 & 1 & -- & K (0.59), S (0.27) & 0.16 & 4.13 & 1.13 & 5.56 & 99 (12) \\
543738 & 1.576 & 2 & 4 & 1.686\,(0.84); 2.957\,(0.73) & M (0.56), K (0.28) & 0.14 & 0.92 & 1.12 & 0.88 & 360 (4) \\
2025 ML24 & 1.623 & 2 & 4 & \shortstack[l]{1.521\,(0.79); 1.740\,(0.82); \\ 3.483\,(3, 0.82)} & C (0.92), K (0.03) & 0.05 & 1.07 & 1.12 & 1.17 & 336 (1) \\
727545 & 1.632 & 2 & 1 & -- & C (0.97), B (0.02) & 0.05 & 1.33 & 1.07 & 1.28 & 424 (1) \\
2015 KR298 & 1.633 & 5 & 1 & -- & C (0.96), P (0.02) & 0.05 & 0.79 & 1.50 & 1.21 & 138 (0) \\
193811 & 1.669 & 6 & 1 & -- & C (0.88), S (0.09) & 0.06 & 2.15 & 1.32 & 1.29 & 173 (36) \\
694717 & 1.681 & 5 & 2 & 5.044\,(6, 0.68) & S (0.73), K (0.25) & 0.20 & 1.82 & 1.41 & 3.67 & 173 (4) \\
866105 & 1.688 & 3 & 4 & \shortstack[l]{1.818\,(0.91); 3.378\,(4, 0.94); \\ 5.452\,(0.84); 5.913\,(0.93); \\ 9.819\,(5, 0.87)} & K (0.34), M (0.23) & 0.17 & 0.64 & 1.18 & 1.10 & 201 (2) \\
2022 QS59 & 1.697 & 4 & 3 & \shortstack[l]{3.395\,(5, 0.55); 4.748\,(6, 0.93); \\ 5.937\,(6, 0.94); 6.786\,(6, 0.70)} & S (0.82), M (0.07) & 0.22 & 0.57 & 1.31 & 1.28 & 310 (6) \\
2025 KP27 & 1.714 & 4 & 4 & 5.141\,(6, 0.60); 9.235\,(5, 0.71) & C (0.60), B (0.40) & 0.05 & 0.93 & 1.49 & 1.06 & 279 (5) \\
2025 MB22 & 1.744 & 4 & 1 & -- & C (0.71), K (0.21) & 0.07 & 0.50 & 1.20 & 1.19 & 373 (0) \\
819328 & 1.755 & 3 & 4 & \shortstack[l]{1.633\,(0.85); 1.897\,(0.88); \\ 3.510\,(4, 0.92)} & C (0.93), B (0.05) & 0.05 & 1.94 & 1.08 & 1.11 & 388 (11) \\
650003 & 1.773 & 4 & 1 & -- & C (0.95), B (0.02) & 0.05 & 1.91 & 1.43 & 1.27 & 307 (2) \\
752203 & 1.797 & 2 & 1 & -- & S (0.44), L (0.37) & 0.21 & 0.70 & 1.14 & 1.15 & 434 (3) \\
314482 & 1.803 & 5 & 2 & 3.606\,(6, 0.46); 5.409\,(6, 0.83) & K (0.75), M (0.11) & 0.13 & 3.17 & 1.46 & 2.62 & 274 (13) \\
798479 & 1.811 & 6 & 1 & -- & S (0.33), L (0.33) & 0.22 & 0.82 & 1.26 & 1.35 & 344 (1) \\
2020 GJ27 & 1.826 & 5 & 1 & -- & C (0.85), S (0.14) & 0.07 & 0.88 & 4.14 & 3.45 & 128 (10) \\
56632 & 1.877 & 3 & 2 & 3.756\,(4, 0.17) & C (0.95), B (0.04) & 0.05 & 3.24 & 1.13 & 1.45 & 319 (14) \\
2025 MX50 & 1.940 & 5 & 1 & -- & L (0.69), S (0.14) & 0.19 & 0.34 & 1.21 & 1.33 & 479 (0) \\
2025 MQ65 & 1.952 & 4 & 1 & -- & C (0.94), B (0.06) & 0.05 & 0.96 & 1.93 & 1.13 & 86 (1) \\
2019 RB90 & 1.988 & 4 & 1 & -- & Q (0.48), S (0.27) & 0.21 & 0.37 & 1.30 & 1.08 & 469 (1) \\
2025 MF76 & 2.026 & 4 & 2 & 4.054\,(5, 0.83) & L (0.45), K (0.22) & 0.19 & 0.51 & 1.11 & 1.14 & 268 (1) \\
4476 & 2.046 & 4 & 1 & -- & X (0.58), S (0.28) & 0.14 & 5.15 & 1.16 & 6.52 & 90 (6) \\
422054 & 2.080 & 6 & 2 & 4.159\,(8, 0.31) & X (0.47), S (0.45) & 0.15 & 0.89 & 1.84 & 6.71 & 105 (5) \\
110253 & 2.103 & 8 & 1 & -- & K (0.39), C (0.33) & 0.10 & 2.26 & 1.07 & 1.51 & 421 (3) \\
93341 & 2.108 & 5 & 1 & -- & S (0.83), Q (0.14) & 0.22 & 1.39 & 1.35 & 9.97 & 289 (72) \\
2015 PF40 & 2.156 & 2 & 3 & 2.063\,(0.87) & S (0.56), L (0.21) & 0.20 & 0.56 & 1.41 & 1.13 & 189 (0) \\
2021 GY121 & 2.188 & 6 & 1 & -- & S (0.71), Q (0.13) & 0.23 & 0.50 & 1.18 & 1.24 & 330 (1) \\
\enddata
\end{deluxetable*}

\textit{{2025~AG16.}} This main-belt asteroid has a single, unambiguous
period of $0.123752$~h, i.e. $7.43$~minutes, and, like 566130, narrowly
misses our ultra-fast rotator threshold. Its taxonomic class is
confidently C-type (probability $0.95$), and for the corresponding albedo
$p_V=0.047$ we obtain a diameter of $0.94$~km. Its light-curve amplitude
implies the elongation  $a/b\simeq1.40$.

Two further objects, 158412 and 240070, share a similar profile: both
have relatively few observations, a confidently determined taxonomic
class, and a reduced $\chi^2$ below unity that is nonetheless consistent
with the number of degrees of freedom once the floor of
Eq.~(\ref{eq:chi2floor}) is taken into account, so that neither case
indicates a problem with the fit. \textit{{158412}} rotates in
$0.1059$~h ($6.35$~minutes), based on $80$ observations with $7$
rejected; although the periodogram shows secondary minima, adopting any
of them would require excluding substantially more observations, so the
$6.35$-minute solution remains the only viable one. Its light-curve
coverage is $83\%$, close to the $89\%$ ceiling expected for this number
of observations (Eq.~\ref{eq:ceiling}), with good group coverage and a
maximum uncovered gap of only $10\%$. The most probable class is S-type
(probability $0.84$), and from $H_V=16.26$ we obtain a diameter of
$1.57$~km ($a=1.66$~km, $b=1.55$~km), with a modest elongation of
$a/b\simeq1.07$. \textit{{240070}} rotates in $0.159$~h, based on $87$
observations with $7$ rejected at the second Fourier order; its coverage
is likewise base-level ($84\%$ against a $92\%$ ceiling) but fully
confirmed by groups. The class is confidently C-type (probability
$0.93$), with a diameter of $2.12$~km ($a=2.25$~km, $b=2.08$~km,
$a/b\simeq1.08$).

Given that both 240070 and 158412 are over  a kilometer in size super-fast rotating asteroids with a possible rotation period less than 10 minutes, we recommend dedicated follow-up observations to confirm their periods.

Two more objects, 2025~MP39 and 2025~MW11, illustrate that a large
individual photometric uncertainty does not by itself prevent a secure
period determination, provided the number of observations and the
light-curve amplitude are large enough. \textit{{2025~MP39}} has a period
of $0.1913$~h ($11.48$~minutes); its individual measurements carry a
weighted-mean uncertainty of $0.16$~mag, yet $285$ observations combined
with an amplitude of $0.32$~mag and full phase and group coverage isolate
a single, unambiguous periodogram minimum. Its taxonomic class is not
securely determined. \textit{{2025~MW11}} follows the same pattern at a
period of $0.240$~h: a per-point uncertainty of $0.11$~mag is compensated
by $187$ observations, full coverage, and a confident C-type
classification (probability $0.83$), giving a diameter of $0.76$~km
($a=0.91$~km, $b=0.71$~km, $a/b\simeq1.27$) assuming $0.057$ weighted albedo.

Finally, \textit{{2024~FK27}} shows that even a modest number of
observations can yield a secure period if the amplitude is large relative
to the photometric noise. With only $80$ observations ($1$ rejected) and
a per-point uncertainty of $0.077$~mag, the amplitude of $0.49$~mag is
still large enough to isolate a unique periodogram minimum, at
$0.242873$~h ($14.57$~minutes, second Fourier order); the light-curve
coverage of $84\%$ is close to its $92\%$ ceiling, confirmed by groups
over $99\%$ of the phase. The taxonomic class is C-type with $99.99\%$
confidence, giving a diameter of $1.47$~km ($a=1.90$~km, $b=1.33$~km,
$a/b\simeq1.43$).

\subsubsection{Objects with a unique period}
\label{sec:uniqueperiod}

\begin{figure*}
\centering
\includegraphics[width=0.23\textwidth]{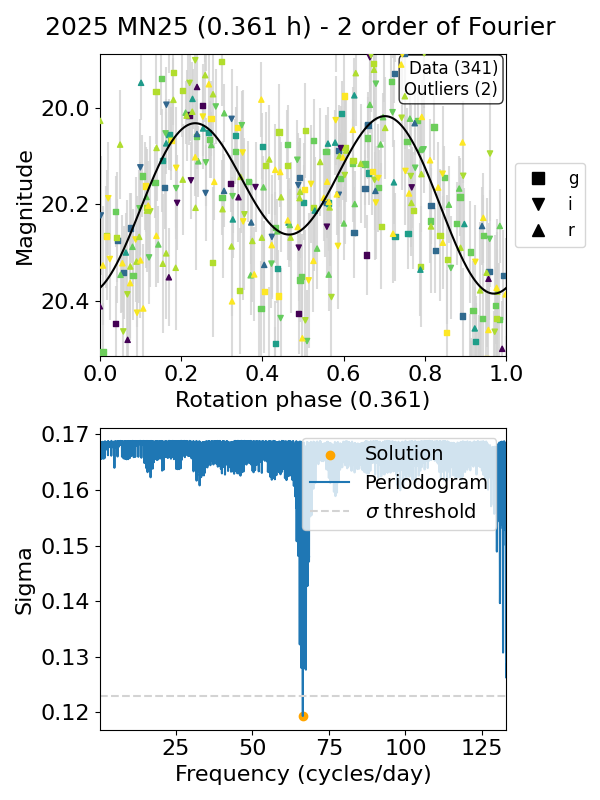}   \includegraphics[width=0.23\textwidth]{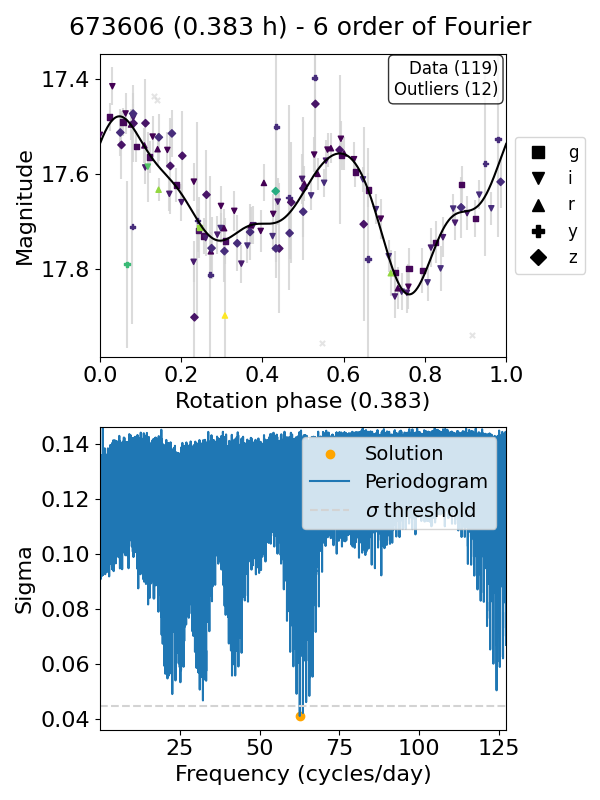} 
\includegraphics[width=0.23\textwidth]{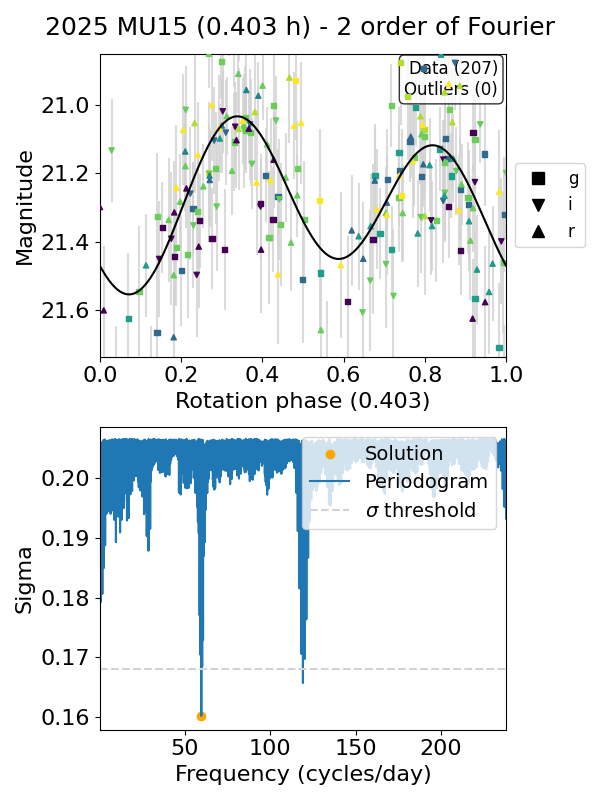} 
\includegraphics[width=0.23\textwidth]{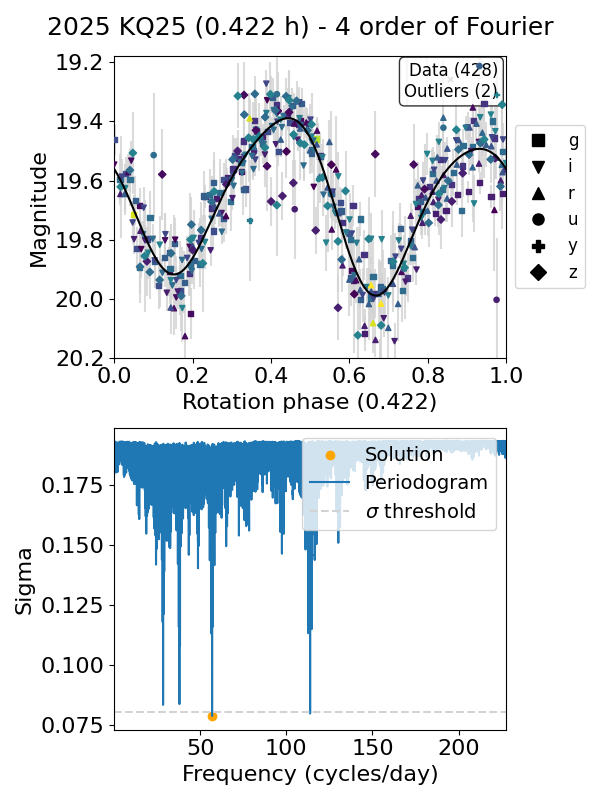}    \\ 
\includegraphics[width=0.23\textwidth]{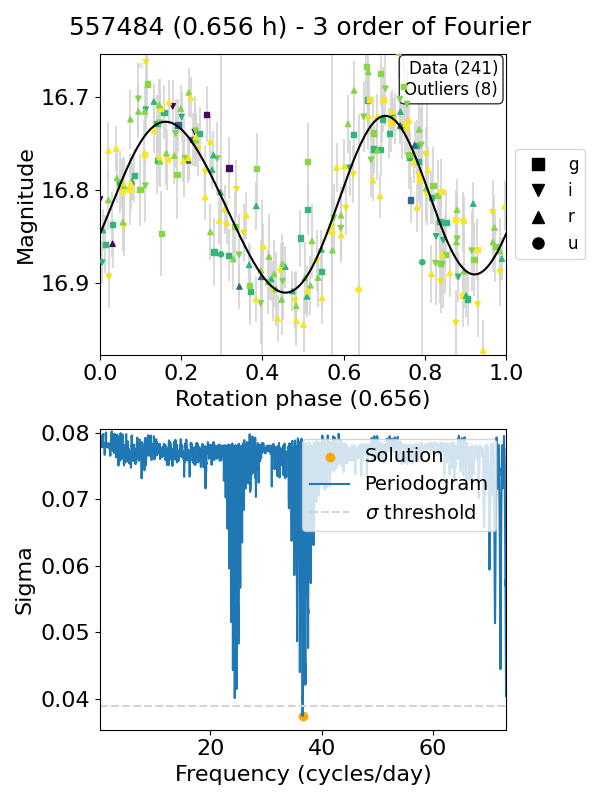}   \includegraphics[width=0.23\textwidth]{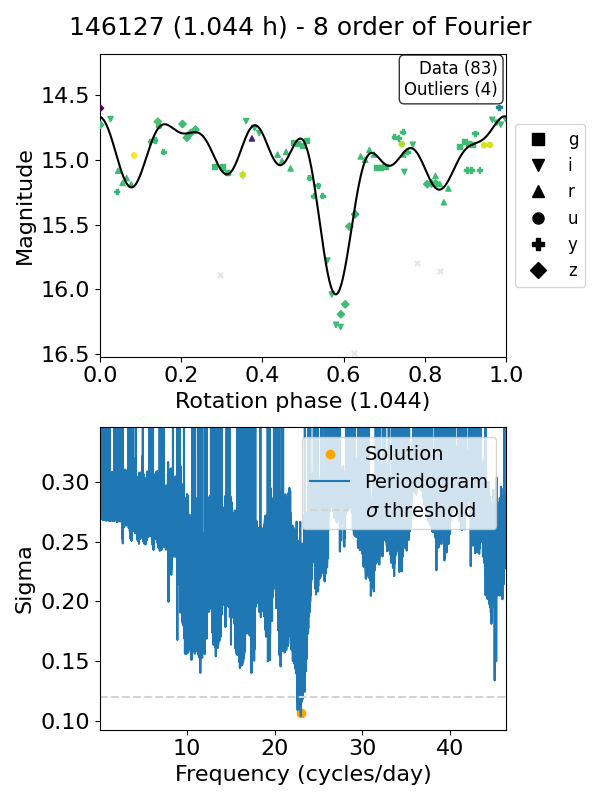} 
\includegraphics[width=0.23\textwidth]{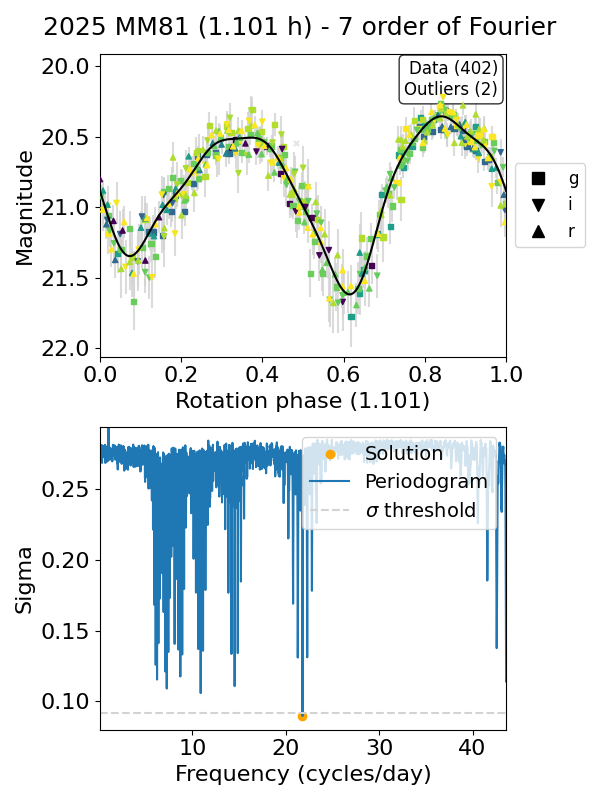} 
\includegraphics[width=0.23\textwidth]{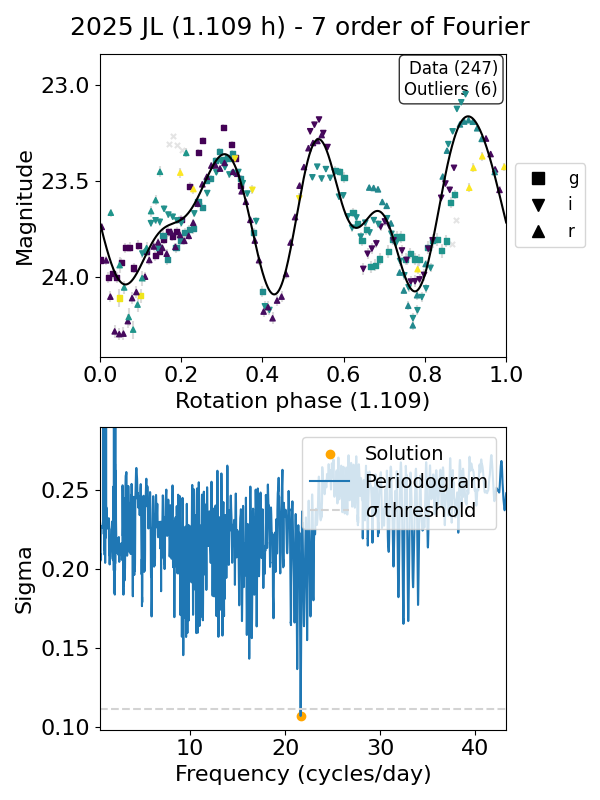}    \\ 
\includegraphics[width=0.23\textwidth]{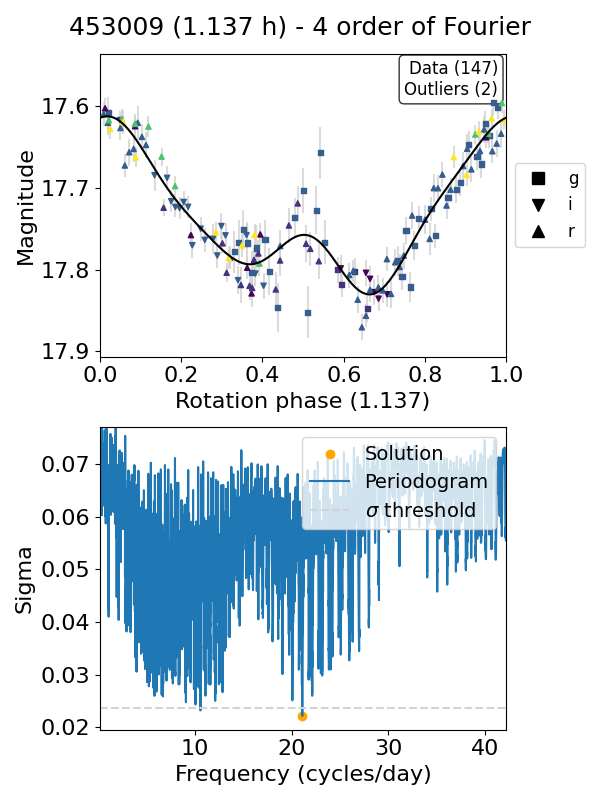}   \includegraphics[width=0.23\textwidth]{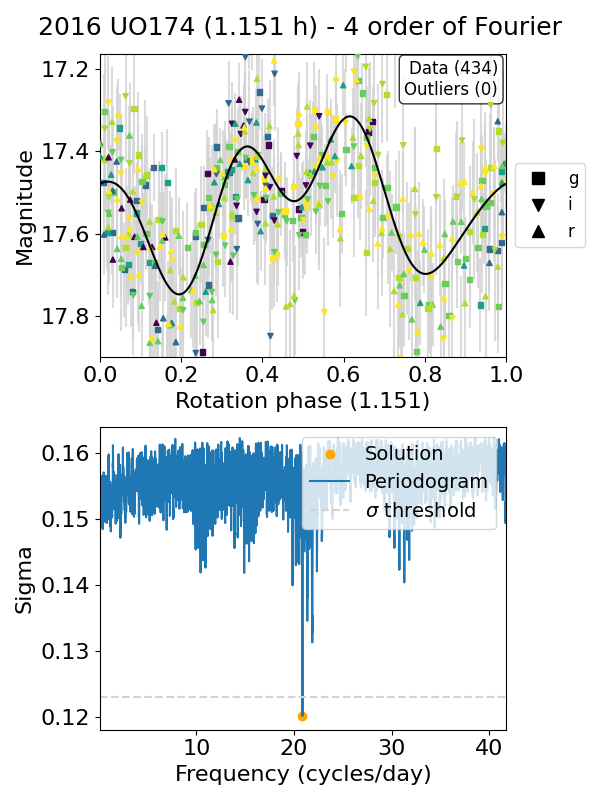} 
\includegraphics[width=0.23\textwidth]{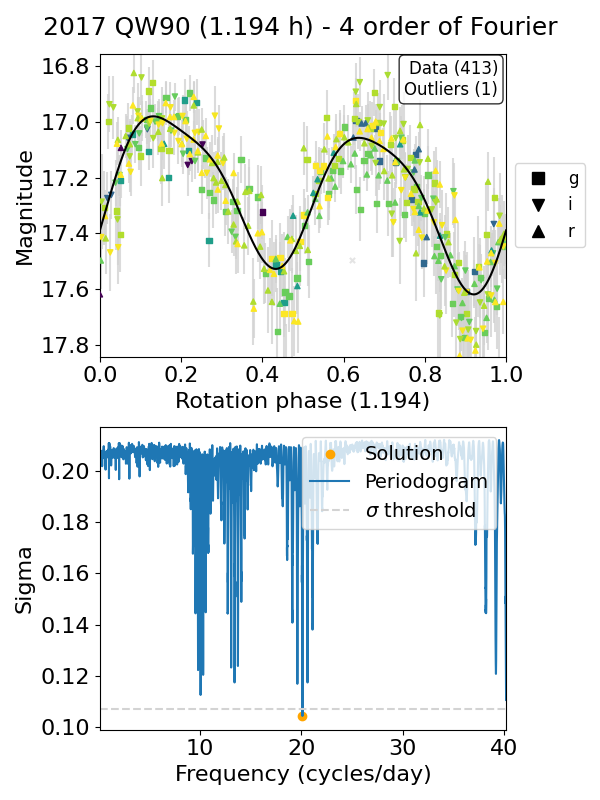} 
\includegraphics[width=0.23\textwidth]{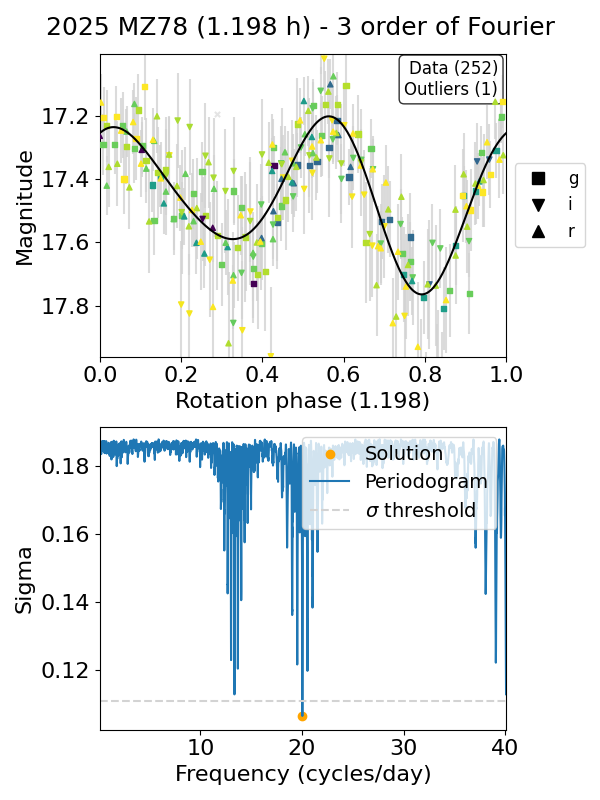}    \\ 
\includegraphics[width=0.23\textwidth]{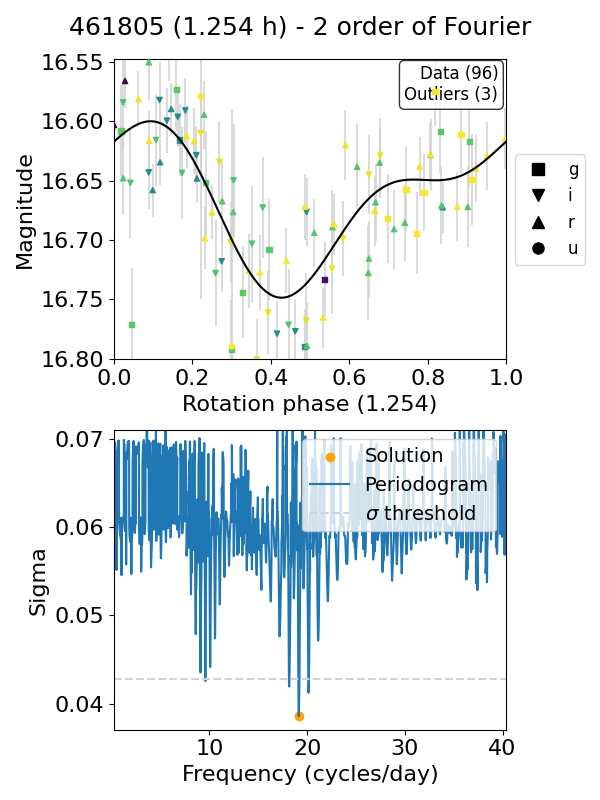}   \includegraphics[width=0.23\textwidth]{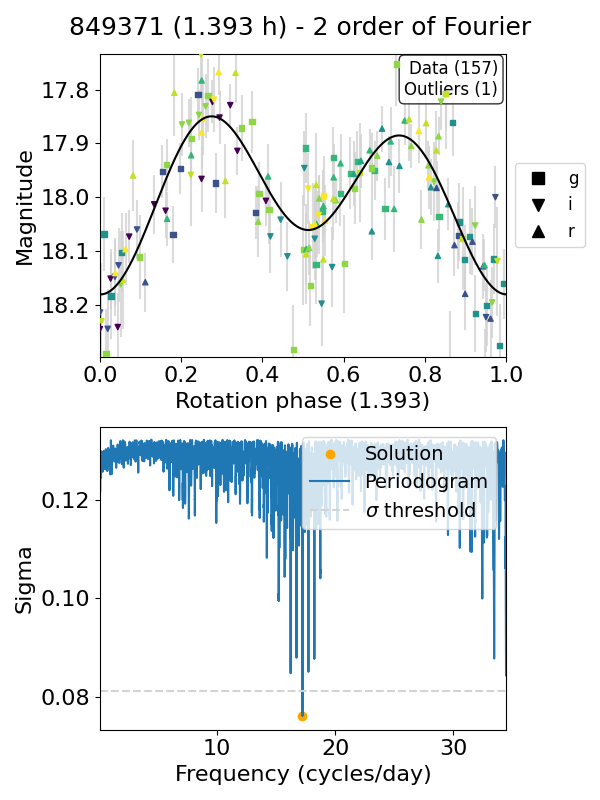} 
\includegraphics[width=0.23\textwidth]{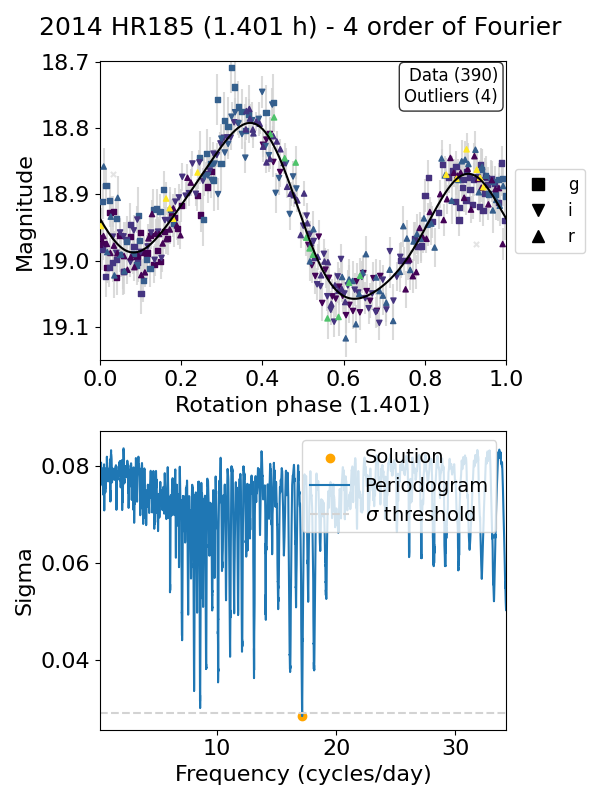} 
\includegraphics[width=0.23\textwidth]{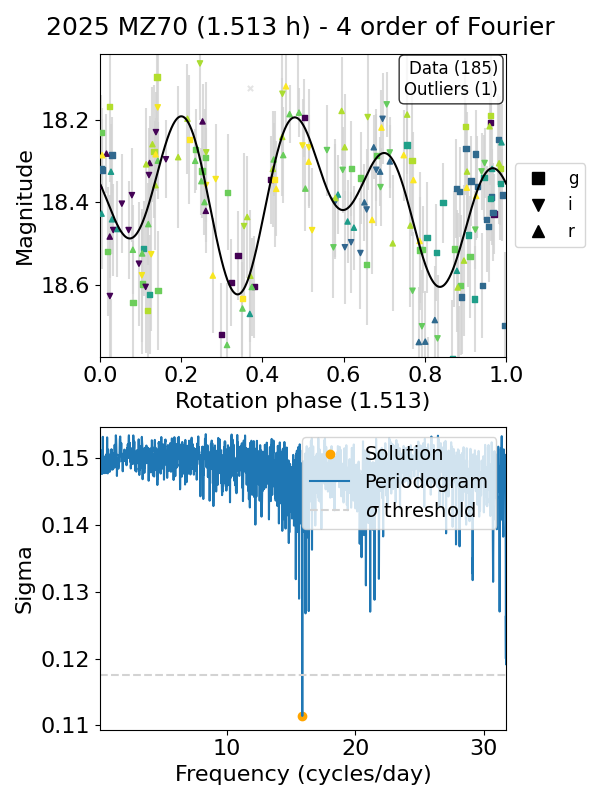}    
\caption{Same as for Fig.~\ref{fig:fastests} but for 32 super-fast rotators with uniquely determined period. Some of these objects were reported by \citet{greenstreet2026}, and our results agree with theirs. Some are questionable (see text). }
\label{fig:unique}
\end{figure*}

\begin{figure*}

\centering
\includegraphics[width=0.23\textwidth]{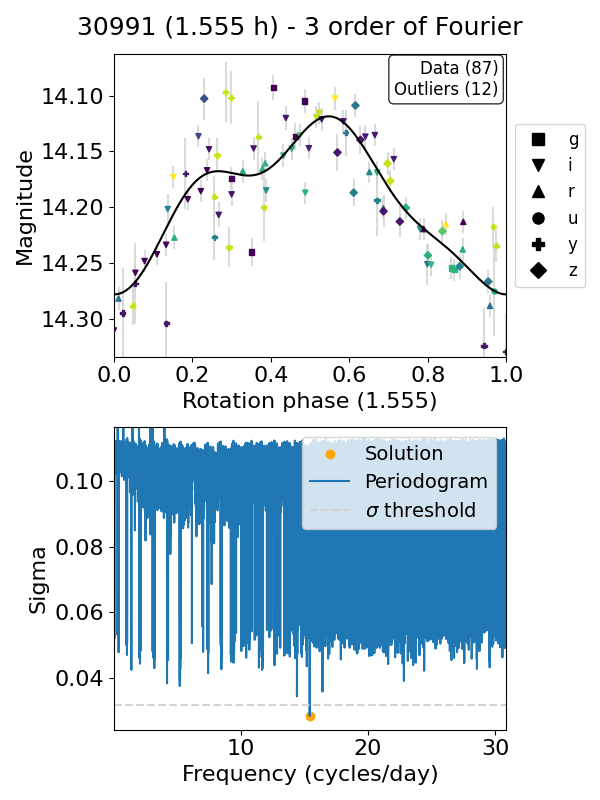}   \includegraphics[width=0.23\textwidth]{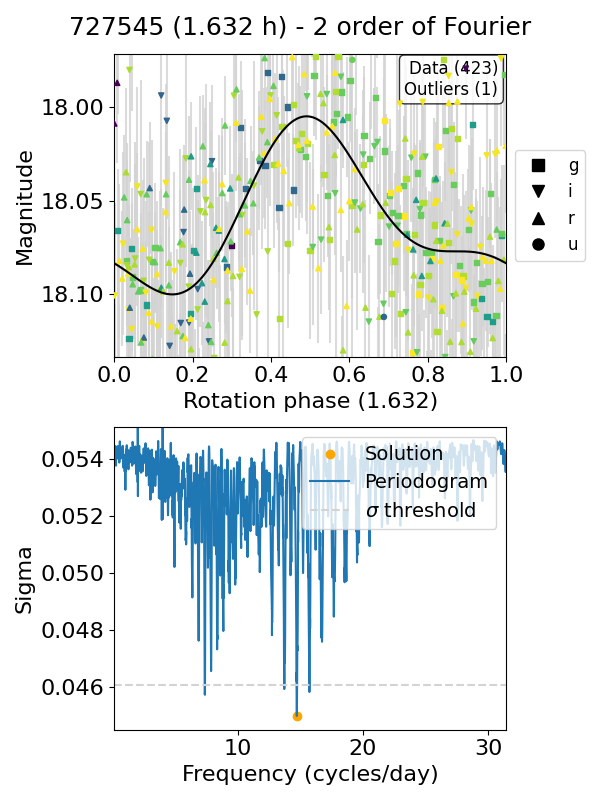} 
\includegraphics[width=0.23\textwidth]{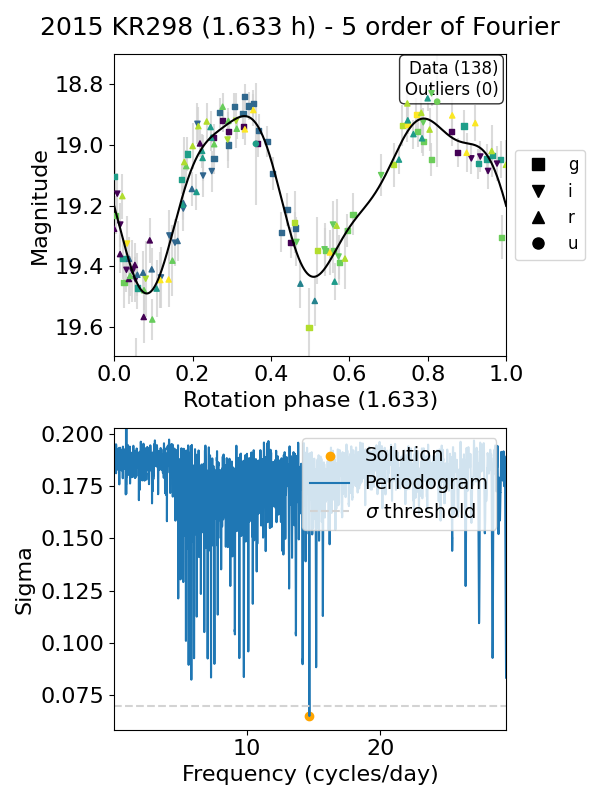} 
\includegraphics[width=0.23\textwidth]{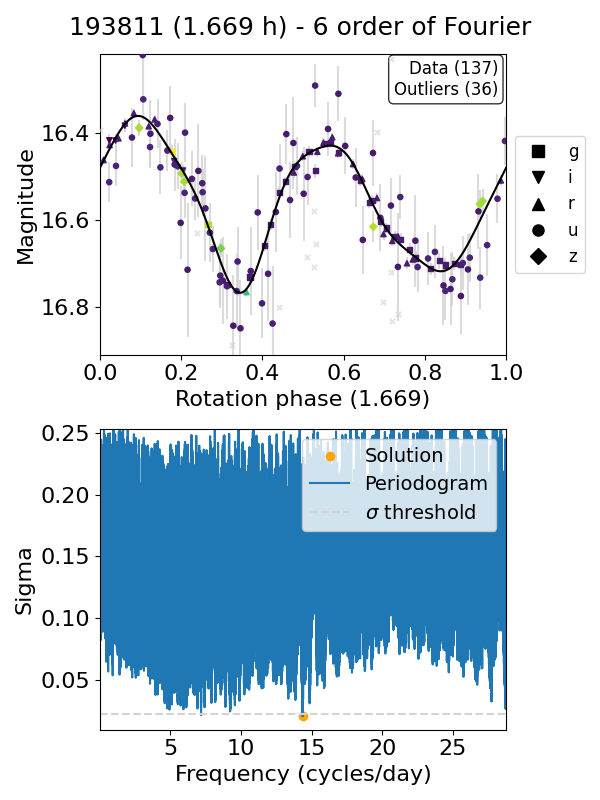}    \\ 
\includegraphics[width=0.23\textwidth]{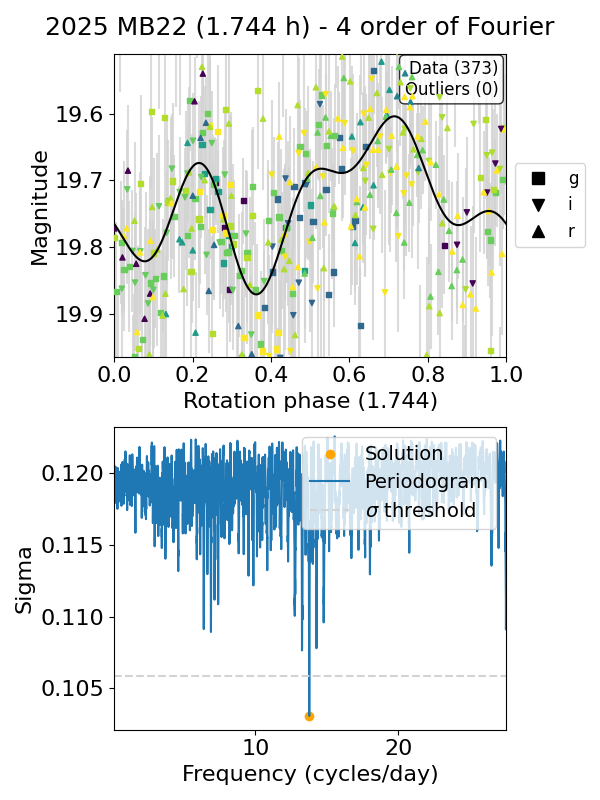}   \includegraphics[width=0.23\textwidth]{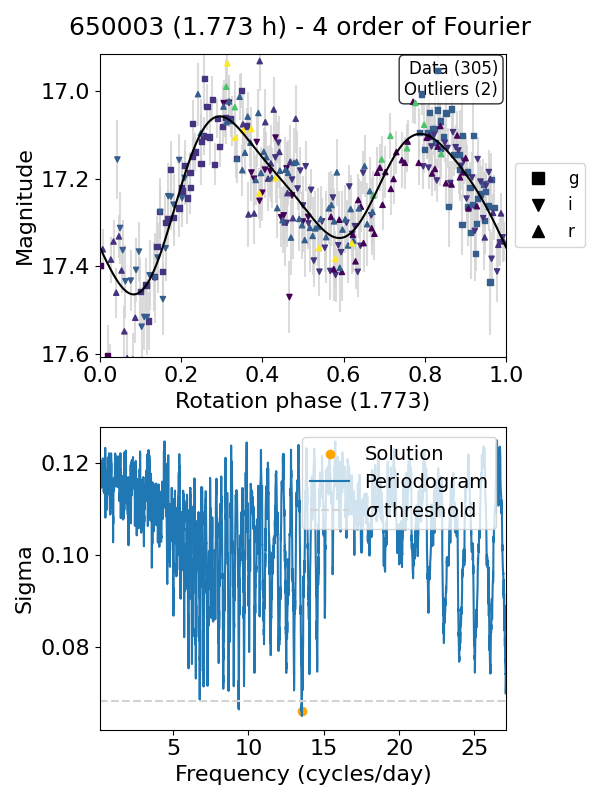} 
\includegraphics[width=0.23\textwidth]{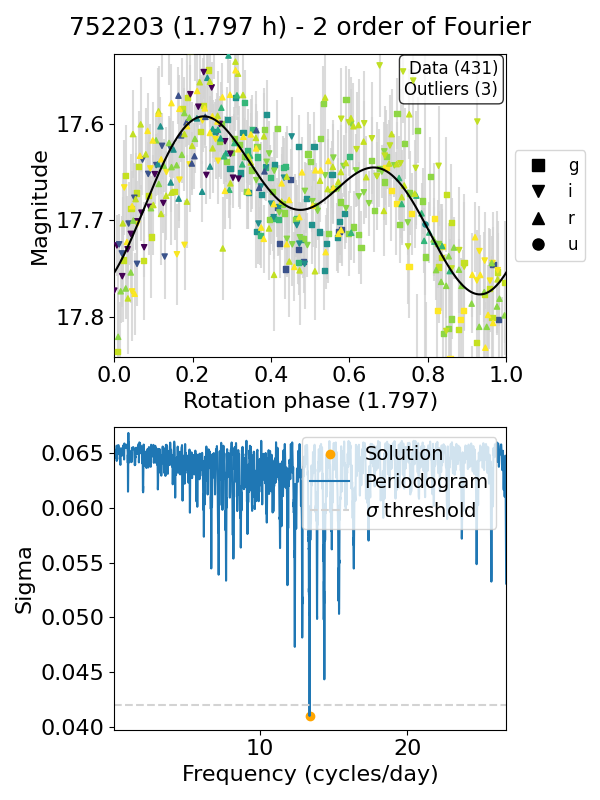} 
\includegraphics[width=0.23\textwidth]{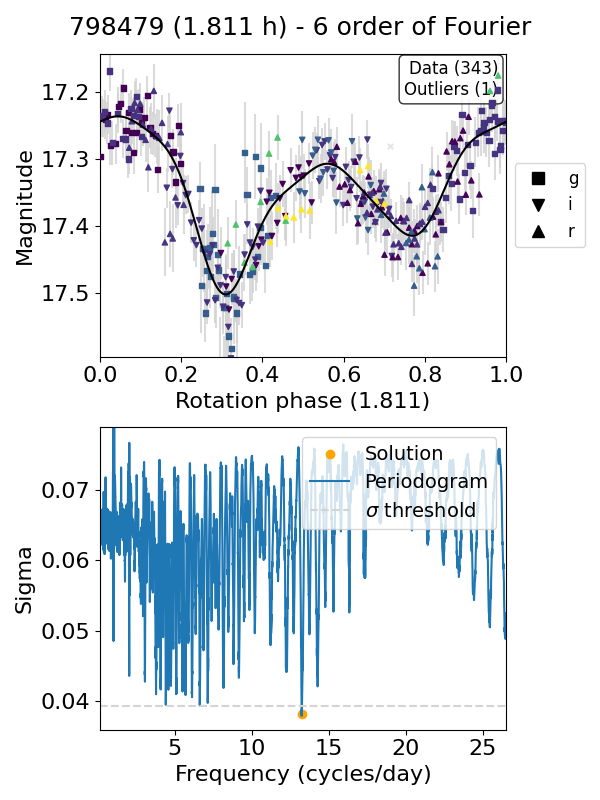}    \\ 
\includegraphics[width=0.23\textwidth]{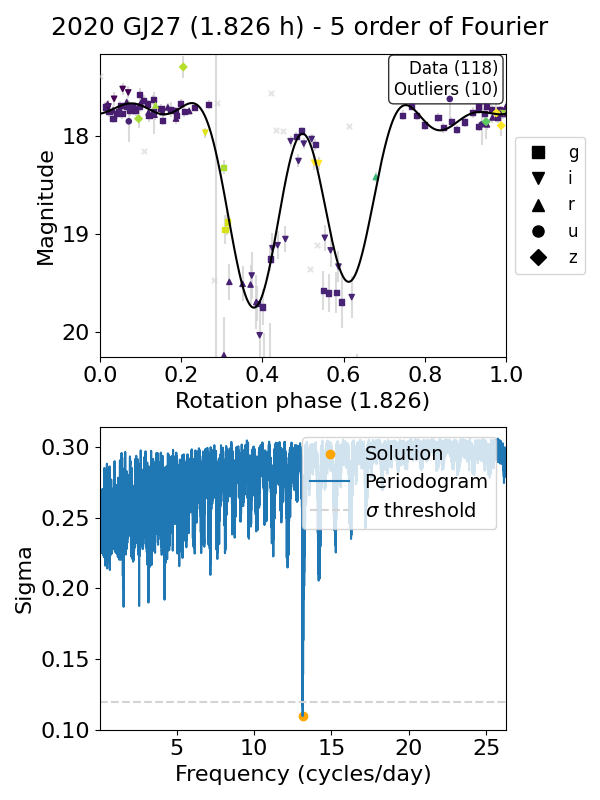}   \includegraphics[width=0.23\textwidth]{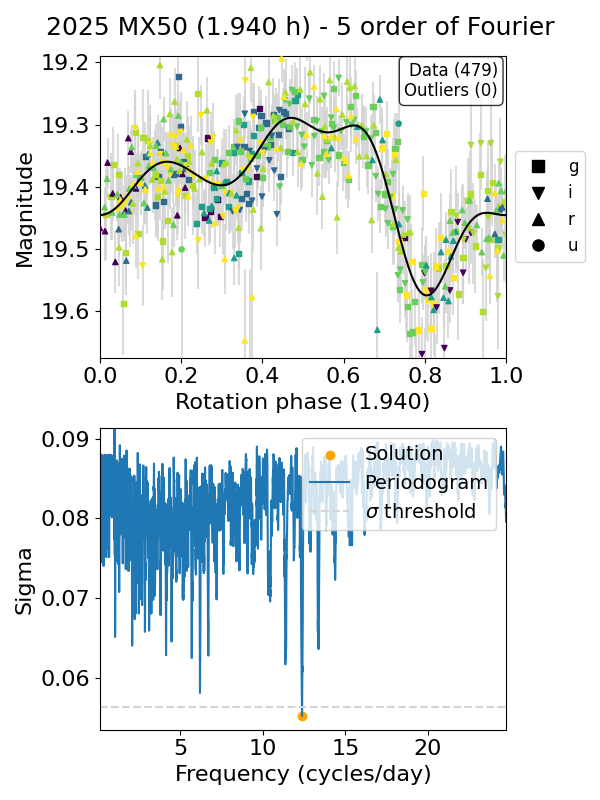} 
\includegraphics[width=0.23\textwidth]{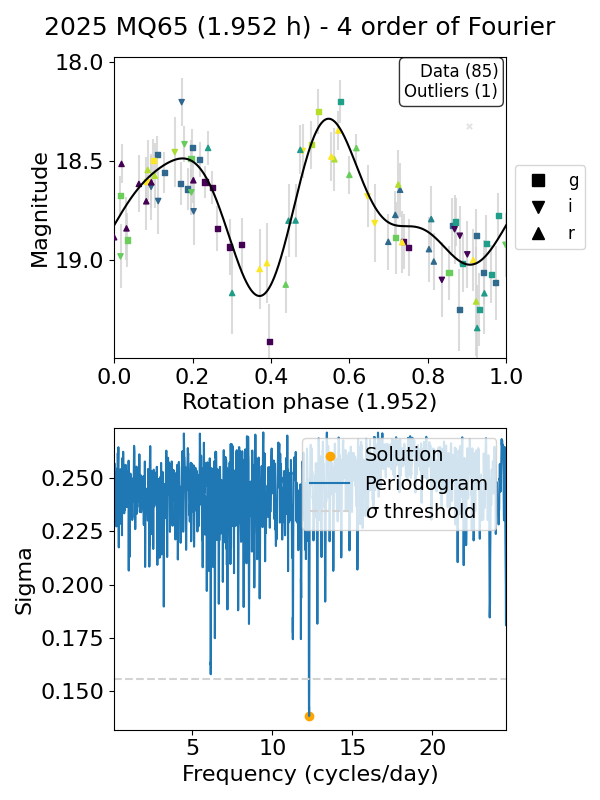} 
\includegraphics[width=0.23\textwidth]{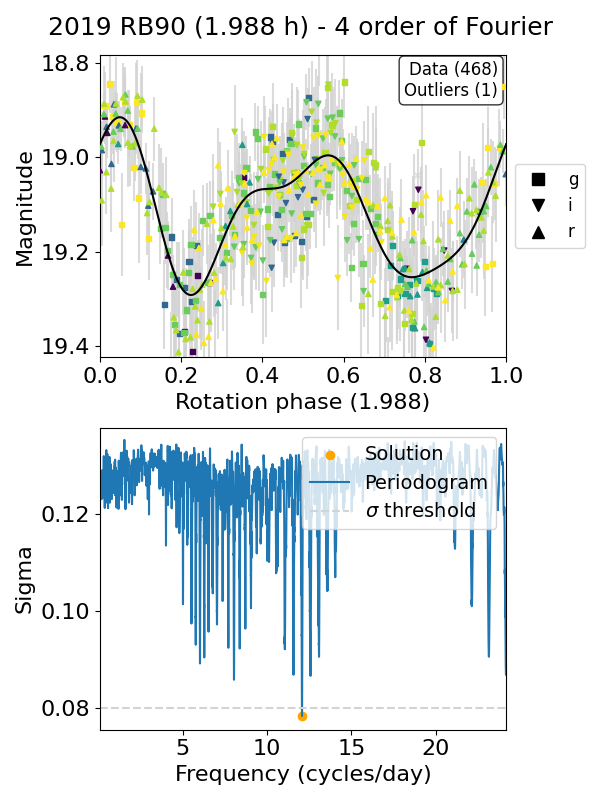}    \\ 
\includegraphics[width=0.23\textwidth]{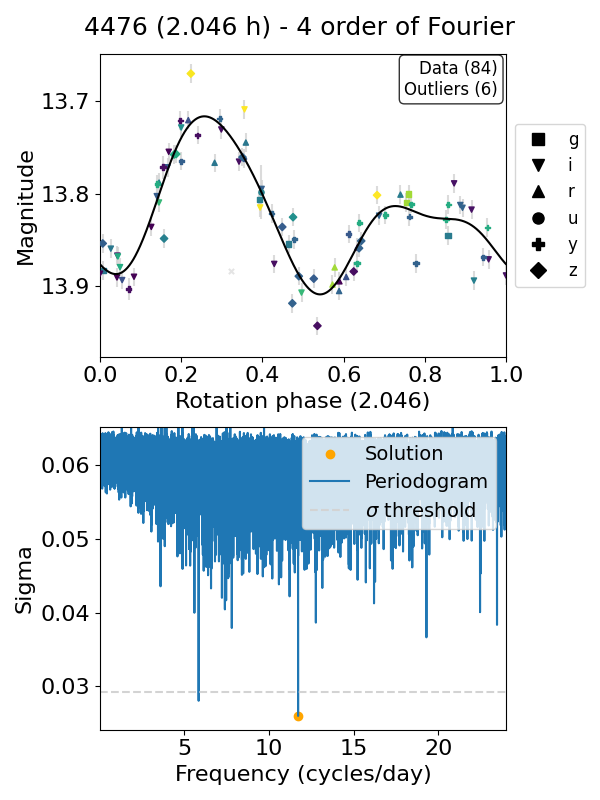}   \includegraphics[width=0.23\textwidth]{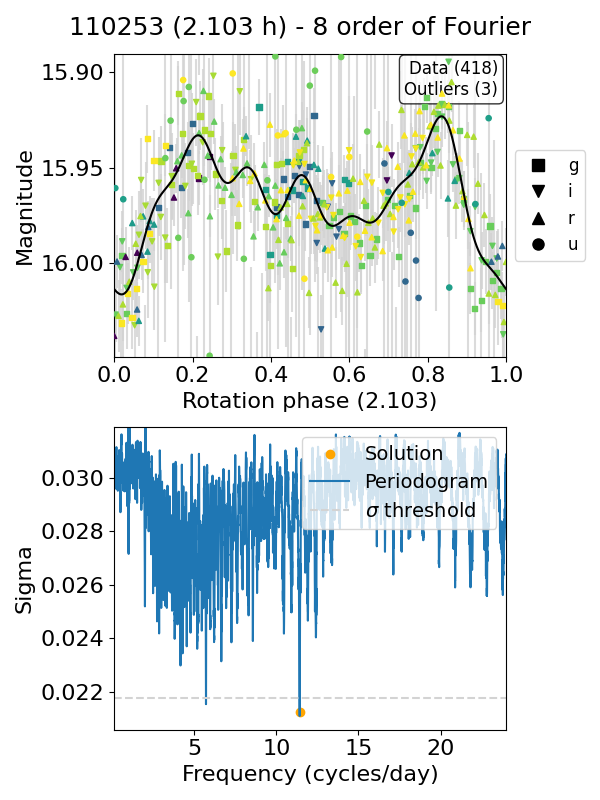} 
\includegraphics[width=0.23\textwidth]{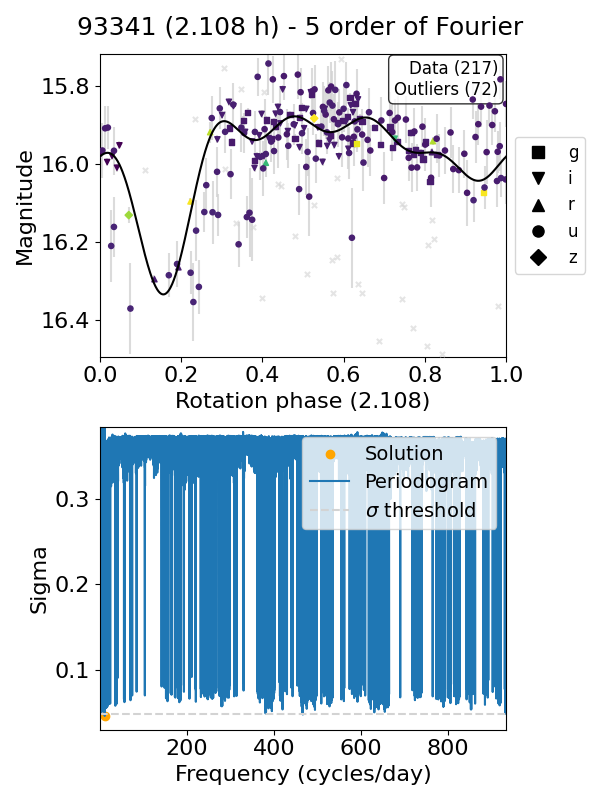} 
\includegraphics[width=0.23\textwidth]{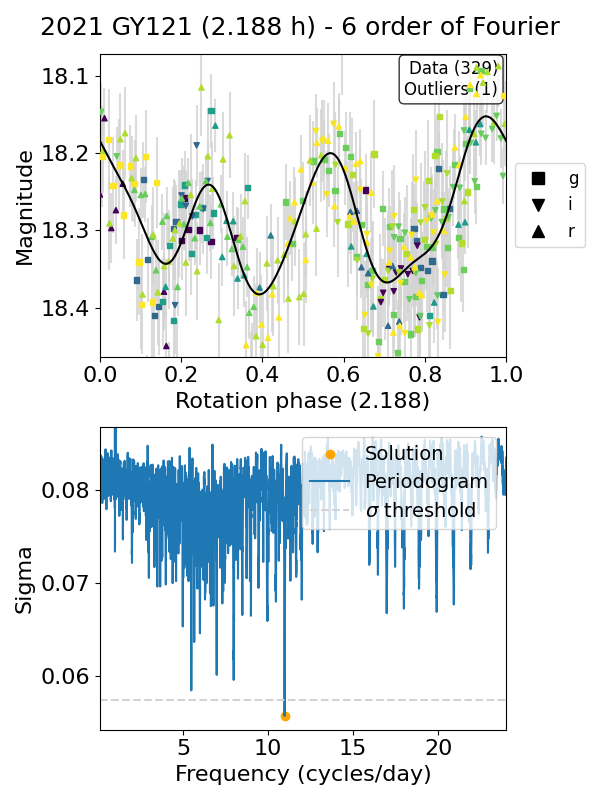}   

\addtocounter{figure}{-1}
\caption{(continued) Same as for Fig.~\ref{fig:fastests} but for 32 super-fast rotators with uniquely determined period. Some of these objects were reported by \citet{greenstreet2026}, and our results agree with theirs. Some are questionable (see text). }

\end{figure*}

A further 32 objects of the sample have a uniquely determined period,
with adopted values ranging from $22$~minutes to the $2.2$~h spin
barrier; their properties are listed in
Table~\ref{tab:super_fast_rotators}. We do not discuss every object
individually, but highlight several of particular interest.

Five of these objects were already reported by \citet{greenstreet2026}:
2025~MN25, 2025~MU15, 2025~MM81, 2025~MZ78, and 2025~MX50. The adopted
periods agree with theirs in every case, and the present DP2-based sample
provides more photometry for all five: from $318$ to $343$ observations
for 2025~MN25, $182$ to $207$ for 2025~MU15, $390$ to $404$ for
2025~MM81, $167$ to $253$ for 2025~MZ78, and $425$ to $479$ for
2025~MX50.

Among the objects of this group, 2025~MM81 has the largest light-curve
amplitude and the correspondingly largest elongation, $a/b\simeq2.26$.
Two objects discussed below, 146127 and 2020~GJ27, formally have an even
larger nominal elongation ($a/b\simeq2.86$ and $4.14$, respectively), but
given the problems with their fits described next, we do not consider
these values reliable, and 2025~MM81 remains the most robustly elongated
object of the unique-period group.

For three objects -- 461805, 30991, and 727545 -- a single dominant
maximum is visible in the folded light curve, and a secondary maximum is
present but too weak or too poorly sampled to establish with confidence
whether the true period is the adopted one or twice its value; we
therefore flag these periods as possibly requiring doubling, without
being able to decide the question with the present data.

Three further objects -- 93341, 2020~GJ27, and 146127 -- show a light
curve that is visibly poorly described by the model, and all three share
the same underlying symptom: a subset of their observations, offset in
magnitude from the bulk of the data, that the rejection procedure
struggles to fully remove. For 93341, $72$ of the $289$ observations were
excluded, yet the reduced $\chi^2$ of the remaining fit is still
$9.97$, well above unity, and the periodogram minima adopted at different
Fourier orders do not agree with one another, which is itself a sign of
an unreliable solution. 2020~GJ27 has a more moderate reduced $\chi^2$ of
$3.45$ and rejects only $10$ of its $128$ observations, but its
light-curve coverage is just $83\%$ -- low for $118$ used
observations -- and its formally large elongation should be treated with
caution given the fit quality. 146127 is the most extreme case: with only
$87$ observations and $4$ rejected, the reduced $\chi^2$ reaches $81.7$.
In all three cases, a plausible explanation is that some epochs are
affected by problematic observations (Sect.~\ref{sec:baddata} discusses
this in more detail), though an unresolved binary or otherwise
non-single-body light curve cannot be excluded either.

Two more objects show an elevated reduced $\chi^2$ without being quite as
extreme as the three above, and are worth flagging alongside them: 2025~JL
($\chi^2_\nu=28.1$, comparable in severity to 93341) and 4476
($\chi^2_\nu=6.5$); 30991, already mentioned above as a possible
period-doubling candidate, also has a somewhat elevated $\chi^2_\nu=5.6$.
None of these reach the extremes of 146127 or 93341, but their fits should
likewise be regarded with some caution. Separately, 193811 rejects an
unusually large fraction of its observations ($36$ of $173$, or
$21\%$), the highest rejection fraction in this group after 93341; the
adopted period nonetheless remains the unique periodogram minimum and the fit looks very good.

\subsubsection{Remaining candidates}
\label{sec:remaining}

\begin{figure*}
\centering
\includegraphics[width=0.19\textwidth]{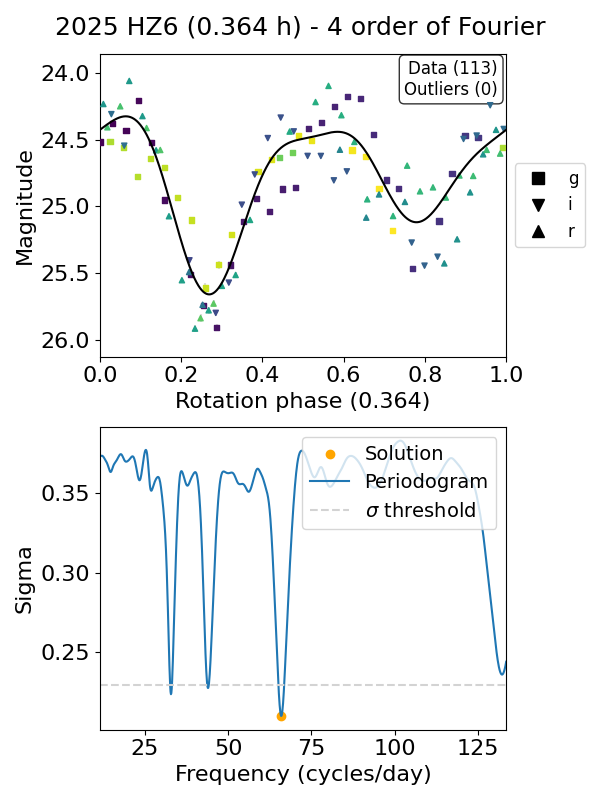}   \includegraphics[width=0.19\textwidth]{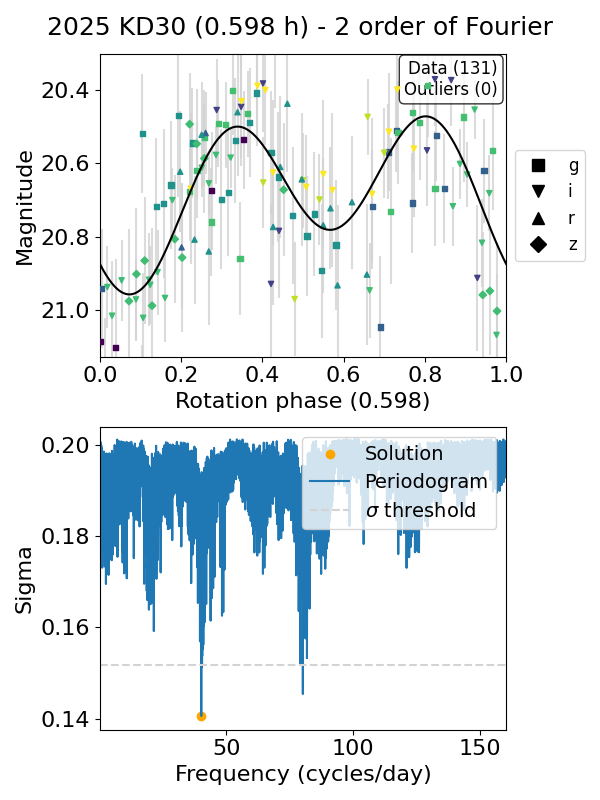} 
\includegraphics[width=0.19\textwidth]{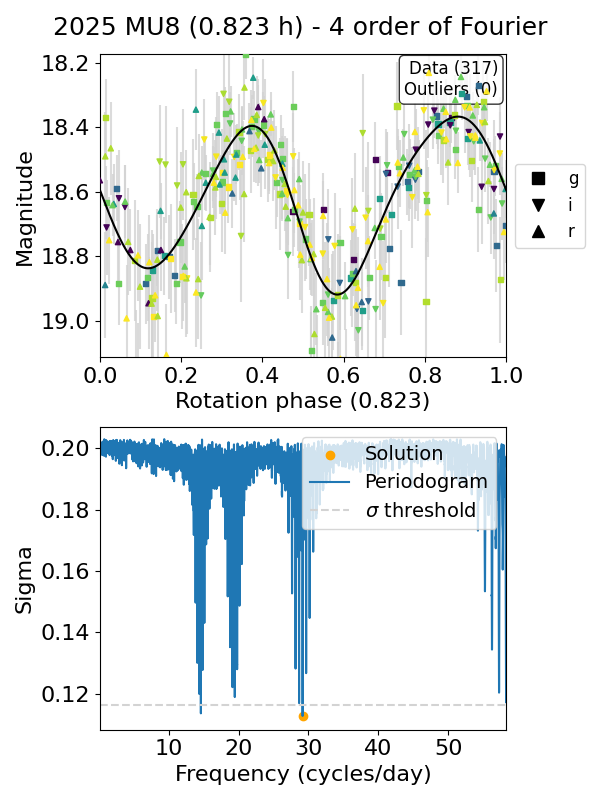} 
\includegraphics[width=0.19\textwidth]{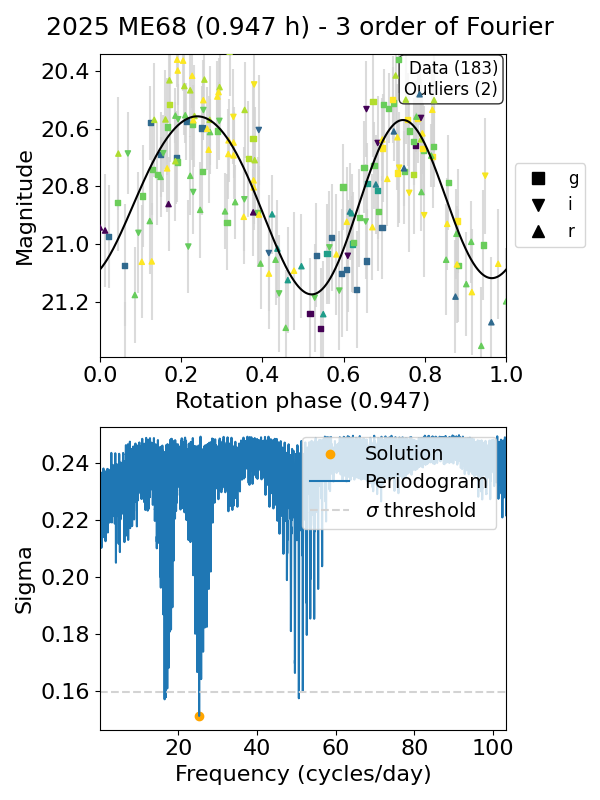}
\includegraphics[width=0.19\textwidth]{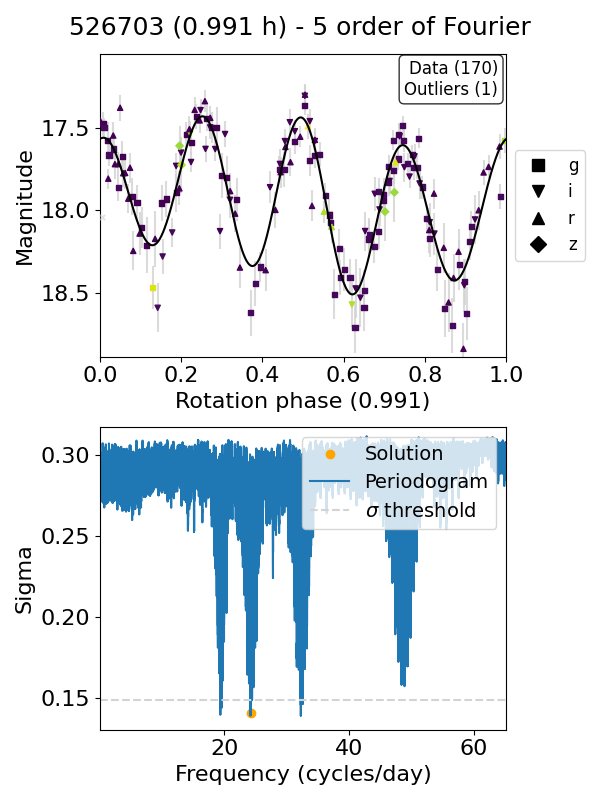} \\ 

\includegraphics[width=0.19\textwidth]{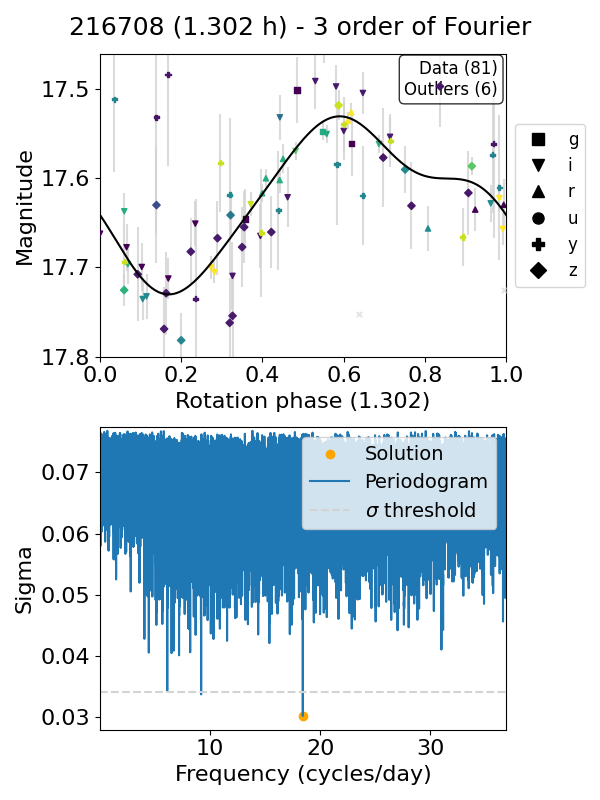}   \includegraphics[width=0.19\textwidth]{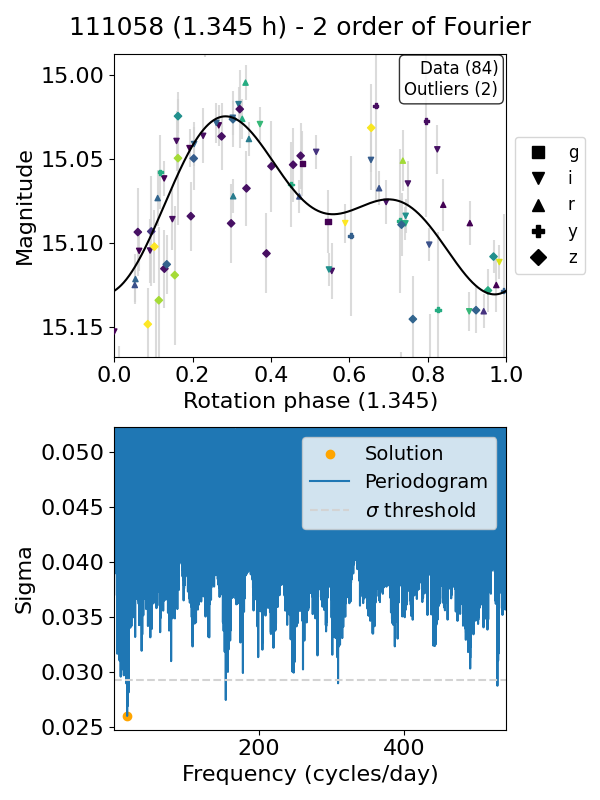} 
\includegraphics[width=0.19\textwidth]{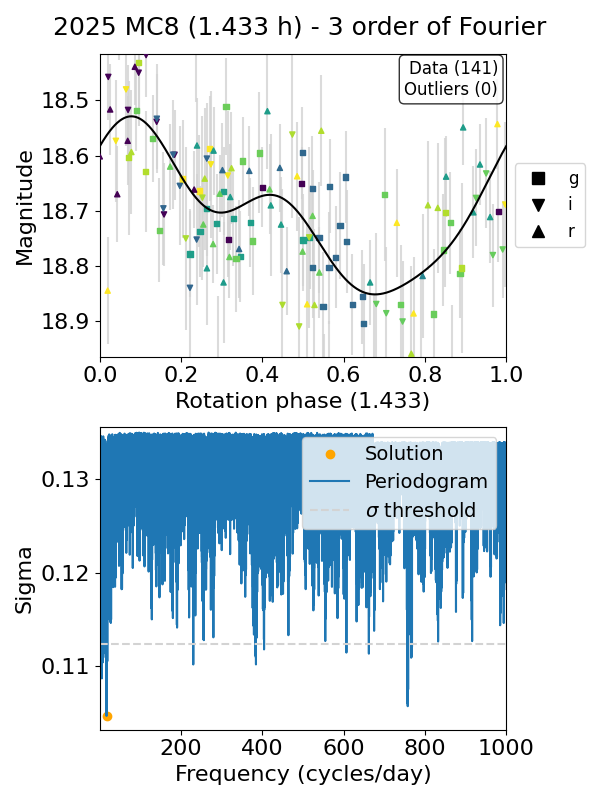} 
\includegraphics[width=0.19\textwidth]{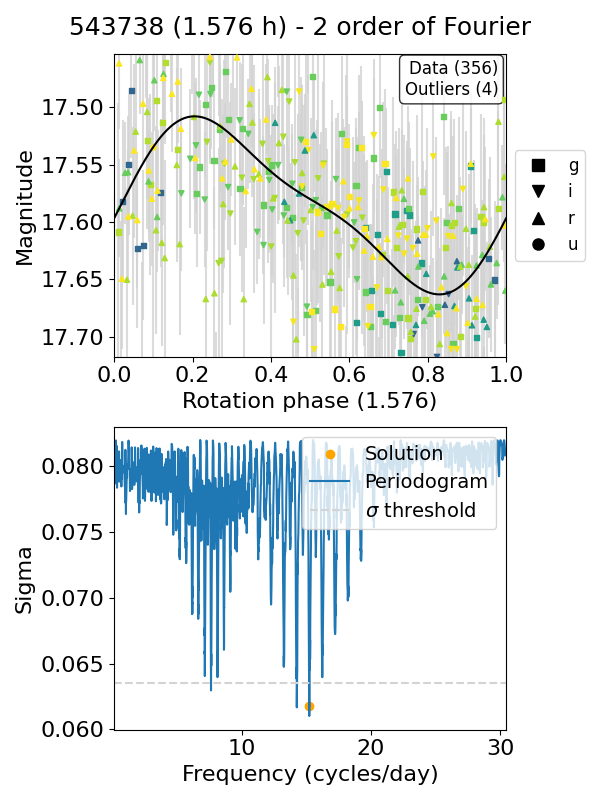}
\includegraphics[width=0.19\textwidth]{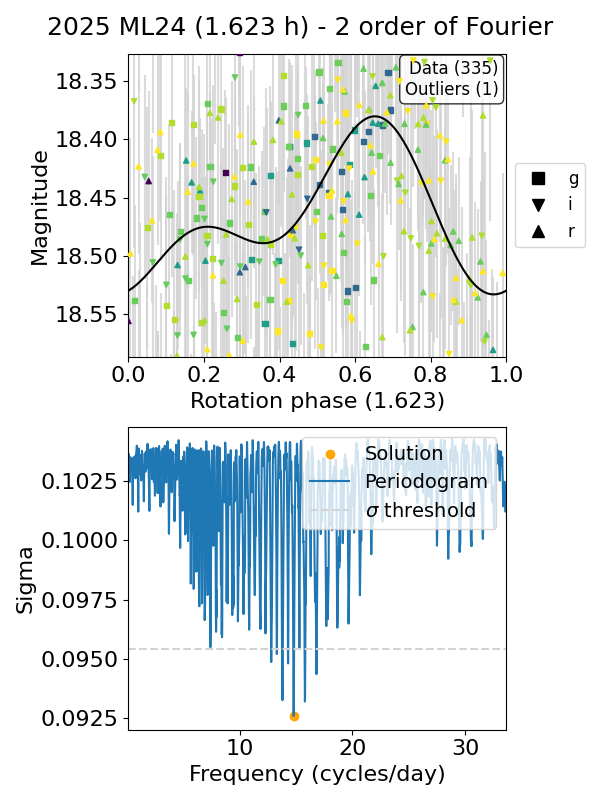} \\ 

\includegraphics[width=0.19\textwidth]{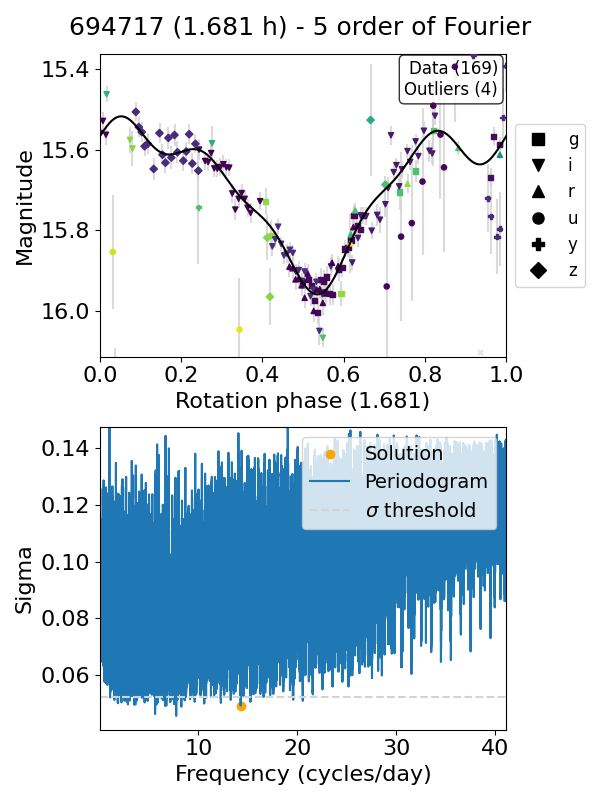}   \includegraphics[width=0.19\textwidth]{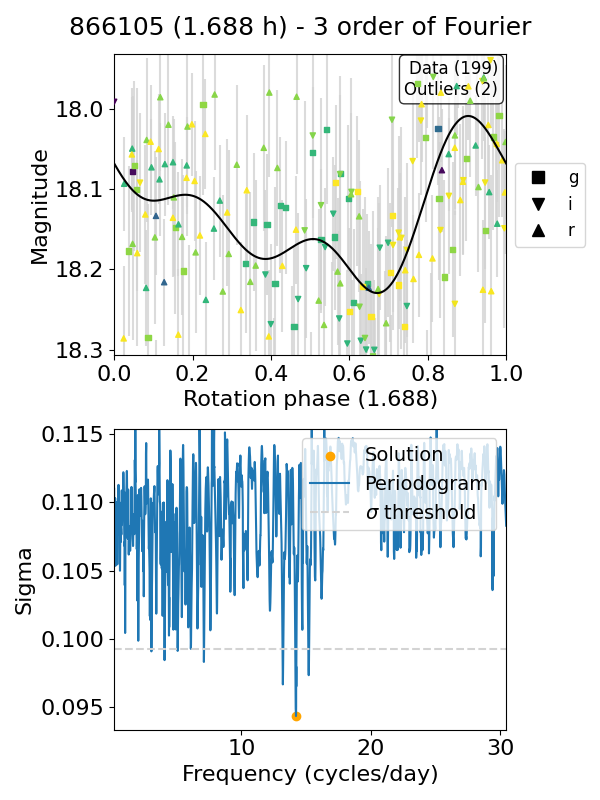} 
\includegraphics[width=0.19\textwidth]{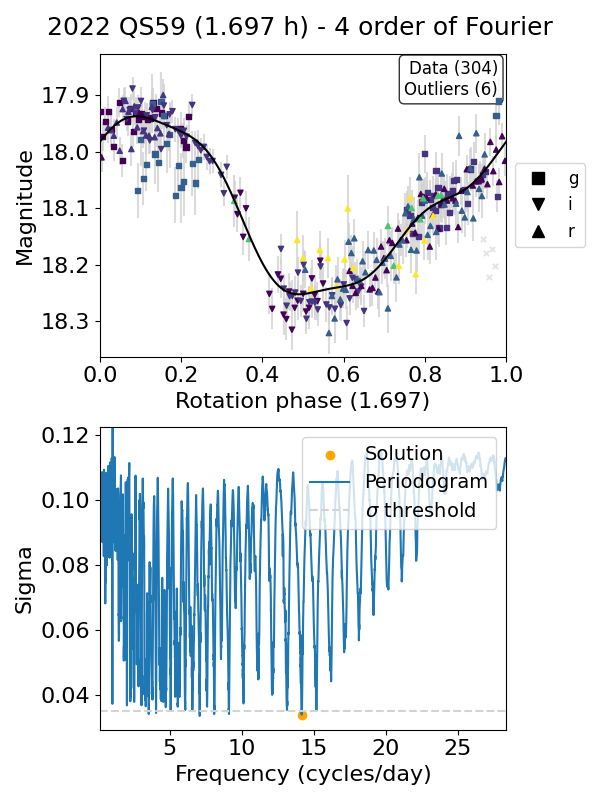} 
\includegraphics[width=0.19\textwidth]{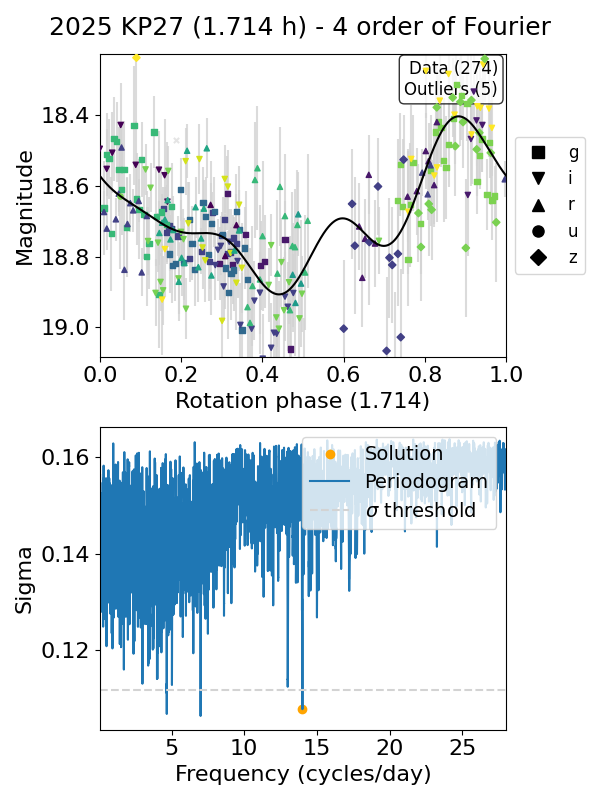}
\includegraphics[width=0.19\textwidth]{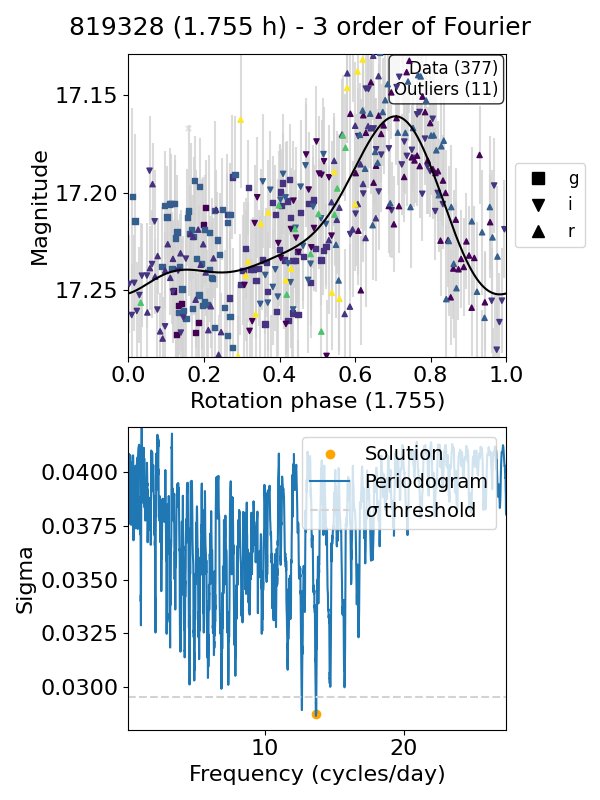} \\ 

\includegraphics[width=0.19\textwidth]{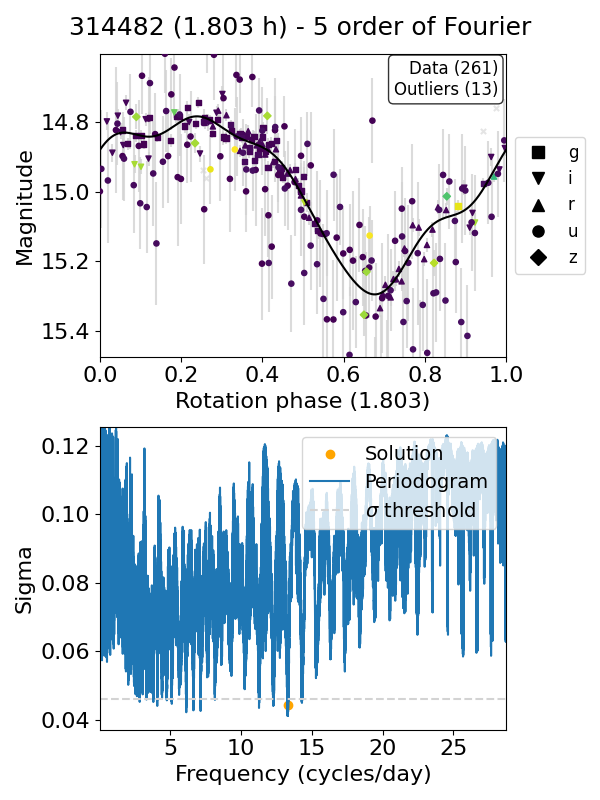}   \includegraphics[width=0.19\textwidth]{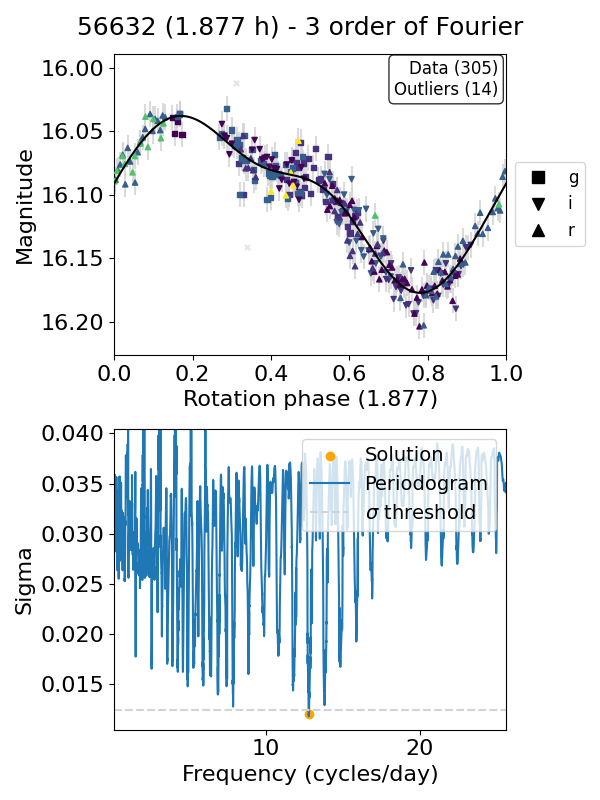} 
\includegraphics[width=0.19\textwidth]{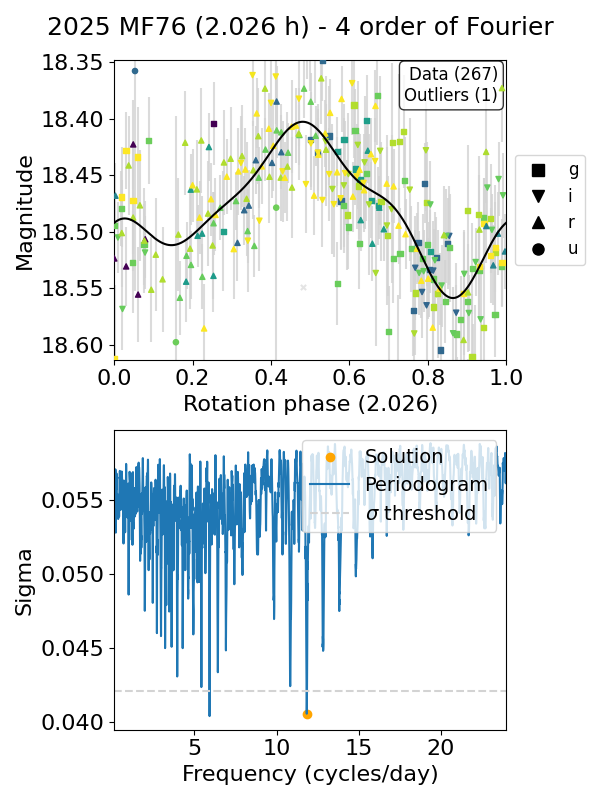} 
\includegraphics[width=0.19\textwidth]{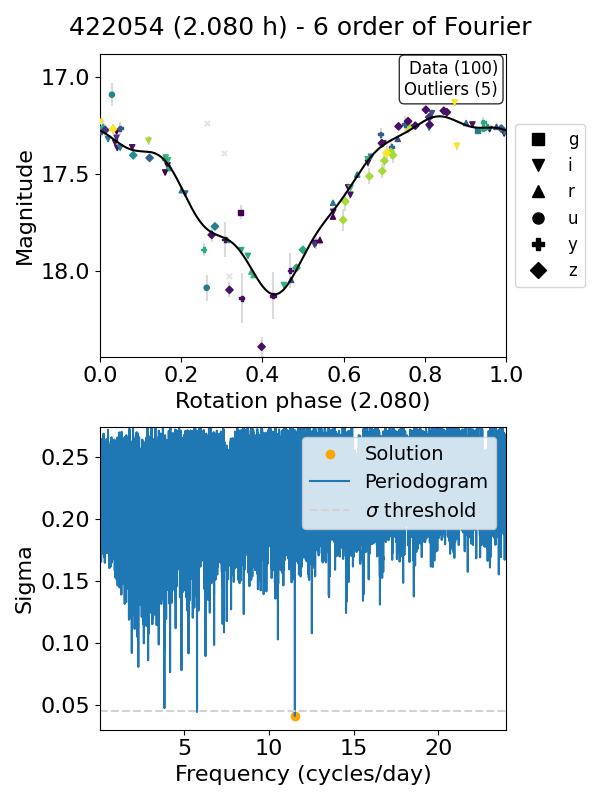}
\includegraphics[width=0.19\textwidth]{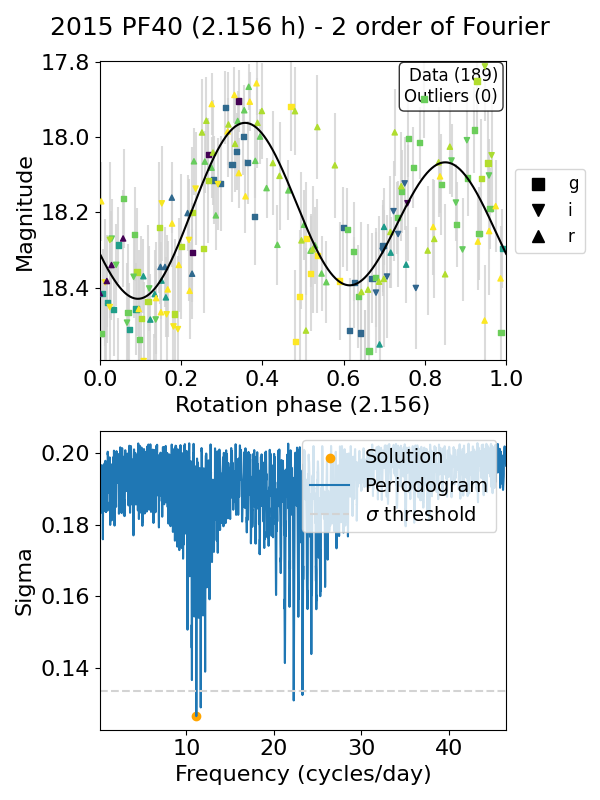} \\ 
\caption{Same as for Fig.~\ref{fig:fastests} but for 20 super-fast rotators whose periods can not be determined with 100\% certainty but have high chances of being super-fast rotators.}
\label{fig:others}
\end{figure*}

The final 20 objects of the sample have a period that is not uniquely
determined, either because a harmonic alternative survives with a
non-negligible probability or because a genuine, non-harmonic rival
remains (see Fig.~\ref{fig:others} for their light-curves and periodogramms). For 11 of them the folded light curve already hints at the
issue directly: a second local maximum is present but comparatively weak,
so that doubling the adopted period  (which would restore a light curve
with the expected number of extrema) remains a real possibility.

For six of these eleven, the adopted, shorter period is nonetheless
clearly preferred: 111058, 543738, 694717, 2022~QS59, 819328, and
2025~MF76 all beat their harmonic alternative with a probability of
$\sim$70--90\% or higher (819328, for instance, beats its doubled period
with $92\%$ probability, and 2022~QS59 beats each of its several
harmonics with $55$--$94\%$ probability). We regard the weak secondary
maximum in these cases as intrinsic to the light-curve shape rather than
as evidence for doubling, and adopt the shorter period. Two of them are
comparatively large bodies despite their fast, harmonically ambiguous
rotation: 111058 (probably a K-type, $D=2.85$~km) and, among the group discussed
next, 314482 and 56632 are larger still.

For the remaining five, the evidence favors the opposite conclusion.
\textit{{216708}} is preferred over a rival at four times its adopted
period, $5.208$~h (eighth Fourier order), with only $1\%$ probability,
i.e.\ the longer period is preferred with $99\%$ probability. However, 216708 has only 87 available observations and taking a 4 times longer rotation period would make the coverage extremely poor. We believe that the lack of observations is creating such a scenario.
\textit{{2025~MC8}} is beaten by a period twice as long ($2.867$~h,
fourth order) with $86\%$ probability. \textit{{56632}}
($D=3.24$~km, a confident C-type at $95\%$ probability, the largest
object in this subsection) is beaten by its double, at the fourth order,
with $83\%$ probability. \textit{{422054}} is beaten by its double
(eighth order, versus the sixth for the adopted solution) with $69\%$
probability. \textit{{314482}} ($D=3.17$~km, K-type) is beaten by its
double (sixth order) with a more modest $54\%$ probability -- still
favoring the longer period, but only slightly. 

The remaining nine objects each present a distinct picture.

\textit{{2025~HZ6.}} All surviving alternative periods have a probability
below $50\%$ of being preferred and are themselves shorter than
$1.1$~h, so the object remains a super-fast rotator regardless of which
solution is correct. The unusually high reduced $\chi^2$ of $309.6$
 is most likely a sign of tumbling rotation,
visible directly in the light curve, rather than of poor photometry; a
single-period Fourier fit is not expected to describe such a light curve
well, which also makes the period comparisons above less certain than
for a simple rotator. The diameter of $26$~m implied by our fit is small
but not implausible, since 2025~HZ6 is a near-Earth object.

\textit{{2025~KD30.}} A single, very close rival period, $0.594$~h
versus the adopted $0.598$~h, survives with only $10\%$ probability, so
the adopted solution is effectively unique in practice.

\textit{{2025~MU8.}} A doubled period (sixth order, versus the fourth
for the adopted solution) is preferred with $88\%$ probability, which
would imply four maxima per rotation instead of two. However, doubled period is 1.646~h, hence the object still remains a super-fast rotator.

\textit{{2025~ME68.}} The adopted solution loses to its doubled period,
fitted at the fourth Fourier order rather than the third, which is
preferred with $66\%$ probability and would imply four maxima per
rotation instead of two. Either way, since even the doubled period stays
below $2.2$~h, the object remains a super-fast rotator.

\textit{{526703.}} The adopted solution already has four local maxima
per rotation and is preferred over the two-maximum solution at the same
period. A further alternative at $1.24$~h survives with $19\%$
probability; being a 5:4 harmonic of the adopted frequency, it would
imply five maxima, and, like the adopted solution, is itself fast enough
to remain a super-fast rotator. The object is also notable for its
elongation, $a/b\simeq2.13$, unusually large for this sample, though its
taxonomic class is not securely determined (X-type, probability $0.61$).

\textit{{2025~ML24.}} The light curve has two maxima. Two genuine,
non-harmonic alternatives survive with small probabilities, $1.521$~h
($21\%$) and $1.740$~h ($18\%$), together with a weaker harmonic at
$3.483$~h (third order, $18\%$); none of these probabilities is high
enough to be considered established, but neither do they overturn the
adopted solution decisively.

\textit{{866105.}} This object has by far the largest number of
surviving candidate periods, but all are clearly disfavored, with the
adopted solution winning against each of them with probabilities starting from
$0.84$; we consider its period well established despite the crowded
periodogram.

\textit{{2025~KP27.}} The two surviving alternatives, at the fifth and
sixth Fourier orders, would imply substantially more structure in the
light curve than currently observed and are both preferred with lower
probability than the adopted solution, which we therefore retain. Its
taxonomic classification is essentially split between C- and B-type
($60\%$ and $40\%$), the most ambiguous case among the objects discussed
in this section.

\textit{{2015~PF40.}} The single surviving alternative, at $2.063$~h, is
disfavored (the adopted solution wins with $87\%$ probability) and, being
itself shorter than the adopted period, would not change the object's
status as a super-fast rotator even if it were correct.

\subsection{The need for a variable Fourier order}
\label{sec:order_need}

\begin{figure*}
\centering
\includegraphics[width=0.3\textwidth]{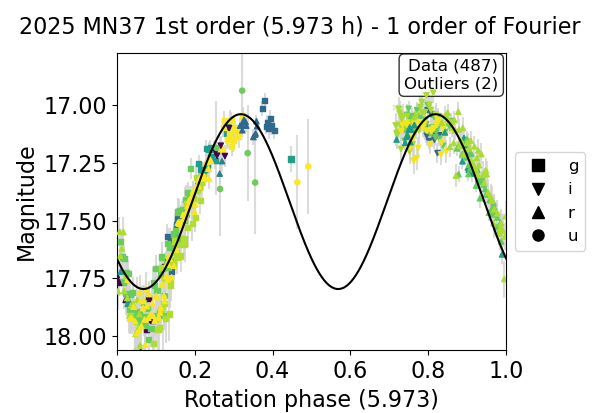} \includegraphics[width=0.3\textwidth]{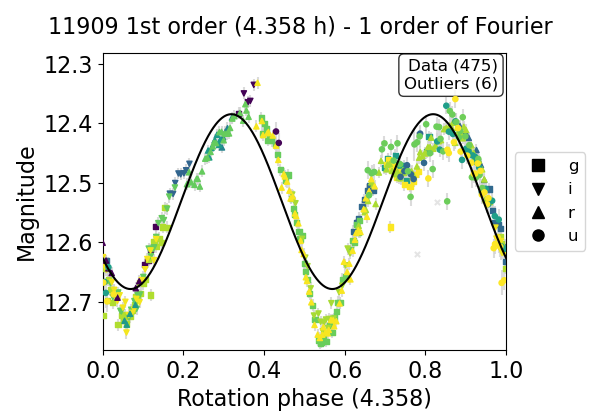}
\includegraphics[width=0.3\textwidth]{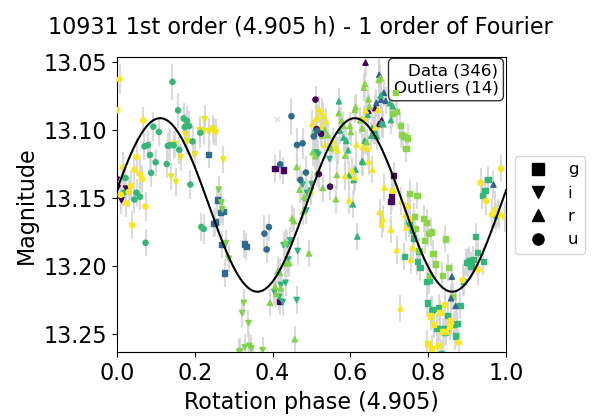} \\
\includegraphics[width=0.3\textwidth]{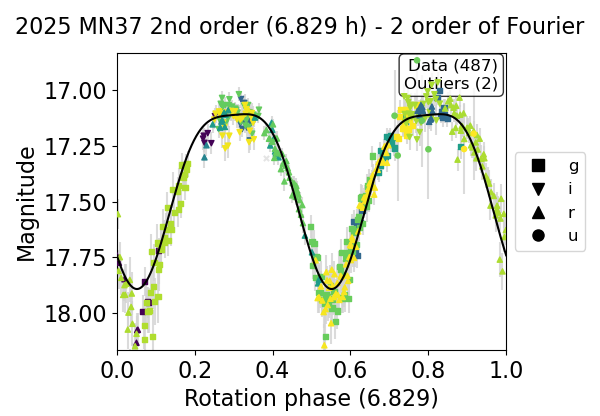} \includegraphics[width=0.3\textwidth]{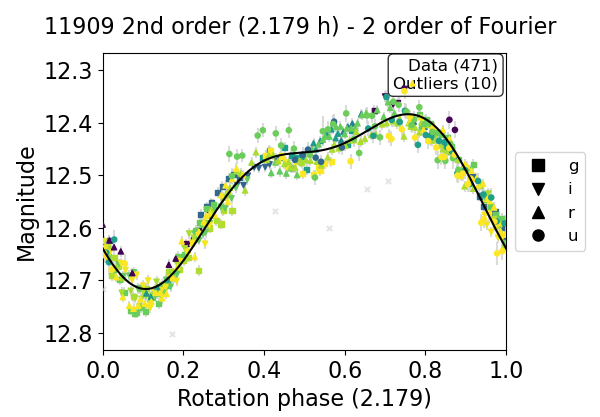}
\includegraphics[width=0.3\textwidth]{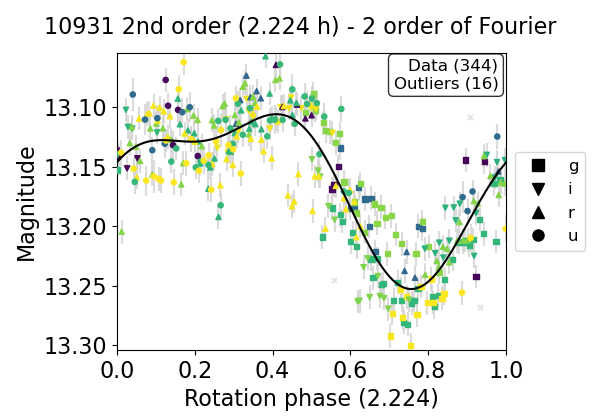} \\
\includegraphics[width=0.28\textwidth]{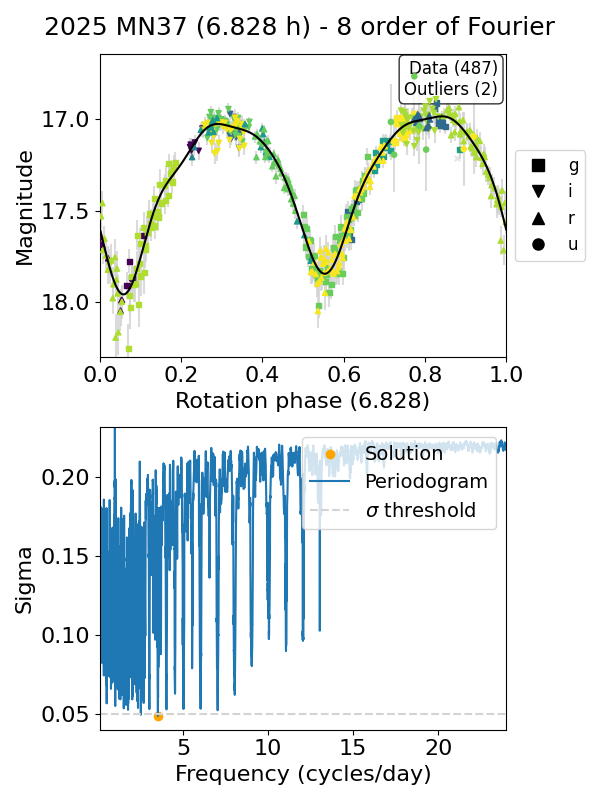} \includegraphics[width=0.28\textwidth]{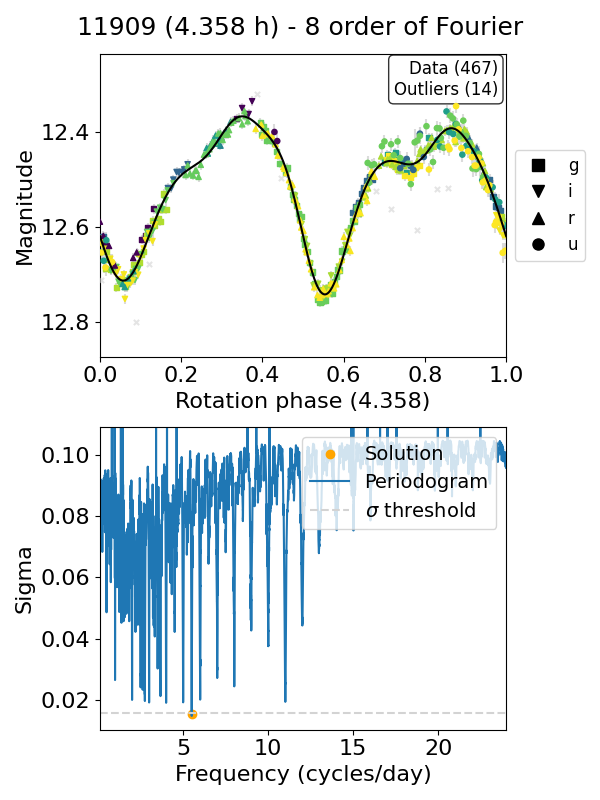}
\includegraphics[width=0.28\textwidth]{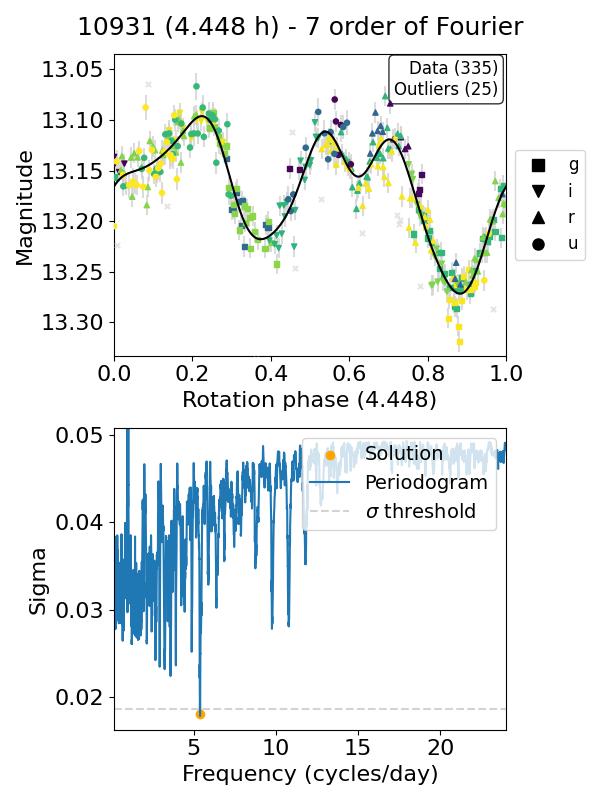}

\caption{Same as for Fig.~\ref{fig:fastests} but for 3 examples where the low-order Fourier fit mismatches the true result, with two extra rows added for each object: from top to bottom, the light curve folded at the first-order period, the light curve folded at the second-order period, the light curve folded at the adopted period, and the periodogram of the adopted
solution. For asteroid 180611 the first order gives the wrong period, while the second recovers it correctly; for asteroid 11909 the situation is reversed; for asteroid 10931 both the first and second orders give incorrect periods. }
\label{fig:mismatch}
\end{figure*}

A central feature of our method is that the Fourier order is not fixed in
advance but selected for each object (Sect.~\ref{subsec:diff_order_selection}). This distinguishes our approach from the classical Lomb--Scargle
periodogram \citep{1976Ap&SS..39..447L, 1982ApJ...263..835S}, which fits a
single sinusoid at each trial frequency and is therefore equivalent to a
first-order Fourier model. As the examples below show, this lowest order
returns the wrong period for a large fraction of the best-observed
objects, so a plain Lomb--Scargle analysis cannot be relied upon for this
kind of data.
It is therefore natural to ask how often the adopted period actually differs
from the one that a low, fixed order would return. Among the 260 objects with a
unique period and full, group-confirmed coverage -- the cleanest part of the
sample, where the answer is least ambiguous -- the adopted period differs from
the first-order (single-cosine) solution for 88 objects (39 of which the second order gives the correct answer), from the second-order
solution for 68 (for 19 of which 1st order is correct), and from both for 49. In other words, for a large fraction of
even the best-observed objects a fit restricted to one or two harmonics would
return the wrong period.

The reason is that the quality of the data controls how much light-curve
structure can be resolved. When the photometry is dense and precise, the light
curve reveals secondary extrema and departures from a simple double-peaked shape
that a low-order fit cannot represent; the low-order model then locks onto a
harmonic of the true period instead of the period itself. This is not a failure
of the low order at fixed data quality but a consequence of the data being good
enough to show more than a low order can describe. We illustrate this with three
objects (Fig.~\ref{fig:mismatch}), chosen so that in one the first order
misses while the second recovers the period, in one the first recovers while the
second misses, and in one both miss.


\textit{2025~MN37.} At first order, using photometry that covers only $70\%$ of
the rotational phase, the fit follows the period-doubling rule and returns
$5.97$~h (twice the single-cosine period); this period, however, is not a
harmonic of the true one at all. The second order already recovers essentially
the correct value, $6.83$~h, again with two maxima per rotation -- the same
number of maxima the doubled first-order solution nominally has. Higher orders
confirm this, and the adopted eighth-order solution gives $6.828$~h. Here the
first order misses but the second succeeds. The object's spectral class is not
fully resolved; it is tentatively an L-type ($0.61$) of about $0.89$~km, with
$H=17.70$. Its reduced $\chi^2$ of $1.18$, close to unity, confirms that the fit
is good and that the period itself is secure.

\textit{11909.} The situation is reversed. The first order, being a coarse
single-cosine approximation, doubles its period to $4.36$~h and happens to land
on the correct value. However, the fit doesn't look good and the amplitude computed from it is underestimated. The second order, however, already resolves two maxima and
therefore does \emph{not} double, returning $2.18$~h -- half the true period. Only
at higher orders does the fit distinguish the two unequal maxima and settle again
on $4.36$~h, adopted at eighth order. Here the first order succeeds by
coincidence and the second fails. The object is a C-type ($0.94$) of about
$16$~km, with a reduced $\chi^2$ of $2.08$.

\textit{10931.} In this case both low orders miss. The first order returns
$4.90$~h and the second $2.23$~h, while the true period is $4.45$~h. The
light curve has four maxima per rotation, two of them close together, which only
a higher-order fit resolves; the adopted seventh-order solution recovers
$4.45$~h. The object is a C-type ($0.84$) of about $11$~km, with a reduced
$\chi^2$ of $3.14$.


These examples show why the Fourier order cannot be fixed at a low value. A
low-order fit can miss the true period in two distinct ways. The first is an
error in the harmonic multiple: a single-cosine fit has only one maximum and is
doubled by construction, so it recovers the period only when the light curve is
nearly symmetric, while a low-order fit that resolves fewer maxima than the curve
actually has -- two instead of three or four, for example -- treats several cycles
as one and settles on a fraction of the true period. The number of maxima a fit
can represent is limited by its order, so the correct multiple emerges only once
the order is high enough to resolve every extremum of the curve. The second way
is more basic: a low order may simply lock onto the wrong frequency altogether, as
for 2025~MN37, whose first-order period ($5.97$~h) is not a harmonic of the true one
($6.83$~h) at all, because the coarse single-cosine model fits a different
periodogram minimum. Neither type of error is visible without going to higher
orders, and which one occurs depends on the detailed shape of each light curve.
Selecting the order object by object, and letting the resolved structure of the
curve set the period, is therefore necessary to recover the correct rotation from
well-sampled data.

The same effect matters for the most interesting new object of this
sample, 491384 -- a body of more than 2~km rotating in under five minutes
(Sect.~\ref{sec:veryfast}). For this asteroid the first-order fit returns
a period twice as long as the true one, and only a higher-order model
recovers its genuine, ultra-fast rotation. A fixed low-order analysis,
including a plain Lomb--Scargle search, would therefore have missed
exactly the kind of object this study is designed to find.

\subsection{Flagging problematic visits}
\label{sec:baddata}

As described in Sect.~\ref{sec:methods}, our pipeline iteratively rejects
observations that deviate from the fit by more than $\theta=3$ estimated
standard deviations. Most such rejections are presumably genuine outliers
specific to a single asteroid's photometry, but some may instead reflect
a problem with a particular Rubin visit itself -- a cosmic ray, a
calibration issue, clouds, or some other instrumental artifact affecting the
exposure as a whole. If a visit is at fault, this should show up as
several unrelated asteroids independently rejecting an observation taken
in that same visit, and we use this signature to flag visits worth
treating with caution.

For this cross-check we restrict the underlying sample to objects whose
light-curve coverage exceeds $80\%$, whose reduced $\chi^2$ exceeds
$0.7$ (excluding solutions that overfit the noise, in the same spirit as
the floor of Eq.~\ref{eq:chi2floor}), and whose period uncertainty
interval lies within a factor of two of the adopted period on either
side, as in the base filter of Sect.~\ref{sec:credibility}. This leaves
$1{,}221$ objects. The logic mirrors that used elsewhere in this paper
for quantities other than the period itself: a reasonably narrow
uncertainty interval and adequate phase coverage are enough to trust a
fit's other outputs even when the period is not perfectly pinned down (as
in Sect.~\ref{sec:taxonomy} for the colors), and here the same holds for
the fit's rejected observations. We do not exclude objects with a very
high reduced $\chi^2$, such as those discussed in
Sect.~\ref{sec:uniqueperiod}: an inflated $\chi^2$ correspondingly
inflates the fit's own estimated scatter, which raises the effective
rejection threshold rather than lowering it, so such solutions are, if
anything, more conservative in what they flag rather than less.

Within this sample we identify, for every rejected observation, the
Rubin visit it belongs to, and count how many distinct asteroids reject
an observation from that same visit. To further guard against picking up
ordinary statistical fluctuations, we only count a rejection here if the
observation deviates from the fit by more than $4\sigma$, appreciably
stricter than the $3\sigma$ threshold used by the pipeline itself. For
every visit flagged in this way we record the epoch (JD), the
photometric band, and the designations of the asteroids that rejected an
observation from it; the full table is made available on
Zenodo\footnote{Link to be added.}.

One visit stands out, with $58$ asteroids independently rejecting an
observation from it. More broadly, $127$ visits have more than $10$
asteroids rejecting an observation, $491$ have more than $5$, and
$1{,}415$ have at least $2$. We stress that this list is not exhaustive
-- a visit can certainly be problematic without being flagged here, for
instance if too few asteroids happen to have been observed in it -- but
we recommend particular caution when using data from the visits most
heavily flagged in this way.

\section{Conclusions}
\label{sec:conclusions}

In this work we identify the hereto largest known ultra-fast rotating asteroid (491384) from NSF-DOE Vera C. Rubin Observatory Data Preview 2 (DP2). This novel result was enabled by a substantially improved version of the varying-order Fourier method of \citet{2025A&A...693A..66V}, the same approach that led
to the original discovery of ultra-fast rotators in the Rubin First Look by
\citet{greenstreet2026}. The main changes in the spin period determination concern model selection. Within
a given Fourier order, the best-fitting frequency is now chosen by a
Bayesian information criterion that includes a clustered rejection
penalty of observations, rather than by the residual scatter alone, so
that a model can no longer lower its apparent quality simply by
discarding inconvenient points. Across different orders and frequencies,
nested models (same frequency, different order) are compared with the
classical F-test, non-nested models (no shared frequency content) with a
bias-corrected Vuong test, and the intermediate, overlapping case (models
sharing only part of their frequency content) with a further,
$\chi^2$-based variant of the same test -- all three accounting for both
the difference in the number of free parameters and the difference in
the number of rejected observations. Beyond the single adopted solution,
the pipeline identifies the surviving alternative periods and reports
the probability that the adopted solution is preferred over each of
them, which defines a formal uncertainty interval and a two-axis
credibility scheme that grades every solution by the reliability of the
period and by the phase coverage of the observations, together with a
simple overfitting check based on the reduced $\chi^2$. We also extended
the resulting color analysis by propagating the full covariance matrix of
the fitted band magnitudes into the taxonomic classification with
\texttt{classy}, which removes most of the common uncertainty introduced
by the poorly constrained phase-function slope on our short observing
arcs.

We applied this method to the Rubin Observatory Data Preview 2 (DP2)
catalog \citep{10.71929/rubin/3382528, RTN-115}, supplemented with observations published in the Asteroid Institute
(B612) archive to cover objects and epochs not yet included in DP2,
selecting all asteroids with at least 80 photometric measurements. 
Excluding isolated, information theoretically irrelevant clusters of observations
(Sect.~\ref{sec:methods}), yielded a sample of 9,417 asteroids. Of
these, 1,221 passed our base reliability filter. For 273 of the former we could determine a
taxonomic class with a confidence above 90\%: 215 C-type,
30 S-type, 7 L-type, 6 each of V- and Q-type, 4 A-type, 3 X-type, and one
each of B- and R-type.

Among the reliable solutions for spin period, our pipeline identifies 66 super-fast
rotators with a period shorter than the $\approx2.2$ hour spin barrier -- compared
to 1,487 objects with a nominal period below this limit. A total of 899 asteroids had spin periods below
five minutes. However, after strict reliability filtering was applied, these numbers were reduced to the aforementioned 66 super-fast rotators and 5 ultra-fast rotators. We group these 66
objects by the strength of the evidence supporting our estimates of their spin period: 14 rotate in
under 16~minutes, a further 32 have a uniquely determined period, and the
remaining 20 retain some ambiguity, most often a competing harmonic of
comparable or higher likelihood. Two objects among the fastest are
particularly striking. 491384 completes a rotation in only $4.18$~minutes
and, with an estimated size of $2.94\times2.40$~km ($D\simeq2.53$~km,
confidently C-type), is the largest ultra-fast rotator reported to date. Asteroid 566130 rotates in $5.34$~minutes, and is also comparably large ($2.27\times1.90$~km,
$D\simeq1.99$~km, also C-type). Neither body could survive such rapid
spin without substantial internal cohesion
\citet{holsapple2007}. Preventing centrifugal disruption requires a
cohesive strength of $\sim$2.3~MPa for 491384 and $\sim$0.8~MPa for
566130, comparable to the cohesion of lightly cemented regolith. Given
how exceptional both of those objects are, we recommend dedicated follow-up
observations to confirm their periods independently.

A central feature of our method is that the Fourier order is selected
for each object individually rather than fixed in advance
(Sect.~\ref{subsec:diff_order_selection}), and we used the cleanest part
of the sample -- the 260 objects with a unique period and full,
group-confirmed phase coverage -- to quantify how often a fit restricted
to a fixed, low order would have failed. The adopted period differs from
the first-order (single-cosine) solution for 88 objects, from the
second-order solution for 68, and from both for 49. This confirms, on a
substantially larger sample than before, that letting the order vary
from object to object is necessary to recover the correct period once
the data are good enough to resolve more structure than a low-order model
can represent. As the examples of 491384 and 216708 (among others)
illustrate, this is often precisely the case for the most interesting,
best-observed objects.

Working with real survey data at this scale also exposed a more mundane
but practically important issue: individual Rubin visits are
occasionally affected by an artifact that causes several unrelated
asteroids to independently reject an observation from the same visit
(Sect.~\ref{sec:baddata}). Among the 1,221 objects with the best-behaved
fits, we identify one visit rejected by 58 asteroids at once, 127 visits
rejected by more than 10 asteroids, 491 by more than 5, and 1,415 by at least 2.
We make the full list of these flagged visits available so that other
users of Rubin photometry can treat them with appropriate caution.

\begin{acknowledgments}
This research is supported by the National Science Foundation under grant Nos. 2307569 and 2307570 and is
partially funded by a generous gift of Charles Simonyi to the NSF Division of Astronomical Sciences. The award is made in recognition of significant contributions to Rubin Observatory’s Legacy Survey of Space and Time. Any opinions, findings, and conclusions or recommendations expressed in this material are those of the author(s) and do not necessarily reflect the views of the National Science Foundation.

This work uses the NSF-DOE Vera C. Rubin Observatory Data Preview 2 (DP2) dataset \citep{10.71929/rubin/3382528}.

This research has made use of data and services provided by the IAU Minor Planet Center and by the Asteroid Institute, a program of the B612 Foundation.

The work of S.G. is supported by NOIRLab, which is managed by the Association of Universities for Research in Astronomy (AURA) under a cooperative agreement with the US National Science Foundation.

D.E. Vavilov is grateful to his wife, Elizaveta Vavilova, whose original
suggestion to search for asteroids with minute-scale rotation periods
remains the motivation behind this work as well: without it we would not
have devoted the order-of-magnitude larger computational effort that
searching for such short periods requires, and that led to the ultra-fast
rotators reported here.

This research has made use of LTE's SsODNet VO service (https://ssp.imcce.fr/webservices/ssodnet/).

\end{acknowledgments}

\begin{contribution}

\textbf{D.E.V.}: Conceptualization (lead), Methodology, Software, Visualization, Writing -- original draft (lead). 
\textbf{S.E.}: Formal cohesion analysis, Visualization, Funding acquisition, Project administration, Supervision, Writing -- review \& editing. 
\textbf{S.G.}: Funding acquisition, Project administration, Supervision, Writing -- review \& editing.

\end{contribution}

\bibliography{sample701}{}
\bibliographystyle{aasjournalv7}
\end{document}